**Anchoring Convenience Survey Samples to a Baseline Census for Vaccine Coverage Monitoring in Global Health**

**Authors:**
Nathaniel Dyrkton,[a] Shomoita Alam,[a] Susan Shepherd,[b] Ibrahim Sana,[b] Kevin Phelan,[b] and Jay JH Park,[a,c]

**Affiliations:**

a. Core Clinical Sciences, Vancouver, BC, Canada
b. The Alliance for International Medical Action (ALIMA), Dakar, Senegal
c. Department of Health Research Methods, Evidence, and Impact, McMaster University, Hamilton, ON, Canada

**Corresponding author:**
Jay JH Park
Department of Health Research Methods, Evidence and Impact
McMaster University
1280 Main St W, Hamilton, ON L8S 4L8
+1 604-779-8240
parkj136@mcmaster.ca

## Abstract

While conducting probabilistic surveys is the gold standard for assessing vaccine coverage, implementing these surveys poses challenges for global health. There is a need for more convenient option that is more affordable and practical. Motivated by childhood vaccine monitoring programs in rural areas of Chad and Niger, we conducted a simulation study to evaluate calibration-weighted design-based and logistic regression-based imputation estimators of the finite-population proportion of MCV1 coverage. These estimators use a hybrid approach that anchors non-probabilistic follow-up survey to probabilistic baseline census to account for selection bias. We explored varying degrees of non-ignorable selection bias (odds ratios from 1.0-1.5), percentage of villages sampled (25-75%), and village-level survey response rate to the follow-up survey (50-80%). Our performance metrics included bias, coverage, and proportion of simulated 95% confidence intervals falling within equivalence margins of 5% and 7.5% (equivalence tolerance). For both adjustment methods, the performance worsened with higher selection bias and lower response rate and generally improved as a larger proportion of villages was sampled. Under the worst scenario with 1.5 OR, 25% village sampled, and 50% survey response rate, both methods showed empirical biases of 2% or less, below 95% coverage, and low equivalence tolerances. In more realistic scenarios, the performance of our estimators showed lower biases and close to 95% coverage. For example, at OR≤1.2, both methods showed high performance, except at the lowest village sampling and participation rates. Our simulations show that a hybrid anchoring survey approach is a feasible survey option for vaccine monitoring.

**Funding:** This work was supported by the Gates Foundation, Seattle, WA, USA

# Introduction

Vaccine coverage is an important indicator for global health that can be used to evaluate performance of public health programs (Cutts et al., 2016). Monitoring vaccine coverage can be resource-intensive and pose practical challenges, since they often rely on vaccine surveys being implemented based on probabilistic sampling. The World Health Organization (WHO) recommends probabilistic surveys as the gold standard for assessment of vaccine coverage (World Health Organization, 2018). However, probability surveys increasingly being replaced by non-probability surveys due to high costs and low response rates (Beaumont, 2020). This is especially given the current funding challenges in global health (Brown et al., 2023). Surveys based on convenience or non-probabilistic sampling can be more practical; however, the validity of vaccine coverage estimates obtained convenience surveys can be limited due to selection bias (Lohr, 2021).

If there is a reference probability survey of covariates, estimator bias can be corrected even in the presence of selection bias (Beaumont, 2020; Elliott & Valliant, 2017; Wu, 2022). Such hybrid approach, where statistical adjustment is applied to the convenience surveys for selection bias can be a practical way to monitor vaccine coverage given the theoretical background for selection bias correction (Beaumont, 2020; Elliott & Valliant, 2017; Wu, 2022). The use of this hybrid approach still requires careful considerations in vaccination surveys if assumptions are not met or statistical correction is not carefully done, even in very large sample sizes (Bradley et al., 2021).

The main assumption required to effectively control for selection bias induced by convenience sampling is independence between sample inclusion and outcome given the adjusted covariates (referred to as ignorable selection bias, conditional ignorability, or exchangeability) (Beaumont, 2020; Elliott & Valliant, 2017; Mercer et al., 2017; Wu, 2022). In general, it is neither possible to verify this assumption, nor adjust surveys with non-ignorable selection bias without heavy assumptions, such as instrumental variables or knowledge of the selection process (Heckman, 1979). Limited investigation has been done to evaluate the validity of statistical adjustment methods for this hybrid surveying approach in monitoring vaccine coverage in community-health settings. To our best knowledge, there has not been any investigation examining the validity of the performance of statistical adjustment methods under this hybrid survey approach to monitor childhood vaccine coverage in global health.

In this paper, we conducted a simulation study to assess the performance of two statistical adjustment methods, calibrated survey weights and logistic regression-based imputation model, for estimating the finite-population proportion of children aged 12-24 months who received at least one Measles-Containing Vaccine (MCV1) dose. We designed our simulation study based on an ongoing childhood vaccine monitoring program in rural areas of Ngouri, Chad and Mirriah, Niger. Specifically, we assessed a two-stage cluster sampling design for our simulation. In the first stage, baseline data on population-level covariates and measles vaccine coverage among children aged 12-24 months was obtained through census. In the second stage, we assumed that a follow-up data would be collected using convenience sampling. We used the census baseline data to adjust for the selection bias that exists in the follow-up survey. We examined the performance of the two statistical adjustment methods in estimation of the vaccine coverage under ignorable and non-ignorable selection biases. Our work builds on the current literature that predominantly focused on the theory of non-probabilistic surveys under ignorable selection bias by investigating the sensitivity of these statistical correction methods under various levels of non-ignorable selection bias. We benefit from the unique circumstance of having access to a past census of our entire target population, rather than a typical broad census of a superset of our target population; for example, the census of the entire country.

In the following sections, we review the objectives of the vaccination survey and the analysis of the baseline census data. We then review the sampling design, the candidate models for correcting selection bias, and the simulations for comparing our candidate models. We lastly report on our simulation findings and discuss the implications for future non-probability vaccination surveys.

## Methods

### *Overview of the study and sampling design set-up*

Performing a probability sample or an exhaustive census on such rural remote areas to obtain community-based vaccine coverages can be costly and pose practical challenges. As such, we aim to assess the performance of statistical adjustment methods for selection biases in convenience surveys that are anchored based on a census baseline survey. We aim to test these statistical adjustment methods for such hybrid survey approach that could estimate the childhood vaccine coverage within a reasonable margin of error. Our simulation study was motivated by an existing childhood vaccine monitoring program in remote areas of Ngouri, Chad, and Mirriah, Niger.

The primary endpoint is the proportion of children aged 12-24 months with at least one dose of the MCV1. Our target population is all children aged 12-24 months in the regions of Ngouri and Mirriah. The sampling unit in each region the caregiver of the child, who are nested within a total of 381 villages across both countries (196 in Chad, and 185 in Niger), where the total number of children in the given village was ≥5 in Chad, and ≥20 in Niger. We propose utilizing a two-stage cluster sampling design, where the first stage is a simple random sample of villages (without replacement), and the second stage samples caregivers within each village via a convenience sample. The convenience sample is defined as the caregivers who attend the group-based training program delivered within each village that aims to provide training on how to measure the middle-upper arm circumference (MUAC) to better monitor acute malnutrition (UNICEF, 2025). Each caregiver who attends the training would be asked verbally or via a vaccination card if their child/children are vaccinated. To anchor the estimates of our non-probability survey, a population level baseline census which includes vaccination status, demographic, and location variables was conducted in all villages in both regions between December 2024 to January 2025; see **Figure 1** for a visualization of the study schema.

### *Notation*

We generate the data from a "model-based" approach where we assume the outcome (vaccination status) is a random variable; for a discussion on the different perspectives on survey sampling methodology see Särndal et al., (1978) and Gregoire (1988). Let $Y_{ij}$ be a binary indicator with $Y_{ij} = 1$ representing whether child $i$ in village $j$ is vaccinated. Let $R_{ij}$ be an indicator variable with $R_{ij} = 1$ representing whether a child in village $j$ is included in the follow-up convenience sample. Let $\boldsymbol{X}_{ij}$ be a vector of our covariates that allow us to achieve ignorable selection bias: $P(Y_{ij}|\boldsymbol{X}_{ij}) = P(Y_{ij} = 1|R_{ij} = 1, \boldsymbol{X}_{ij})$. Our goal is to then calculate the population proportion of all children aged 12-24 months who have received MCV1 which we denote as $p$ from a finite population size $N$, which is known and assumed to be constant over time.

### *Methods to Correct Selection Bias*

There are a multitude of methods for correcting selection bias in non-probability surveys. As there are many similarities between survey and causal inference (Mercer et al., 2017), the methodologies for correcting selection bias are similar in both fields, but have slightly different names. The main methods are the use of propensity scores (sometimes referred to as design-based in survey inference) and model-based prediction approaches (Beaumont, 2020) (Wu, 2022). Wu (2022) provides a comprehensive overview of non-probability survey methodology within these two umbrella methods and the many assumptions that underpin their validity. The propensity score methods rely on adjusting the inclusion probabilities (or design weights) of individuals in the sample that were induced by the sampling design. The model-based prediction instead "imputes" or "predicts" the response variable by fitting a model relating the response variable and the auxiliary covariates. For this study, we selected one propensity score-based and one model-based prediction approach each that best aligned with the assumptions derived from the baseline census data and offered simplicity in implementation. We assumed that the demographics and characteristics of the target population would remain relatively stable between the time the baseline census was collected, and the follow-up survey was conducted. We also assume that data-linkage to the census is not possible in our convenience sample, thus we do not know which individuals attend the training and those who do not.

To better align with this assumption, it would be preferable to use summary statistics from the baseline census rather than individual level covariates from the census.

Within the propensity score-based methods, we chose the calibration weighting approach because it is relatively simple to implement in the survey package (Lumley & Lumley, 2020). It takes advantage of the survey weights known from our first-stage survey design, and it only requires the population totals. Within the model-based prediction approach, we chose the simplest option of using a logistic-regression model to impute the missing vaccination values. In both methods we did not use any finite population corrections to maintain a more-conservative estimate of the variance of our estimator. We acknowledge the lack of a finite population correction may lead to a larger-than-nominal level of coverage; especially as the sample size increases.

### *Calibration Weights*

Our first candidate method is a propensity score-based method: calibrated weighting model (also referred to as "Estimating equation based method" in Wu (2022)) (Deville & Särndal, 1992; Särndal, 2007). The general idea is that we calibrate the individual sampling weights such that the sample covariates match the totals of the population given we have ignorable selection bias.

This method was chosen because 1) we have initial knowledge of the sampling weights from the first-stage cluster sampling, and 2) it only requires the true population totals of each covariate, which are not expected to change significantly from the baseline to the follow-up survey. Other propensity score estimation methods are available, but require more granular covariate information, has complex variance estimation procedures, or can be more difficult to justify (Wu, 2022).

Our first stage design weights are $\pi_j = \frac{m}{M}$, where $M$ is the total number of villages, and $m$ is the number of villages sampled. We then assume our second stage weights to be $\pi_{i|j} = \frac{v_j}{V_j}$, where $v_j$ out of $V_j$ children attend the MUAC training in village $j$. Our final design weights are then $d_{ij} = \frac{1}{\pi_j \pi_{i|j}}$. Following Deville and Särndal (1992), we want to find weights $w_{ij}$ close to $d_{ij}$ according to some distance function $f$: $\sum_j \sum_i f(w_{ij}, d_{ij})$ such that the weights match the population totals $\sum_j \sum_i w_{ij} \boldsymbol{x}_{ij} = t_{\boldsymbol{x}}$, where $t_{\boldsymbol{x}}$ is the population totals of our vector of covariates $\boldsymbol{x}_{ij}$. There are several choices for $f$, a common choice presented by Deville and Särndal (1992) is the $\chi^2$-distance function: $f(w_{ij}, d_{ij}) = \frac{(w_{ij} - d_{ij})^2}{2d_{ij}}$.

Other distance functions for $f$ can be chosen if the corresponding minimizers $w_{ij}$ are negative, or extreme. After the weights have been calibrated, we can calculate the Horvitz-Thompson point-estimator proportions: $\hat{p}_{cal} = \frac{1}{N} \sum_j \sum_i w_{ij} y_{ij}$ (Horvitz & Thompson, 1952). The asymptotic variance is a two-stage cluster sampling extension to the variance estimate discussed in Deville and Särndal (1992) which is outlined in Chapter 8 of Särndal et al., (1992). We review this estimator in the Section 1 of the Supplementary Materials 1. The advantage of this model is that it does not require specifying a particular functional form between the outcome and the auxiliary variables.

***Model-Based Prediction***

Our second candidate method uses a model-based prediction approach (Beaumont, 2020; Elliott & Valliant, 2017; Wu, 2022). This builds a model to predict the vaccination status for all individuals in the census using the super-population framework (Elliott & Valliant, 2017; Wu, 2022; Wu & Thompson, 2020). If the selection bias is ignorable, we can generally choose any statistical model. For simplicity, we picked a logistic regression model to those who respond to the data. In this situation we ignore the cluster structure of our sampling design. This model-based estimator is an imputation estimator which aims to impute across our reference data $\boldsymbol{X}_i$. This mass imputation estimator is derived in Kim et al., (2021) for a reference probability sample, and described in Section 3 of Wu (2022) for a census; also see Chapter 5 of Wu and Thompson (2020).

Under the ignorable selection assumption we have $P(Y_i|\boldsymbol{X}_i) = P(Y_i|R_i = 1, \boldsymbol{X}_i)$, but we also make a further assumption of model transportability, where the model is consistent across the respondents and non-respondents: $E[Y_i|R_i = 1, \boldsymbol{X}_i] = E[Y_i|R_i = 0\ \boldsymbol{X}_i]$ (Kim et al., 2021). Our model of choice is a logistic regression, thus $\hat{E}[Y_i|\boldsymbol{X}_i] = \text{expit}\left(\mathbf{x}_i^T \hat{\boldsymbol{\beta}}\right)$, fit on our sample where $\text{expit}(\text{t}) = \frac{1}{1+e^{-t}}$. Using the population level covariates $\boldsymbol{X}_{i_{pop}}$ we impute the vaccination status for each member of the census and take the mean across all $E[Y] = E\left[E\left[Y_i\middle|\boldsymbol{X}_{\boldsymbol{i}_{pop}}\right]\right] \approx \frac{1}{N}\sum_{i=1}^{N} \hat{E}\left[Y_i\middle|\boldsymbol{X}_{\boldsymbol{i}_{pop}}\right]$. Note that under this framework no data-linkage is required between the census and the non-probability sample, as we are imputing all values of the census regardless of known or unknown inclusion in the non-probability sample.

The variance of this estimator is approximated using the delta-method (Dowd et al., 2014) using the superpopulation framework (Elliott & Valliant, 2017; Wu & Thompson, 2020). We use a clustered sandwich estimator in our variance-covariance matrix to account for the effect of clustering (Cameron et al., 2011). See Section 1 of Supplementary Materials 1 for details.

### *Simulation Study*

We followed the ADEMP (Aims, Data-generating mechanisms, Estimands, Methods, and Performance measures) framework to design and report our simulation study (Morris et al., 2019). Our ADEMP simulation protocol is made available as Supplementary Material 2. In brief, the aim of our simulation study was to evaluate the feasibility of two adjustment methods for selection bias under the hybrid survey approach for this child vaccine monitoring program. The assumptions used in our simulations were informed by the census data and consultation with the ALIMA team that had contextual understanding of the childhood vaccine monitoring programs in these rural areas. These included the expected participation rate of the MUAC training and the feasibility of the number of villages sampled by the field team. The analysis of the baseline census data and is available in Section 2 of Supplementary

Material 1. Although less straight forwards, the assumption for the odds ratio of the selection bias was based on previous attendance of training and current vaccination rates.

### ***Data Generating Mechanism***

We used two data generating mechanisms for the outcome model and the selection model. Our notations are as the following. Let $p_j$ be the total population of village $j$, and $d_j$ be the distance of village $j$ to a medical centre. Let be $cg_{ij}$ and $ca_{ij}$ be the gender and age (in months) of child $i$ in village $j$ respectively. Lastly, let $ag_{ij}$ and $aa_{ij}$ be the gender and the age of the caregiver for child $i$ in village $j$. To limit convergence issues in the fitting of mixed effects models, we centred and scaled the village population total $sp_j$.

### ***Outcome Model***

Let $Y_{ij}^t$ be a Bernoulli random variable representing whether for child $i$ in village $j$ at time $t$ is vaccinated. Let $t = 0$ be the baseline census survey and $t = 1$ be the follow-up convenience survey. The vaccination data is generated using a logistic mixed effects model

$$Y_{ij}^1 \sim Bernoulli(p_{ij}^1),$$

$$logit(p_{ij}^1) = \beta_0 + \beta_1 sp_j + \beta_2 d_j + \beta_3 ca_{ij} + \beta_4 aa_{ij} + \beta_5 cg_{ij} + \beta_6 ag_{ij} + logit\left(\frac{p_j^0}{2}\right) + \alpha_j,$$

where $\alpha_j \sim N(0, \sigma^2)$ is the village level random intercept. We suppose the follow-up survey takes place 6 to 12 months later, thus roughly $\frac{p_j^0}{2}$ would be the remaining village level vaccination rate measured from the census.

### ***Selection Model***

We subsequently generate the sample inclusion using a similar model. Let $R_{ij}$ be a binary random variable representing whether child $i$ in village $j$ is for the MUAC training. Then

$$R_{ij} \sim Bernoulli(p_{\text{inc } ij}),$$

$$logit(p_{\text{inc } ij}) = \gamma_0 + \gamma_1 sp_j + \gamma_2 d_j + \gamma_3 ca_{ij} + \gamma_4 aa_{ij} + \gamma_5 cg_{ij} + \gamma_6 ag_{ij} + \log(\xi)\, Y_{ij}^1 + \mu_j + \omega_i,$$

where $\mu_j \sim N(0, \tau^2)$ is the village level random intercept, $\omega_i \sim N(0,1)$, is individual random selection, and $\xi$ is the odds ratio representing the degree of non-ignorable selection bias. When the selection bias is ignorable, $\xi = 1$.

By construction, the logistic regression outcome model has an evident advantage as it is essentially a model mimicking the outcome and selection models and thus is expected to perform better. The logistic regression model allows for the inclusion of covariates to predict the vaccination rate which must also be properly specified. For example, due to the randomization of the first stage cluster sampling, the village level variables provide no benefit to the calibrated model and are not included in the calibration, whereas it does provide information to the logistic regression model being a valid predictor of vaccination. In practice, the logistic regression model relies on many more parametric assumptions than the calibrated model, requires individual level data from the census, and requires more involvement from the practitioner to properly gauge the assumptions. If the calibrated model performs reasonably well compared to the logistic regression model, then it is the clear choice for its robustness to variable and model misspecification.

***Estimands***

Our primary estimand is the finite population proportion of children aged 12-24 months who received at least one dose of the MCV1 across Mirrah, Niger and Ngouri, Chad, written as $p = \frac{1}{N}\sum_{i=1}^{N} Y_i$.

***Performance Metrics***

Let $k = 1, \ldots, n_{rep}$ index the simulation repetition, where $n_{rep} = 1000$ for each of the two methods. For each scenario, we calculated the following four performance metrics.

Bias: The estimated bias is defined as,

$$\widehat{Bias} = \frac{1}{n_{rep}} \sum_{k=1}^{n_{rep}} \hat{p}_k - p_k \, ;$$

95% coverage: The estimated coverage is defined as,

$$\widehat{Coverage} = \frac{1}{n_{rep}} \sum_{k=1}^{n_{rep}} I(\hat{p}_{0.025k} \leq p_k \leq \hat{p}_{0.975k})\,;$$

Equivalence tolerance: We define equivalence tolerance as the estimated proportion of replicates where equivalence was concluded using two one-sided tests (TOST) (Walker & Nowacki, 2011). For our simulations, we used two equivalence margins of 5% and 7.5% while controlling the overall Type I Error of 5% (two 5% one-sided Type I Error Rate). The estimated equivalence is defined as,

$$\widehat{Equivalence} = \frac{1}{n_{rep}} \sum_{k=1}^{n_{rep}} I([\hat{p}_{0.05k}, \hat{p}_{0.95k} \ \subseteq (p_k - \delta, p_k + \delta]),$$

for $\delta = 0.05, 0.075$.

Monte Carlo uncertainty of these performance metrics is discussed in Section 3 in the Supplementary Materials 1.

***Factors of the DGM***

We consider 54 different scenarios with three response rates of 50%, 65%, 80% (the average across all villages), three villages sampling rate (25%, 50%, 75%), and five odds ratios of non-ignorable selection bias (1.0, 1.1, 1.2, 1.3, 1.4, 1.5). These simulation values factors were informed with the discussion with the content experts running the childhood vaccine monitoring program from the ALIMA team. We tuned the intercept parameter of the selection model to achieve the corresponding response rates Supplementary Table 1 of Supplementary Materials 1 outlines all the factors of the DGM.

***Computation***

Our simulations were done using R version 4.5.1 (R Core Team, 2025). Packages used included dplyr and tidyr for data manipulation, ggplot2 and rsimsum for data visualisation, and lme4 for fitting the logistic mixed effects model (Bates et al., 2015; Gasparini, 2018; Wickham, 2011; Wickham et al., 2019).

The survey package was used to fit the survey design and calibrate the weights, and the sandwich package was used to use calculate the cluster robust variance-covariance matrix of the logistic regression (Cameron et al., 2011; Lumley & Lumley, 2020).

## Results

Our simulation results under the high-to-extreme levels of selection biases (odds ratio of 1.3 to 1.5) can be seen in **Table 1**. **Table 2** shows our simulation results under ignorable-to-moderate selection biases (odds ratio of 1.0 to 1.2). Under the high-to-extreme selection biases, both methods performed well across all proportion of villages sampled when the participation rate was high (0.8). With this high participation rate, both methods had an empirical coverage of 91-95% and a bias of approximately 1%. However, this performance deteriorated at lower participation rates (50%), with the empirical coverage declining to approximately 60% and the bias increasing to about 2%. Nevertheless, the proportion of estimates meeting equivalence within ±7.5% margin remained above 90% for all scenarios. Both methods performed optimally, yielding approximately unbiased estimates with 95% empirical coverage, when the participation rate was 0.8.

The village-level participation rate of caregivers to group MUAC training was shown to be an important factor for the performance of our two estimators. Under any given proportion of villages sampled or selection bias, the best performance always had the highest participation rate. For example, the estimated bias **(Figure 2)** and coverage **(Figure 3)** observed under a selection-bias odds ratio of 1.5 with an 80% participation rate were approximately equivalent to those obtained under a selection-bias odds ratio of 1.2 with a 50% participation rate.

Increasing the number of villages sampled did not noticeably affect the bias but substantially improved coverage and the proportion of equivalence by reducing the estimator's variance. For instance, **Figure 3** illustrates that the coverage decreases markedly as the percentage of villages sampled increases under higher levels of selection bias (larger odds ratios). This pattern occurs because the confidence intervals become narrower around the biased point estimates, thereby reducing overall coverage. A similar trend is visible in the equivalence test results (**Figure 4** & **Figure 5**), where the proportion of equivalence is substantially lower when fewer villages are sampled.

Both methods demonstrated strong overall performance, but the logistic regression model had slightly lower standard errors (on average 10% smaller), this led to slightly lower coverage and higher equivalence. The smaller standard errors, combined with a slightly smaller bias, yielded noticeable improvement when calculating equivalence, however this only was practically significant when 25% of villages were sampled. This follows given the logistic regression is effectively mirroring the data-generating-mechanism leading to optimal efficiency of maximum likelihood estimation.

Overall, the participation rate emerged as the most influential factor in determining optimal performance metrics. Increasing the proportion of villages sampled had variable effects, while it generally reduced estimator variance, its benefit depended on the magnitude of non-ignorable selection bias (odds ratio), and in some cases, it led to reduced coverage. Zipper plots for the coverage under all scenarios are available in the Supplementary Figures 1-12 of Supplementary Materials 1; similar plots are available for the equivalence tolerance at 5% and 7.5% in Supplementary Figures 13-36.

## Discussion

In this paper, we have used a simulation-guided hybrid survey sampling design to monitor the vaccination rates of children aged 12-24 months in rural areas of Chad and Niger. We proposed two candidate models representing the two methodological frameworks: a propensity score–based (design-based) approach and a model-prediction (model-based) approach. Our simulations suggest that both methods achieved acceptable empirical bias and coverage, and equivalence when the participation rate was high. The more conservative standard errors of the calibration model resulted in slightly higher empirical coverage, but lower equivalence performance. Based on our simulation results, we chose to use the calibration model as our primary analysis for the future survey work, as it showed robust performance while imposing far fewer assumptions. The calibration model does not require individual-level covariate information for selection bias adjustment, and it does not require a parametric modelling assumption. The calibration weighting method can also be easily extended to other sampling designs such as sampling the primary sampling unit proportional to size (Lohr, 2021).

Under most scenarios including high and extreme levels of selection bias, we can effectively get reliable estimates of the vaccination rate provided the participation rate is high, and under all scenarios with a tolerance of $\delta = 0.075$, the equivalence rate was ≥90% and often reaching 100%. Therefore, we recommend survey practitioners in global health place most of their resources on obtaining a higher participation rate from each sampling unit, rather than aiming only for an overall survey sample size (although a higher participation rate increases the sample size). This finding is consistent with other non-probability vaccination surveys where an extremely large sample size was still significantly biased, where the participation rate was low (5-10%) (Bradley et al., 2021).

Although we cannot directly measure selection bias, indices are available to approximate the degree of non-ignorable selection when estimating proportions (Andridge et al., 2019), and can be used in tandem with our simulation results to better quantify and interpret the magnitude of bias introduced. Another potential option is to use a doubly robust method which combine the model-based prediction with the calibrated design induced by the sampling design (Chen et al., 2020). These methods warrant exploration in future research. Since the motivation of using a convenience survey is to reduce costs, the optimal number of villages can be determined by minimizing total survey cost as a function of the intra-cluster correlation and assumed response rates (Aliaga & Ruilin, 2006).

There are limitations of our study and proposed design that may constrain its broader application. The primary assumption is that our census baseline survey covariates representative of our future convenience sample(s). This becomes less likely over time, especially if there are events that cause populations to shift dramatically. We also generated the data using a linear model with a simple parametric form. The logistic regression model would be most affected, but there are other non-parametric models, such as regression trees that could be explored in such scenarios (Buskirk, 2018). We benefitted substantially from the expectation that the response rates in the convenience survey would be high, which lessens the need for large corrections, extreme weights, and high estimator variance. We also simplified the selection mechanism slightly in our models, by assuming that the probability of attending is modelled for child $i$ in village $j$, when in reality, it is the caregiver, some of whom have multiple children.

Our study provides some promising results for planning surveys and non-probability surveys, specifically applied to vaccine coverage. Our simulations allowed us to assess two common selection bias correction under varying levels of plausible non-ignorable selection to assess the feasibility of conducting such a survey. The cost of running such simulations is quite small, as the computation requirements of correcting survey estimates under these methods is minimal, therefore encouraging its use in the future. The gold standard of survey sampling remains the probability sample, especially in vaccine coverage research, but this study provides grounding that a convenience sample can be sufficiently corrected to estimate vaccine under moderate assumptions.

## Conclusion

In this study we have performed simulations to guide our hybrid survey planning and assess the feasibility of accounting for selection bias that exists in convenience surveys by anchoring it a baseline census in the context of childhood vaccine monitoring in rural areas of Chad and Niger. The simulation-guided design allowed us to effectively plan our convenience survey under realistic and expert-informed parameters. Based on our simulations, we decided to utilize a calibrated weighting approach as the primary analysis, due to its simplicity and its robust ability to achieve empirically desirable results despite an assumed moderate level of non-ignorable selection bias. Our investigation provides survey designers and practitioners another tool to help plan and make inferences with non-probability survey sample.

# Tables

## Table 1: Performance metrics across 1000 simulations by each data generating mechanism factor for high levels of non-ignorable selection bias

| Village Sampling Proportion | Participation Rate | Method | Bias | Coverage (95%CI) | TOST Equivalence ±5% | TOST Equivalence ±7.5% |
|---|---|---|---|---|---|---|
| Odds Ratio for Selection Bias $\xi = 1.3$ | | | | | | |
| 0.75 | 0.8 | Calibrated | 0.007 | 0.985 | >0.999 | >0.999 |
| | | Logistic Regression | 0.007 | 0.979 | >0.999 | >0.999 |
| | 0.65 | Calibrated | 0.010 | 0.932 | 0.999 | >0.999 |
| | | Logistic Regression | 0.009 | 0.923 | >0.999 | >0.999 |
| | 0.5 | Calibrated | 0.013 | 0.854 | 0.997 | >0.999 |
| | | Logistic Regression | 0.012 | 0.853 | 0.997 | >0.999 |
| 0.5 | 0.8 | Calibrated | 0.007 | 0.955 | 0.992 | >0.999 |
| | | Logistic Regression | 0.006 | 0.953 | 0.999 | >0.999 |
| | 0.65 | Calibrated | 0.010 | 0.924 | 0.976 | >0.999 |
| | | Logistic Regression | 0.009 | 0.922 | 0.990 | >0.999 |
| | 0.5 | Calibrated | 0.013 | 0.857 | 0.922 | >0.999 |
| | | Logistic Regression | 0.012 | 0.846 | 0.966 | >0.999 |
| 0.25 | 0.8 | Calibrated | 0.007 | 0.932 | 0.762 | 0.990 |
| | | Logistic Regression | 0.006 | 0.939 | 0.872 | 0.998 |
| | 0.65 | Calibrated | 0.009 | 0.925 | 0.709 | 0.987 |
| | | Logistic Regression | 0.009 | 0.900 | 0.807 | 0.995 |
| | 0.5 | Calibrated | 0.012 | 0.892 | 0.580 | 0.970 |
| | | Logistic Regression | 0.012 | 0.876 | 0.707 | 0.989 |
| Odds Ratio for Selection Bias $\xi = 1.4$ | | | | | | |
| 0.75 | 0.8 | Calibrated | 0.009 | 0.950 | >0.999 | >0.999 |
| | | Logistic Regression | 0.009 | 0.944 | >0.999 | >0.999 |
| | 0.65 | Calibrated | 0.013 | 0.872 | >0.999 | >0.999 |
| | | Logistic Regression | 0.012 | 0.858 | >0.999 | >0.999 |
| | 0.5 | Calibrated | 0.017 | 0.714 | 0.995 | >0.999 |
| | | Logistic Regression | 0.016 | 0.701 | 0.999 | >0.999 |
| 0.5 | 0.8 | Calibrated | 0.009 | 0.960 | 0.989 | >0.999 |
| | | Logistic Regression | 0.008 | 0.937 | 0.998 | >0.999 |
| | 0.65 | Calibrated | 0.012 | 0.871 | 0.954 | >0.999 |
| | | Logistic Regression | 0.012 | 0.863 | 0.983 | >0.999 |
| | 0.5 | Calibrated | 0.016 | 0.803 | 0.897 | >0.999 |
| | | Logistic Regression | 0.015 | 0.787 | 0.952 | >0.999 |
| 0.25 | 0.8 | Calibrated | 0.009 | 0.928 | 0.762 | 0.992 |
| | | Logistic Regression | 0.009 | 0.914 | 0.861 | 0.998 |
| | 0.65 | Calibrated | 0.012 | 0.903 | 0.647 | 0.983 |
| | | Logistic Regression | 0.011 | 0.867 | 0.778 | 0.992 |
| | 0.5 | Calibrated | 0.016 | 0.839 | 0.571 | 0.957 |
| | | Logistic Regression | 0.015 | 0.838 | 0.675 | 0.989 |
| Odds Ratio for Selection Bias $\xi = 1.5$ | | | | | | |
| 0.75 | 0.8 | Calibrated | 0.010 | 0.940 | >0.999 | >0.999 |
| | | Logistic Regression | 0.010 | 0.935 | >0.999 | >0.999 |
| | 0.65 | Calibrated | 0.015 | 0.786 | 0.999 | >0.999 |
| | | Logistic Regression | 0.014 | 0.757 | >0.999 | >0.999 |
| | 0.5 | Calibrated | 0.019 | 0.610 | 0.972 | >0.999 |
| | | Logistic Regression | 0.018 | 0.583 | 0.990 | >0.999 |
| 0.5 | 0.8 | Calibrated | 0.010 | 0.914 | 0.986 | >0.999 |
| | | Logistic Regression | 0.010 | 0.902 | 0.998 | >0.999 |
| | 0.65 | Calibrated | 0.014 | 0.836 | 0.946 | >0.999 |
| | | Logistic Regression | 0.014 | 0.819 | 0.982 | >0.999 |
| | 0.5 | Calibrated | 0.019 | 0.717 | 0.838 | >0.999 |
| | | Logistic Regression | 0.018 | 0.701 | 0.906 | >0.999 |
| 0.25 | 0.8 | Calibrated | 0.010 | 0.906 | 0.716 | 0.991 |
| | | Logistic Regression | 0.010 | 0.886 | 0.839 | 0.997 |
| | 0.65 | Calibrated | 0.014 | 0.856 | 0.623 | 0.979 |
| | | Logistic Regression | 0.014 | 0.838 | 0.750 | 0.996 |
| | 0.5 | Calibrated | 0.018 | 0.817 | 0.528 | 0.953 |
| | | Logistic Regression | 0.017 | 0.794 | 0.638 | 0.981 |

**Table 2: Performance metrics across 1000 simulations by each data generating mechanism factor for ignorable to moderate levels of selection bias**

| Village Sampling Proportion | Participation Rate | Method | Bias | Coverage (95% CI) | TOST Equivalence ±5% | TOST Equivalence ±7.5% |
|---|---|---|---|---|---|---|
| Odds Ratio for Selection Bias ξ = 1.0 | | | | | | |
| 0.75 | 0.80 | Calibrated | 0.000 | 0.999 | >0.999 | >0.999 |
| | | Logistic Regression | 0.000 | 0.999 | >0.999 | >0.999 |
| | 0.65 | Calibrated | 0.000 | 0.996 | >0.999 | >0.999 |
| | | Logistic Regression | 0.000 | 0.997 | >0.999 | >0.999 |
| | 0.50 | Calibrated | 0.000 | 0.997 | >0.999 | >0.999 |
| | | Logistic Regression | 0.000 | 0.996 | >0.999 | >0.999 |
| 0.50 | 0.80 | Calibrated | 0.000 | 0.989 | 0.996 | >0.999 |
| | | Logistic Regression | 0.000 | 0.988 | 0.996 | >0.999 |
| | 0.65 | Calibrated | 0.000 | 0.989 | 0.982 | >0.999 |
| | | Logistic Regression | 0.000 | 0.988 | 0.993 | >0.999 |
| | 0.5 | Calibrated | 0.000 | 0.985 | 0.977 | >0.999 |
| | | Logistic Regression | 0.000 | 0.980 | 0.986 | >0.999 |
| 0.25 | 0.80 | Calibrated | -0.001 | 0.963 | 0.779 | 0.992 |
| | | Logistic Regression | -0.001 | 0.967 | 0.885 | 0.995 |
| | 0.65 | Calibrated | 0.000 | 0.956 | 0.731 | 0.988 |
| | | Logistic Regression | 0.000 | 0.957 | 0.839 | 0.994 |
| | 0.50 | Calibrated | 0.000 | 0.951 | 0.644 | 0.968 |
| | | Logistic Regression | 0.000 | 0.953 | 0.769 | 0.979 |
| Odds Ratio for Selection Bias ξ= 1.1 | | | | | | |
| 0.75 | 0.80 | Calibrated | 0.002 | >0.999 | >0.999 | >0.999 |
| | | Logistic Regression | 0.002 | >0.999 | >0.999 | >0.999 |
| | 0.65 | Calibrated | 0.004 | 0.991 | >0.999 | >0.999 |
| | | Logistic Regression | 0.004 | 0.991 | >0.999 | >0.999 |
| | 0.50 | Calibrated | 0.005 | 0.984 | >0.999 | >0.999 |
| | | Logistic Regression | 0.005 | 0.977 | >0.999 | >0.999 |
| 0.50 | 0.80 | Calibrated | 0.003 | 0.983 | 0.996 | >0.999 |
| | | Logistic Regression | 0.002 | 0.982 | 0.997 | >0.999 |
| | 0.65 | Calibrated | 0.004 | 0.972 | 0.990 | >0.999 |
| | | Logistic Regression | 0.004 | 0.976 | 0.998 | >0.999 |
| | 0.50 | Calibrated | 0.005 | 0.957 | 0.981 | >0.999 |
| | | Logistic Regression | 0.004 | 0.959 | 0.995 | >0.999 |
| 0.25 | 0.80 | Calibrated | 0.002 | 0.966 | 0.779 | 0.991 |
| | | Logistic Regression | 0.002 | 0.966 | 0.878 | 0.992 |
| | 0.65 | Calibrated | 0.003 | 0.946 | 0.728 | 0.992 |
| | | Logistic Regression | 0.002 | 0.956 | 0.853 | 0.992 |
| | 0.50 | Calibrated | 0.004 | 0.939 | 0.675 | 0.978 |
| | | Logistic Regression | 0.004 | 0.946 | 0.781 | 0.990 |
| Odds Ratio for Selection Bias ξ = 1.2 | | | | | | |
| 0.75 | 0.80 | Calibrated | 0.005 | 0.992 | >0.999 | >0.999 |
| | | Logistic Regression | 0.005 | 0.993 | >0.999 | >0.999 |
| | 0.65 | Calibrated | 0.007 | 0.973 | >0.999 | >0.999 |
| | | Logistic Regression | 0.007 | 0.958 | >0.999 | >0.999 |
| | 0.50 | Calibrated | 0.009 | 0.948 | 0.997 | >0.999 |
| | | Logistic Regression | 0.009 | 0.939 | >0.999 | >0.999 |
| 0.50 | 0.80 | Calibrated | 0.004 | 0.972 | 0.995 | >0.999 |
| | | Logistic Regression | 0.004 | 0.970 | >0.999 | >0.999 |
| | 0.65 | Calibrated | 0.007 | 0.955 | 0.983 | >0.999 |
| | | Logistic Regression | 0.006 | 0.957 | 0.996 | >0.999 |
| | 0.50 | Calibrated | 0.009 | 0.936 | 0.962 | >0.999 |
| | | Logistic Regression | 0.008 | 0.929 | 0.990 | >0.999 |
| 0.25 | 0.80 | Calibrated | 0.005 | 0.940 | 0.771 | 0.991 |
| | | Logistic Regression | 0.005 | 0.942 | 0.882 | 0.997 |
| | 0.65 | Calibrated | 0.007 | 0.924 | 0.685 | 0.981 |
| | | Logistic Regression | 0.006 | 0.935 | 0.813 | 0.993 |
| | 0.50 | Calibrated | 0.009 | 0.925 | 0.641 | 0.976 |
| | | Logistic Regression | 0.009 | 0.915 | 0.765 | 0.993 |

## Figures

**Figure 1: The study sampling scheme and study approach to correct the convenience survey via the demographic covariates of the baseline-census survey**

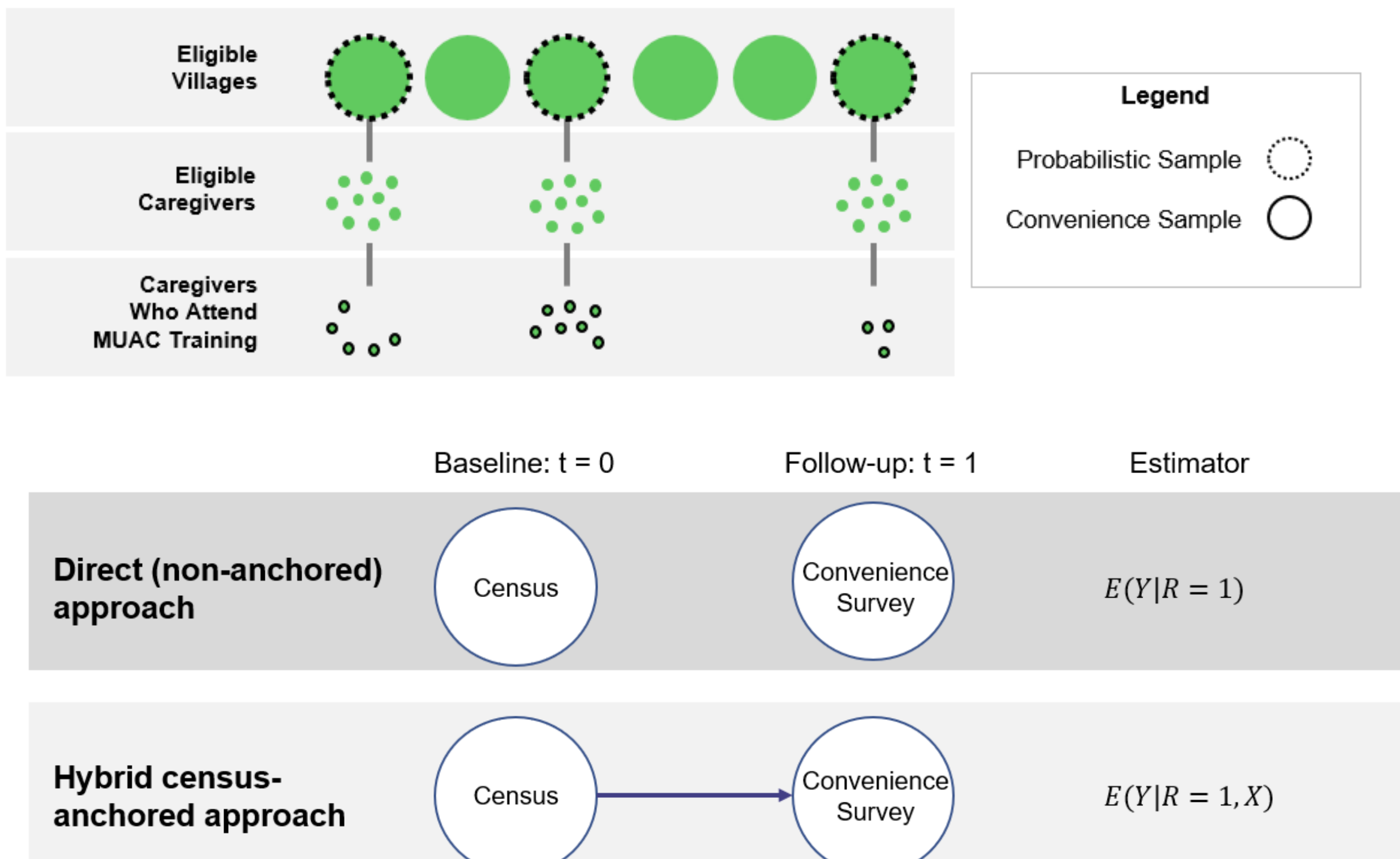

In this diagram $Y$ is the binary random variable representing vaccination status, $X$ is a set of covariates from the baseline census survey present in the follow-up survey, and $R$ is a binary variable indicating sample inclusion.

**Figure 2: Estimated bias for calibrated and logistic regression methods by percentage of villages sampled and assumed response rate**

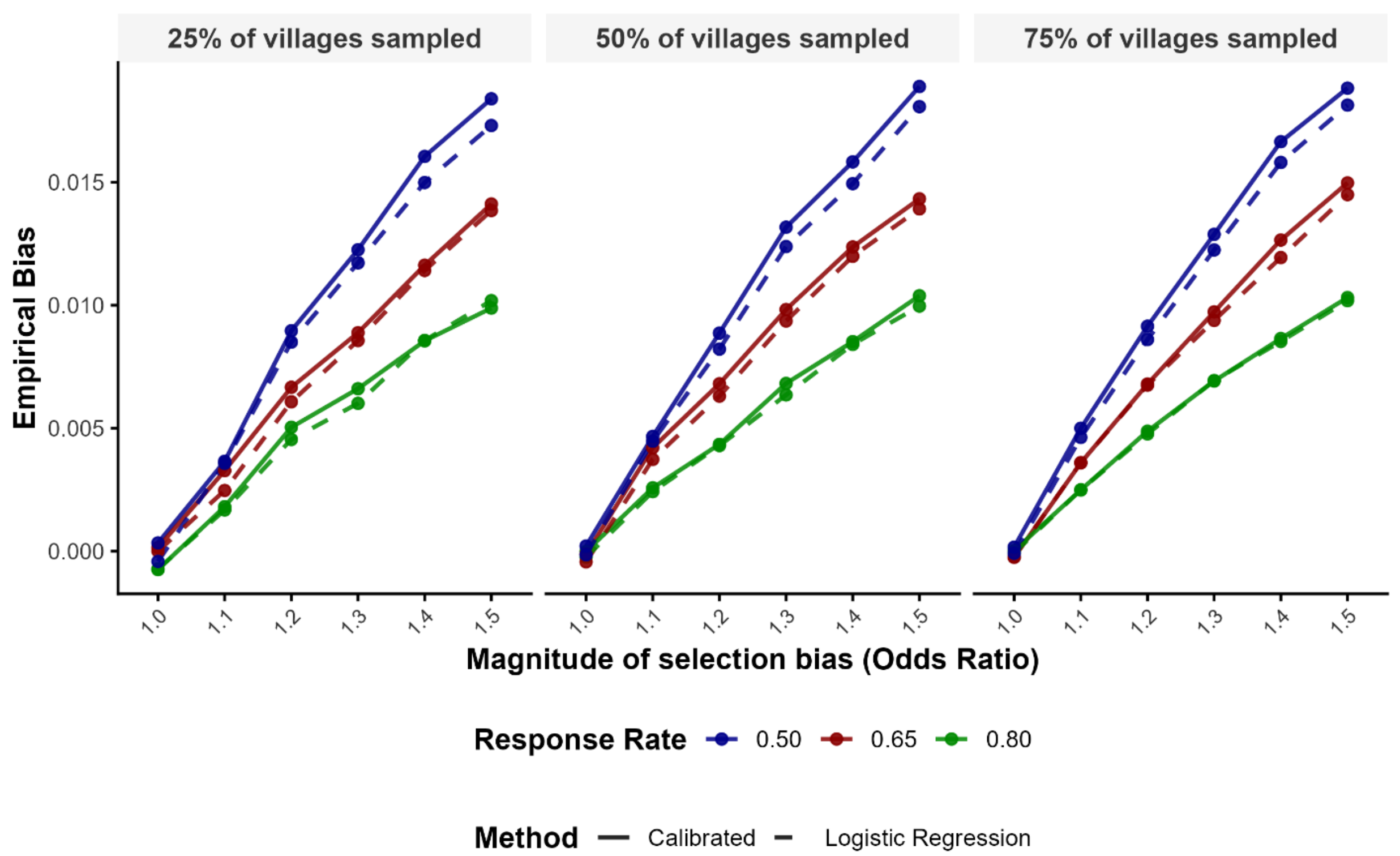

**Figure 3: Empirical coverage probability (95% CI) for calibrated and logistic regression methods by the percentage of villages sampled and assumed response rate**

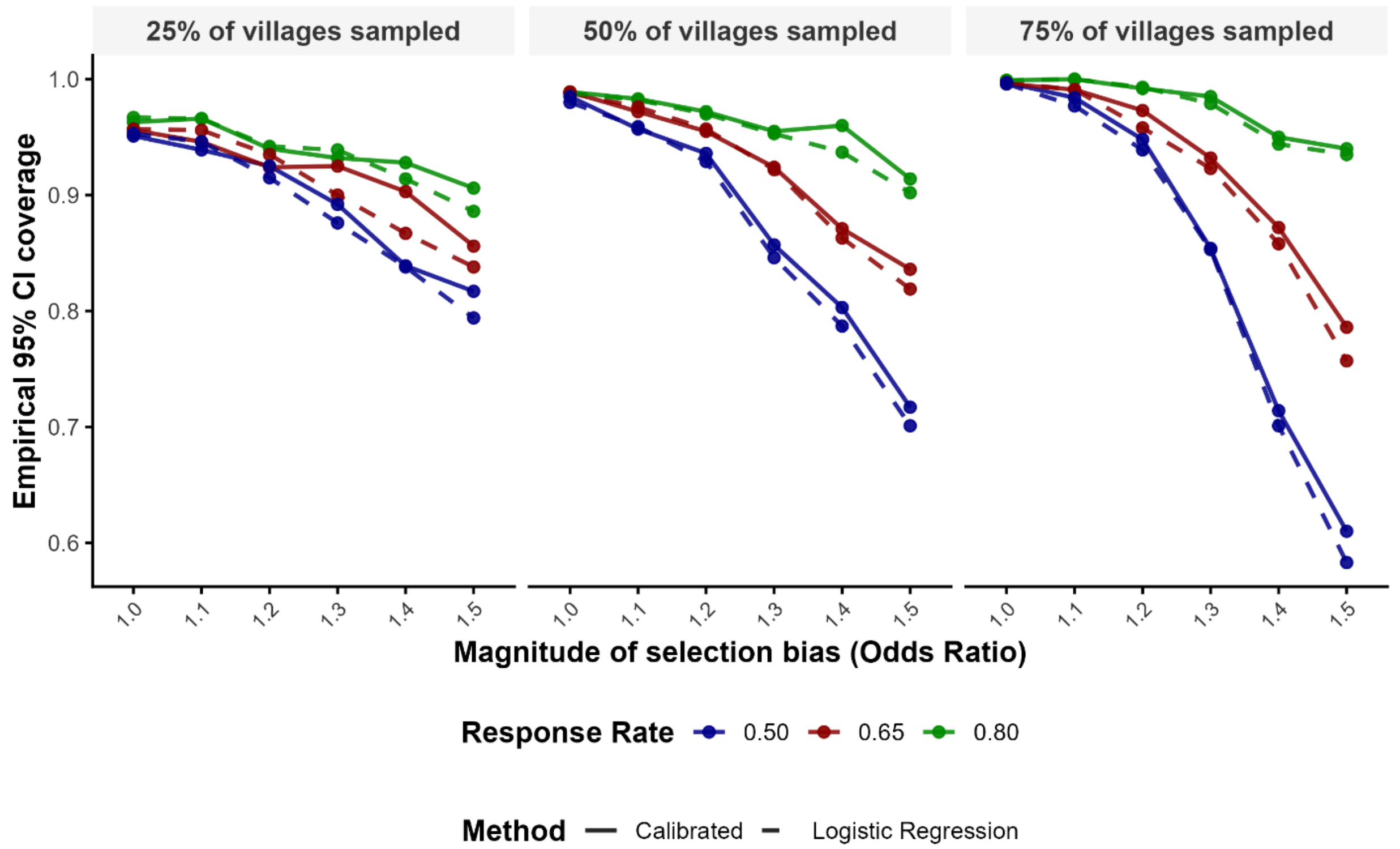

**Figure 4: Estimated probability of concluding equivalence (TOST ±5% tolerance at $\alpha = 0.05$) by the percentage of villages sampled and assumed response rate**

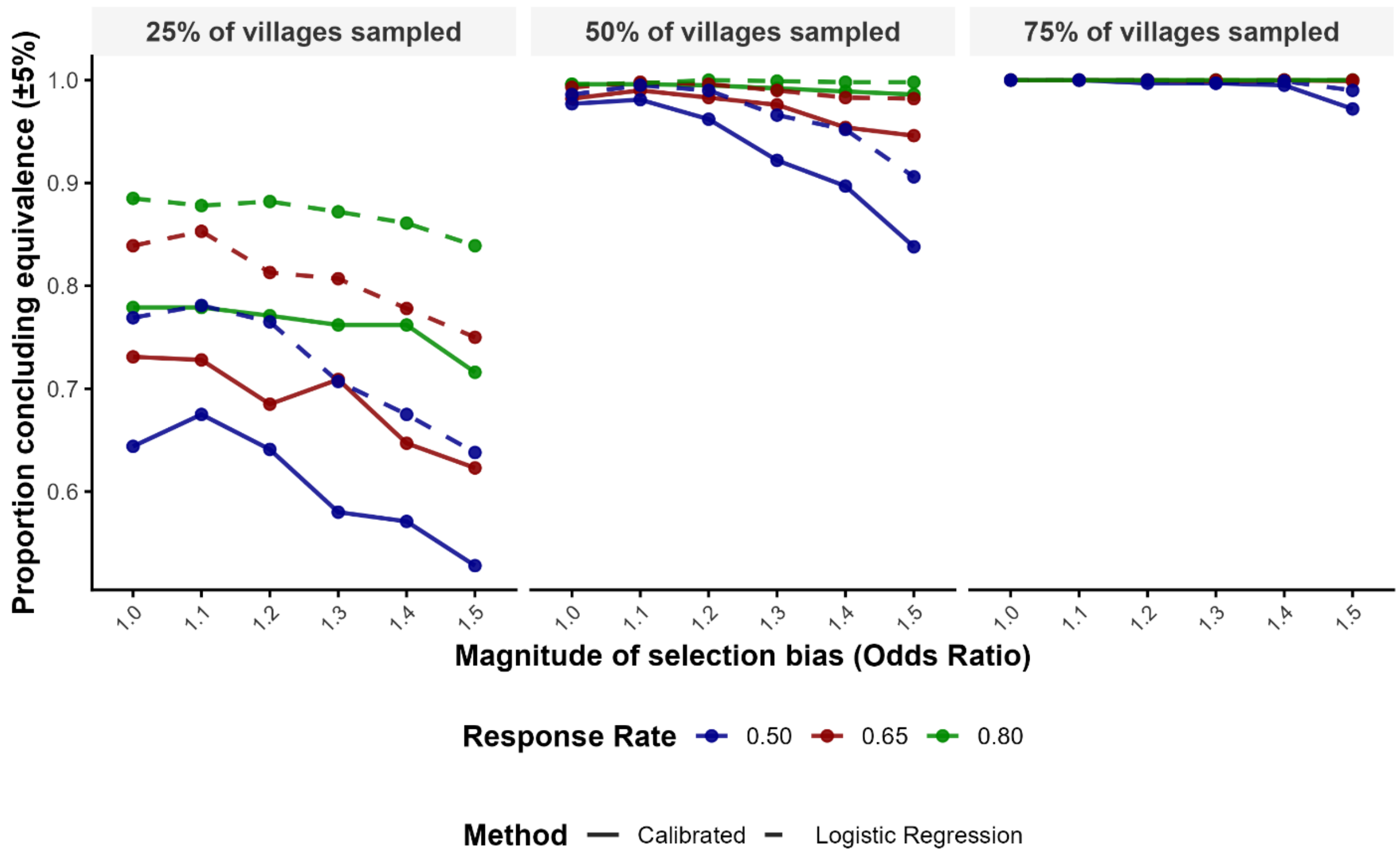

**Figure 5: Estimated probability of concluding equivalence (TOST ±7.5% tolerance at $\alpha = 0.05$) by the percentage of villages sampled and assumed response rate**

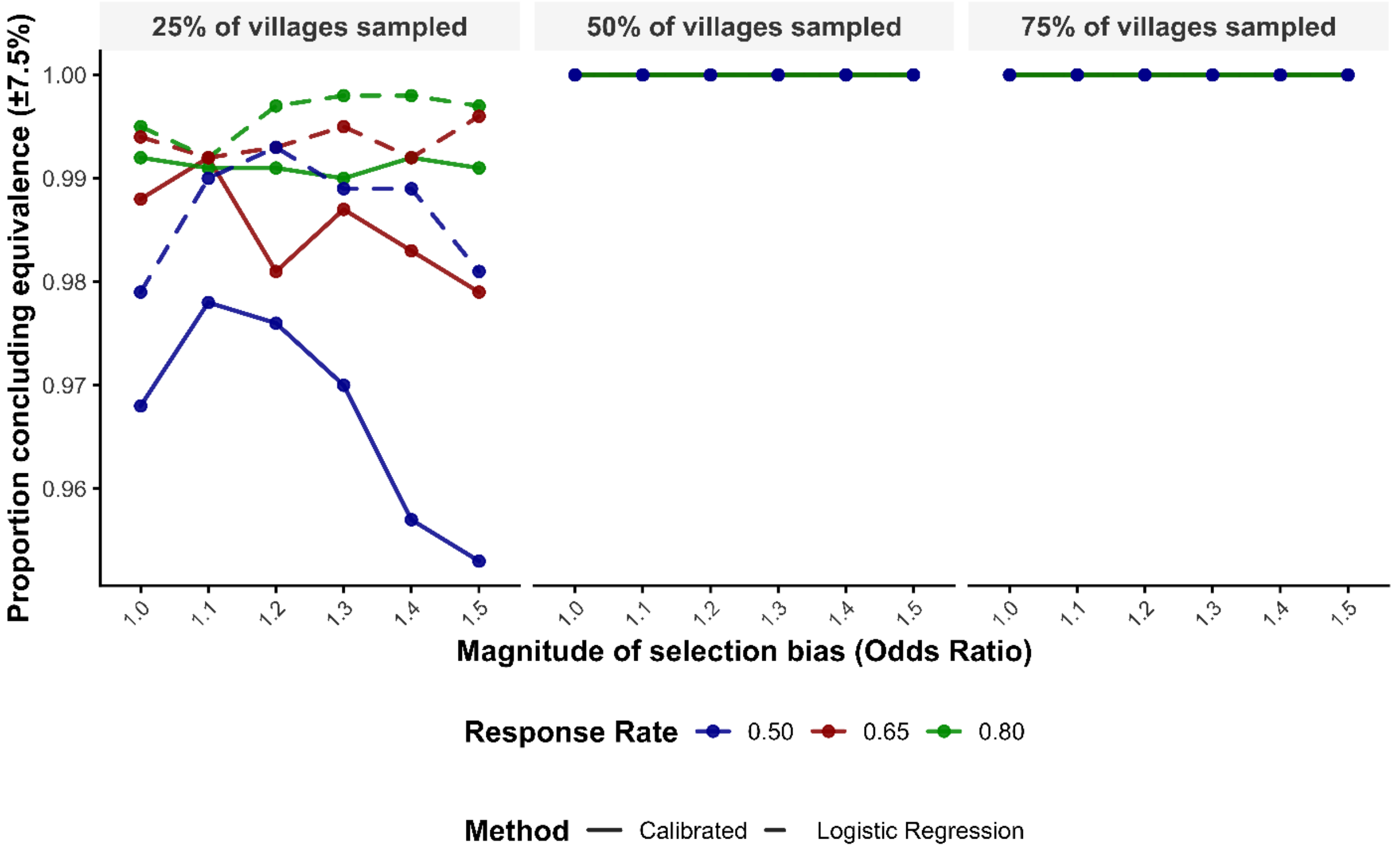

## Supplementary Materials 1:

Supplementary to "Anchoring Convenience Survey Samples to a Baseline Census for Vaccine Coverage Monitoring in Global Health"

# 1. Supplementary Methods

## *Variance Estimation for a Two-Stage Cluster Calibration Estimator*

To estimate the variance of the calibration estimator, we first must note that the distance minimization to calculate the optimal weights $w_{ij}$ discussed in this paper, is asymptotically equivalent to the model-assisted general regression estimator (GREG), but from a different perspective; see (Deville & Särndal 1992). The basic asymptotic variance estimator under a simple random sample for the model-assisted regression is calculated via a first order Taylor series (Särndal et al., 1992). The regression-assisted estimator can be expanded to two-stage cluster sampling scheme. The following result is provided in Särndal et al., (1992): $Var(\hat{p}_{cal}) = Var_{between}(PSU) + Var_{within}(PSU)$. Said another way, the basic decomposition of variance can be applied to the cluster sampling scheme: the total variance is the variance of the primary sampling units (villages) and the variance of the secondary sampling units (attendees of the MUAC training). Chapter 8 of Särndal et al., also derive the Taylor series variance estimates for the PSU and SSU which we review here (Särndal et al., 1992). Let $v_j$ be the sample size in cluster $j$ from a cluster size of $V_j$, $N = \sum_j^J V_j$ be the total population size and let $J$ be total number of clusters (villages), and $y_k$ is the vaccination status of individual $k$. In the two-stage sampling scheme we define $s_{psu}$ as our sample of PSUs, $s_{ssu\ j}$ is the sample of SSU within each PSU $j$. The inclusion probability for the SSU is denoted as $\pi_{i|j}$ and let $\hat{\Delta}_{kl} = \frac{\pi_{kl} - \pi_k \pi_l}{\pi_{kl}}$ for some sampled units $k$ and $l$. Lastly, we let $e_k = y_k - \boldsymbol{x}_k^T \hat{\boldsymbol{B}}$, which are the residuals produced by minimizing the squared distance between $w_{ij}$ and $d_{ij}$ and is equivalent to the GREG procedure. In this case, $w_{ij} = \frac{1}{\pi_{ij}}$.

The variance is thus

$$\widehat{Var}(\hat{p}_{Cal}) = \frac{1}{N^2}\left(\sum_{k\in s_{psu}}\sum_{l\in s_{psu}}\Delta_{\mathrm{kl}}\frac{e_l}{\pi_{Jl}}\frac{e_k}{\pi_{Jk}} + \sum_{j\,\in\, s_{psu}}\frac{\widehat{Var}_{within}(PSU)_j}{\pi_{Jj}}\right),$$

and the variance within PSUs is given by

$$\widehat{Var}(PSU)_j = \left(\sum_{k\in s_{ssu\,j}}\sum_{l\in s_{ssu\,j}}\Delta_{\mathrm{kl|j}}\ \frac{y_{k|j}}{\pi_{k|j}}\frac{y_{l|j}}{\pi_{l|j}}\right),$$

where $k$ and $l$ are indices for any two elements in either the PSU or SSU. This linearized variance estimation is done automatically in the survey package (Lumley & Lumley, 2020).

### *Variance Estimation for the Logistic Regression Imputation Method*

Here we present the delta method or a linearization to approximate the variance of the model-based correction. In practice, a much simpler method would be to use a bootstrap to calculate the variance, as this would avoid deriving any variance estimator. However, for the purposes of simulating, using an asymptotic variance approximator is much more computationally efficient.

The model-based prediction approach treats the finite population parameter $p$ as a random variable, which is sampled from a super-population (Wu & Thompson, 2020) (Elliott & Valliant, 2017). Following this framework, we aim to calculate the variance as $V(\hat{p}_{LR} - p)$ , where both quantities are random variables. In the super-population model: $V(p) = V(\frac{1}{N}\sum_{i=1}^{N} Y_i) = \frac{\mu_p(1-\mu_p)}{N}$, where $\mu_p$ is the super-population proportion. Assuming independence between our sample and the population we have:

$$V(\hat{p}_{LR} - p) = V(\tfrac{1}{N}\sum_{i=1}^{N} expit(\mathbf{x}_i^T\widehat{\boldsymbol{\beta}}) - \tfrac{1}{N}\sum_{i=1}^{N} Y_i),$$

$$= \quad V(\tfrac{1}{N}\sum_{i=1}^{N} expit(\mathbf{x}_i^T\widehat{\boldsymbol{\beta}})) + V(\tfrac{1}{N}\sum_{i=1}^{N} Y_i).$$

We can then estimate each of the variations. We can estimate $V(\frac{1}{N}expit(\mathbf{x}_i^T\widehat{\boldsymbol{\beta}}))$ using the delta method, and $V(\frac{1}{N}\sum_{i=1}^{N} Y_i)$ is estimated as: $\frac{1}{N^2}\sum_{i=1}^{N}(\hat{p}_{i_{LR}}(1-\hat{p}_{i_{LR}}))$.

To estimate $V(\frac{1}{N} expit(\mathbf{x}_i^T \widehat{\boldsymbol{\beta}}))$ using the delta method, we follow directly from the guide provided in (Dowd et al., 2014). Let $\mathbf{x}_i$ be a column vector of census covariates for child $i$, and $\widehat{\boldsymbol{\beta}}$ is the vector of our estimated coefficients from the logistic regression model, then our estimator is $\hat{p}_{LR} = \frac{1}{N}\sum_{i=1}^{N} expit\left(\mathbf{x}_i^T \widehat{\boldsymbol{\beta}}\right)$, where $expit(t) = \frac{1}{1+e^{-t}}$, the expit function. The estimator of the variance is given by (Dowd et al., 2014), and is

$$Var(\hat{p}_{LR}) = \left(\frac{1}{N}\frac{\sum_{i=1}^{N} \partial\, expit\left(\mathbf{x}_i^T \widehat{\boldsymbol{\beta}}\right)}{\partial \widehat{\boldsymbol{\beta}}}\right)^T \hat{\Sigma} \left(\frac{1}{N}\frac{\sum_{i=1}^{N} \partial\, expit\left(\mathbf{x}_i^T \widehat{\boldsymbol{\beta}}\right)}{\partial \widehat{\boldsymbol{\beta}}}\right),$$

where $\hat{\Sigma}$ is the estimated variance-covariance matrix of the logistic regression model. Note we use the cluster-robust variance-covariance matrix (Cameron et al., 2011). It follows that $\frac{\partial expit(\mathbf{x}_i^T \widehat{\boldsymbol{\beta}})}{\partial \widehat{\boldsymbol{\beta}}} = \frac{e^{-\mathbf{x}_i^T \widehat{\boldsymbol{\beta}}}}{\left(1+e^{-\mathbf{x}_i^T \widehat{\boldsymbol{\beta}}}\right)^2}\mathbf{x}_i = expit(\mathbf{x}_i^T \widehat{\boldsymbol{\beta}})\left(1 - expit(\mathbf{x}_i^T \widehat{\boldsymbol{\beta}})\right)\mathbf{x}_i$. The final variance estimate is then

$$V(\hat{p}_{LR} - p) = \left(\frac{1}{N}\sum_{i=1}^{N} \hat{p}_{i_{LR}} \mathbf{x}_{\mathrm{i}}\right)^T \hat{\Sigma} \left(\frac{1}{N}\sum_{i=1}^{N} \hat{p}_{iLR} \mathbf{x}_{\mathrm{i}}\right) + \frac{1}{N^2}\sum_{i=1}^{N}(\hat{p}_{i_{LR}}(1 - \hat{p}_{i_{LR}})).$$

## 2. Baseline Census Data

### *Missing Values*

There were <3% of villages had missing values for the total population in the baseline census data. We imputed these values by dividing number of children aged 6-59 months by the average non-missing proportion of children aged 6-59 months in each country.

### *Vaccination Status*

We analysed the baseline census data to inform our choice of the parameters for the data generating mechanism. A logistic mixed effects model was fit to the baseline census vaccination data with six covariates including: village population, distance to nearest health centre (in km), age of the child, gender of the child, age of the caregiver, and gender of the caregiver. We calculated the intra-cluster correlation using the definition for binary outcomes (Yepnang et al., 2021)

$$ICC_v = \frac{\sigma^2}{\sigma^2 + \frac{\pi^2}{3}}.$$

where $\sigma^2$ is the variance of the village level random intercept.

***Selection Bias***

We fit a similar model to assess proxy for the selection bias of attending the MUAC training. In the baseline census data, the variable "attended previous training" as a proxy for sample inclusion. We fit a logistic mixed effects model on the attendance of previous training against the same covariates as in the vaccination model. The intra-cluster correlation for selection was also calculated $ICC_s = \frac{\tau^2}{\tau^2 + \frac{\pi^2}{3}}$ , where $\tau^2$ is the variance of the village level random intercept.

***Results***

Supplementary Table 2 provides a basic breakdown of all the descriptive statistics of the baseline census survey. In the baseline census the MCV1 rate was 0.73, most demographic variables were similar between those who attended the previous training and those who were vaccinated. There were substantially more female than male caregivers, with approximately 95% of caregivers being female.

The $ICC_v$ estimate was 0.21, and the $ICC_s$ estimate was 0.56. The $ICC_v$ estimate was within the range of the of the WHO guideline, whereas the $ICC_s$ was highly inflated due to many villages having no previous participation. We opted to set $ICC_v$ of 1/3, which is the higher end of the WHO recommendation for vaccination surveys. In general, the $ICC$ values should not greatly affect our results but may slightly change the variance of the estimators. In general, a larger $ICC_v$ would favour sampling more villages and fewer caregivers/village which would reduce estimator variance and vice-versa. A larger $ICC_s$ value can be beneficial or detrimental, depending on if caregivers within villages prefer similarly to attend the training, or prefer to similarly not attend. Parameter estimates from the baseline census used to inform the data generating mechanism are presented in Supplementary Tables 3 & 4.

## 3. Monte Carlo Uncertainty

In this subsection we evaluate the Monte Carlo uncertainty of our performance metrics discussed in the paper. We evaluated 3 performance metrics: bias, coverage, and Equivalence tolerance, each under 54 scenarios and 2 methods. We evaluate here a conservative estimate of the Monte Carlo uncertainty across these scenarios. Morris et al., (2019) (Morris et al., 2019) provide the formulas for calculating the Monte Carlo Standard Error (MCSE) many common performance metrics which we will review here.

### *Bias*

The MCSE of the estimated bias is given by $\sqrt{\frac{1}{n_{rep}(n_{rep}-1)}\sum_{k=1}^{n_{rep}}(\widehat{p_k}-\bar{p})^2}$. Given $p_k$ changes for each iteration, we can set $\hat{d}_k = \widehat{p_k} - p_k$, thus the MCSE is then $\sqrt{\frac{1}{n_{rep}(n_{rep}-1)}\sum_{k=1}^{n_{rep}}\left(\hat{d}_k-\bar{d}\right)^2}$. This value is maximized when $\hat{p}_k - p_k$ (the bias) is large. With a maximum bias of 2% in our simulations, the MCSE is 0.05%.

### *Coverage*

The estimated coverage follows a binomial random variable and thus the MCSE is $\sqrt{\frac{\hat{c}(1-\hat{c})}{n_{rep}}}$ where $\hat{c}$ is the estimated coverage. This standard error is maximized when $\hat{c}$ is closest to 0.5. This was reached when the selection bias odds were 1.5, and participation rate of 0.5 and 75% of villages sampled. The MCSE under this scenario is approximately 1.5%. Under more realistic conditions where $\hat{c} = 0.9$, the MCSE is about 1%.

### *Equivalence*

The equivalence tolerance is identical to the coverage. The MCSE is $\sqrt{\frac{\hat{e}(1-\hat{e})}{n_{rep}}}$ where $\hat{e}$ is the estimated proportion of iterations declared equivalent with our specified margin. Most of all our equivalence tolerance proportions are >0.9, and thus the MCSE would be <1%. In the worst-case scenario, the MCSE would be 1.5%.

We believe the values of the MCSE for each performance metric are acceptable for the purpose of assessing the robustness of the survey selection bias correction methods under various levels of non-ignorable selection bias.

## 4. Supplementary Tables

**Supplementary Table 1. Descriptive Statistics for the Baseline Census Data**

| Factor | Values | Justification |
| --- | --- | --- |
| Percentage of villages sampled | (25%, 50%, 75%) | Feasibility for the ALIMA field team |
| Odds ratio of selection bias | (1.0, 1.1, 1.2, 1.3, 1.4 ,1.5) | Informed by the ALIMA team and previous attendance |
| Participation Rate | (50%, 65%, 80%) | Informed by the ALIMA team and previous attendance |

**Supplementary Table 2. Descriptive Statistics for the Baseline Census Data**

| | Attended Previous Training | | MCV1 Vaccination Status | | |
|---|---|---|---|---|---|
| | **0** (N=6399) | **1** (N=3073) | **0** (N=2561) | **1** (N=6911) | **Overall** (N=9472) |
| Age of Child (Months) | | | | | |
| Mean (SD) | 17.9 (3.99) | 17.8 (3.99) | 17.9 (4.12) | 17.9 (3.95) | 17.9 (3.99) |
| Median [Min, Max] | 18.0 [12.0, 24.0] | 18.0 [12.0, 24.0] | 18.0 [12.0, 24.0] | 18.0 [12.0, 24.0] | 18.0 [12.0, 24.0] |
| Gender of Child | | | | | |
| Female | 3449 (53.9%) | 1679 (54.6%) | 1408 (55.0%) | 3720 (53.8%) | 5128 (54.1%) |
| Male | 2950 (46.1%) | 1394 (45.4%) | 1153 (45.0%) | 3191 (46.2%) | 4344 (45.9%) |
| Gender of Guardian | | | | | |
| Female | 5993 (93.7%) | 2820 (91.8%) | 2359 (92.1%) | 6454 (93.4%) | 8813 (93.0%) |
| Male | 406 (6.3%) | 253 (8.2%) | 202 (7.9%) | 457 (6.6%) | 659 (7.0%) |
| Age of Guardian (Years) | | | | | |
| Mean (SD) | 29.4 (9.90) | 29.3 (9.21) | 29.6 (10.4) | 29.3 (9.42) | 29.4 (9.68) |
| Median [Min, Max] | 27.0 [15.0, 85.0] | 28.0 [15.0, 86.0] | 27.0 [15.0, 85.0] | 27.0 [15.0, 86.0] | 27.0 [15.0, 86.0] |

**Supplementary Table 3. Covariate estimates with 95% confidence intervals for the vaccination model (association with MCV1 vaccination status)**

| Covariate | Log Odds Estimate | 95% CI |
|---|---|---|
| Intercept | 1.38 | [1.03, 1.73] |
| Population size (for every 1 SD increase) | 0.06 | [-0.03, 0.16] |
| Distance to nearest health center (for every 1km increase) | -0.08 | [-0.11, -0.04] |
| Age of the child (for every 1-month increase) | 0.01 | [0.00, 0.02] |
| Age of the guardian (for every 1-year increase) | 0.00 | [-0.01, 0.00] |
| Sex of the child (reference group: female) | 0.07 | [-0.03, 0.17] |
| Gender of the guardian (reference group: female) | -0.24 | [-0.43, -0.05] |
| $ICC_v$ | 0.21 | |

**Supplementary Table 4. Covariate estimates with 95% confidence intervals for the selection model (Association with "did attend previous training")**

| Covariate | Log Odds Estimate | 95% CI |
|---|---|---|
| Intercept | 0.15 | [-0.38, 0.69] |
| Population size (for every 1 SD increase) | -0.33 | [-0.52, 0.14] |
| Distance to nearest health center (for every 1km increase) | -0.08 | [-0.14, -0.01] |
| Age of the child (for every 1-month increase) | -0.01 | [-0.03, 0.00] |
| Age of the guardian (for every 1-year increase) | 0.01 | [0.00, 0.02] |
| Sex of the child (reference group: female) | -0.02 | [-0.13, 0.09] |
| Gender of the guardian (reference group: female) | -0.01 | [-0.22, 0.20] |
| $ICC_s$ | 0.57 | |

# 5. Supplementary Figures

***Zipper plot of empirical coverage (95% CI) ignorable selection, Logistic Regression***

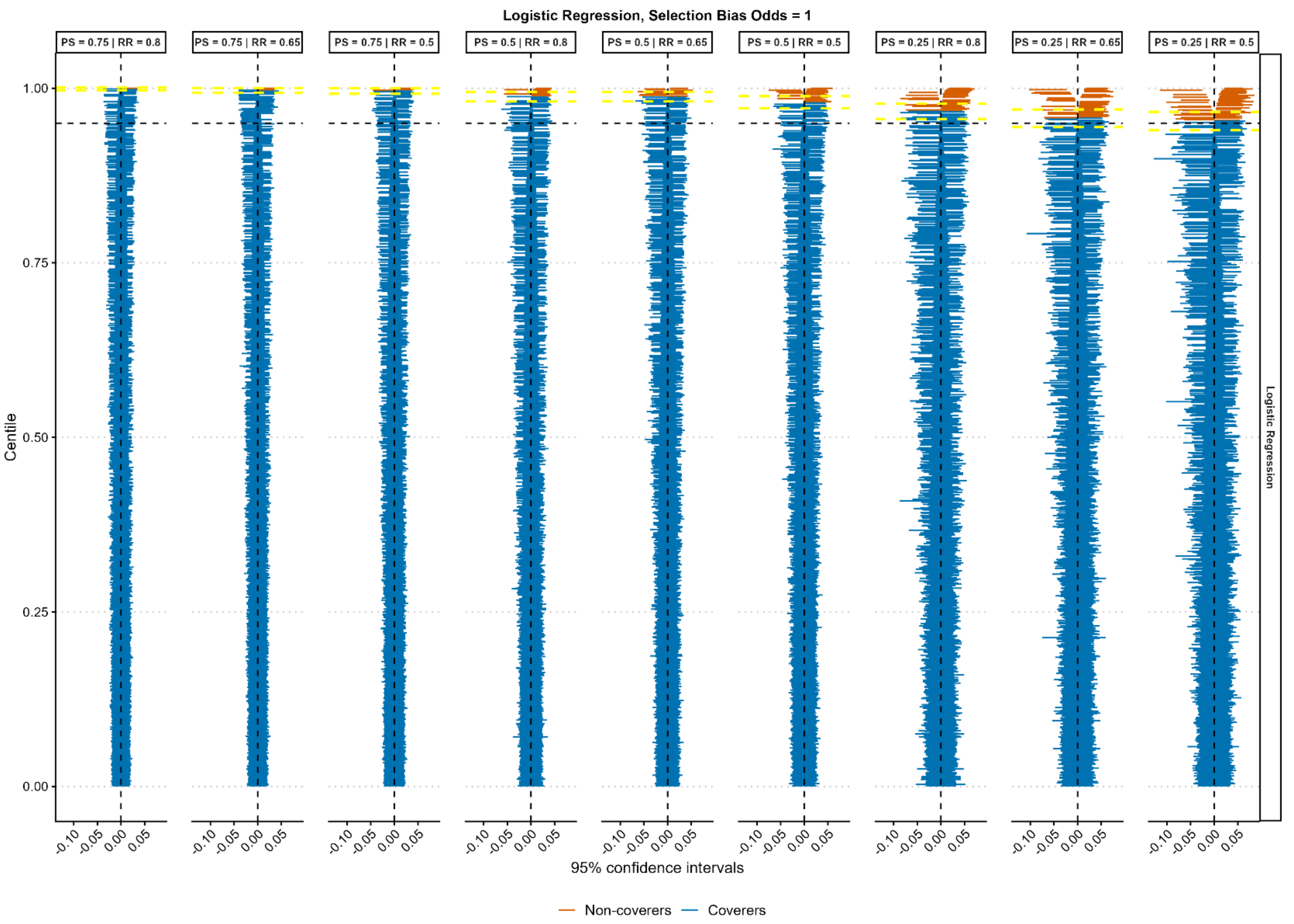

* Intervals ordered thinnest at the bottom to widest at the top. Orange intervals do not cover the bias of 0. PS = Percentage of villages sampled, RR = Response rate.

***Zipper plot of empirical coverage (95% CI) selection bias odds of 1.1, Logistic Regression***

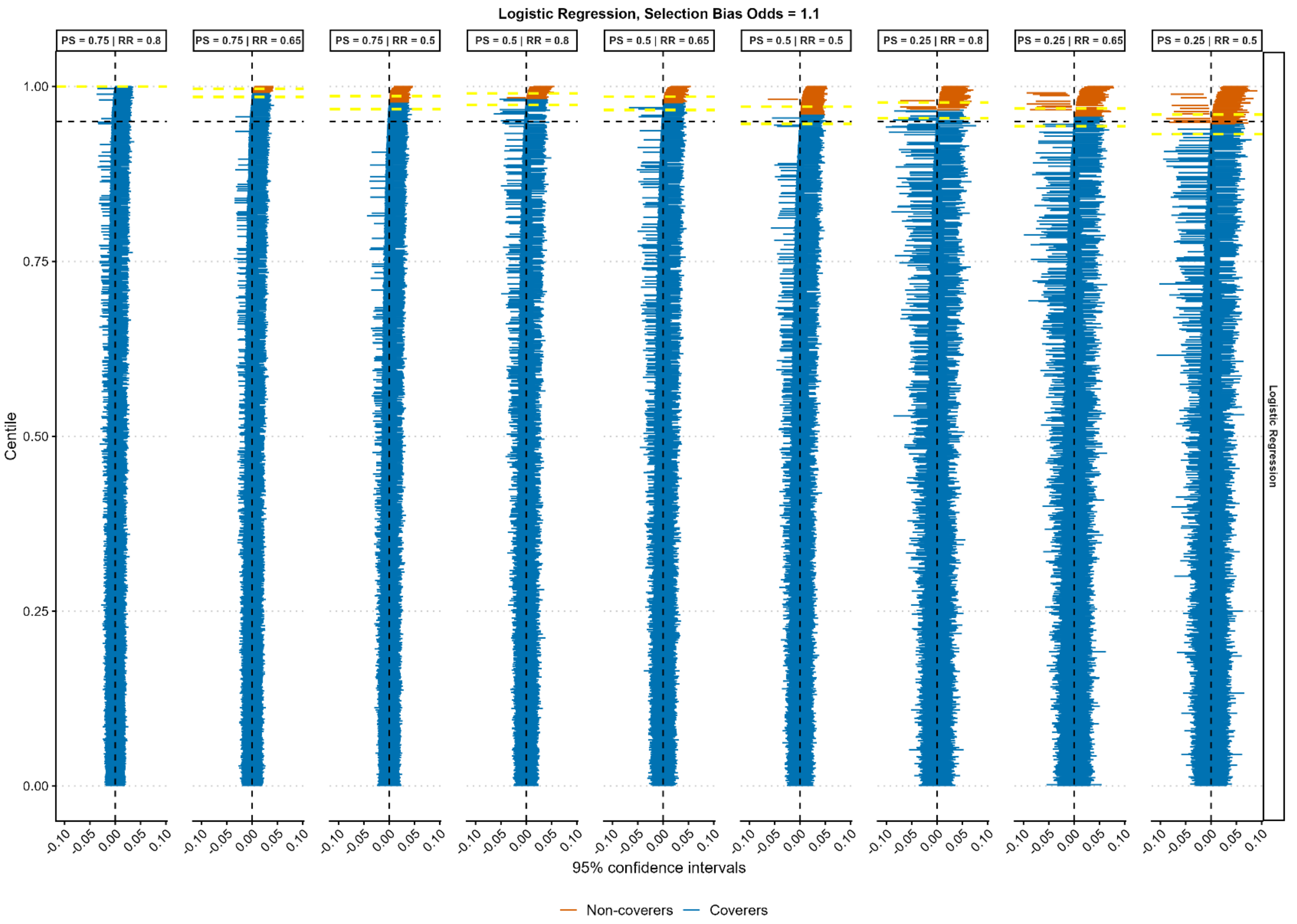

***Zipper plot of empirical coverage (95% CI) selection bias odds of 1.2, Logistic Regression***

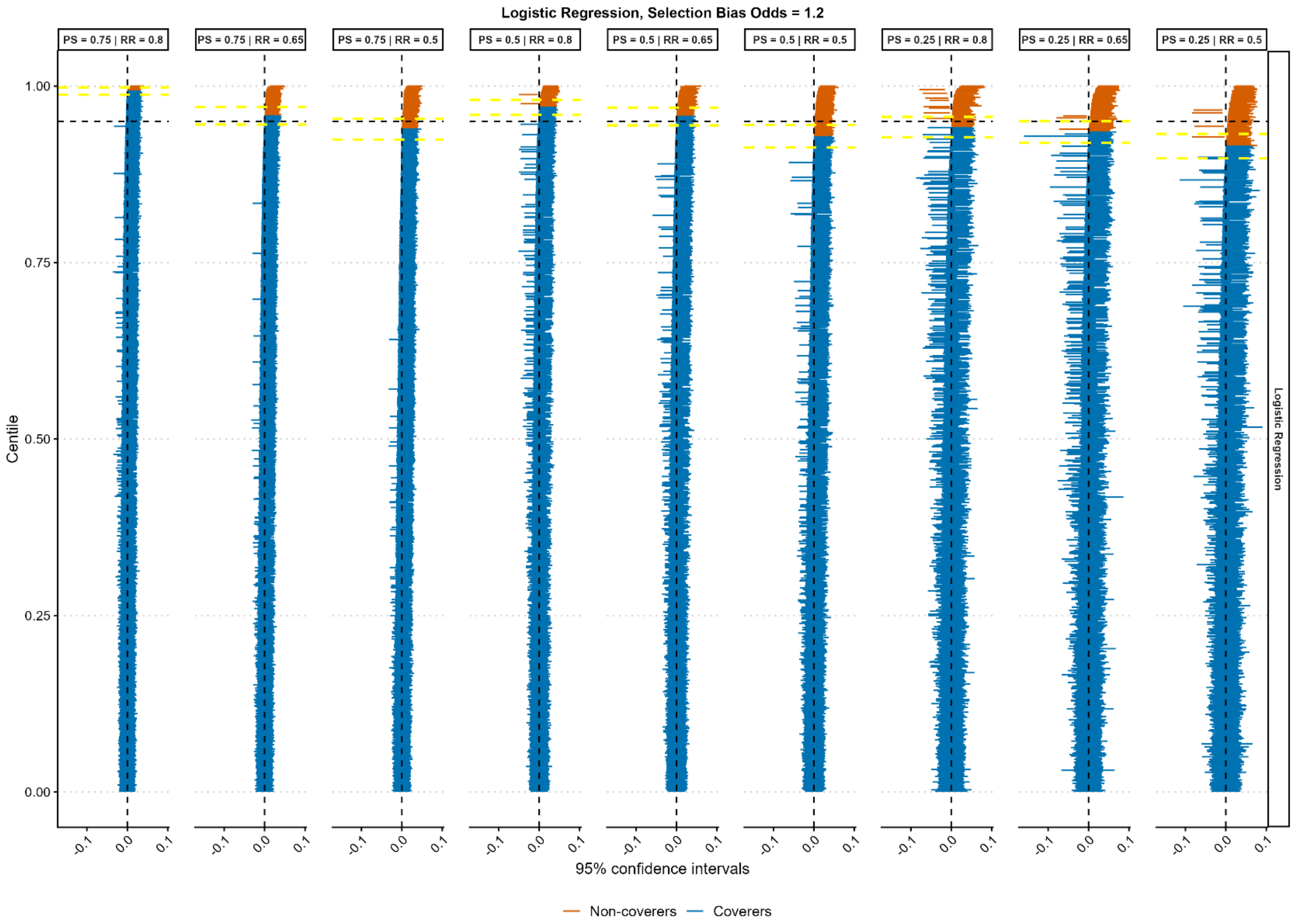

* Intervals ordered thinnest at the bottom to widest at the top. Orange intervals do not cover the bias of 0. PS = Percentage of villages sampled, RR = Response rate.

***Zipper plot of empirical coverage (95% CI) selection bias odds of 1.3, Logistic Regression***

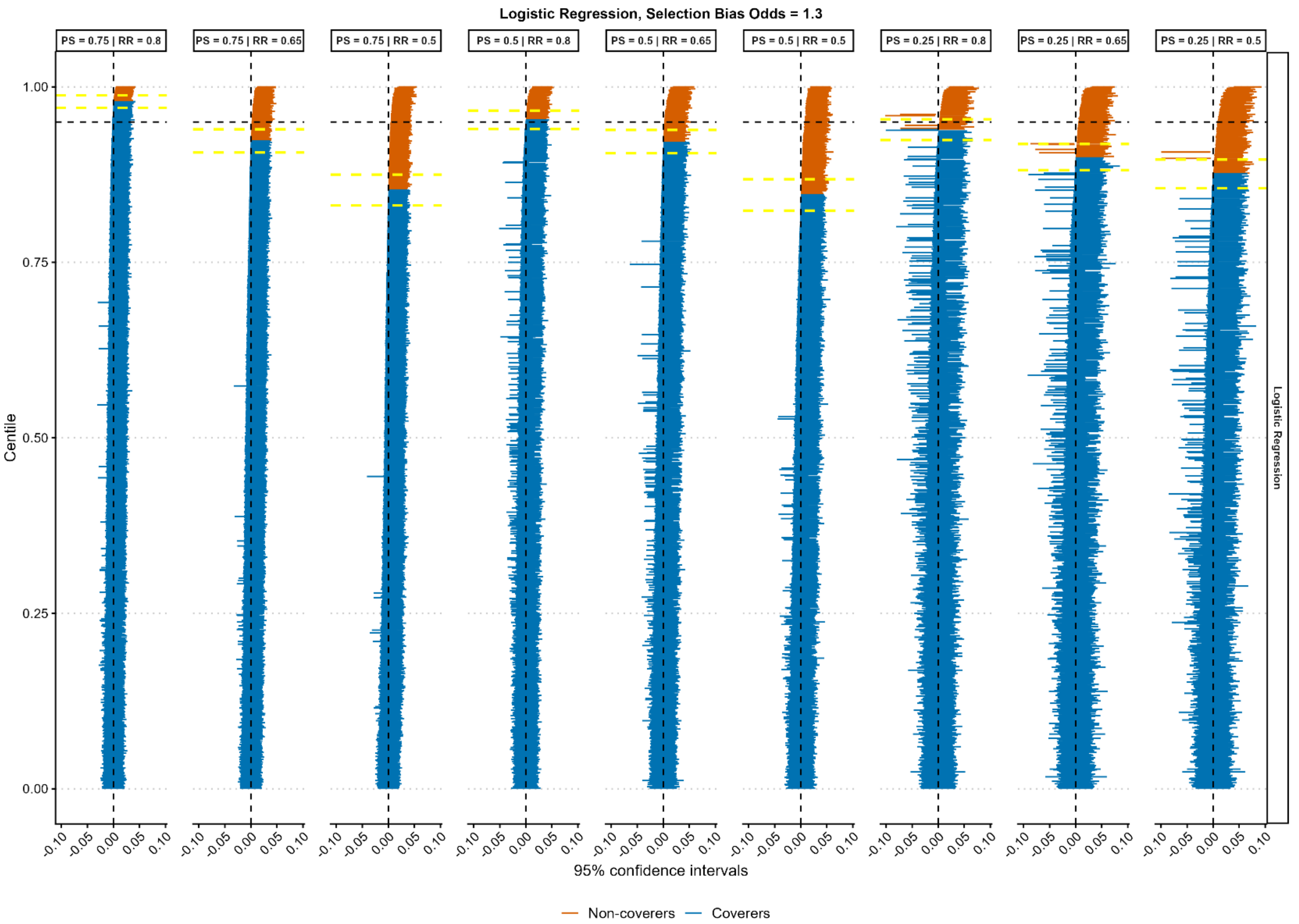

### *Zipper plot of empirical coverage (95% CI) selection bias odds of 1.4, Logistic Regression*

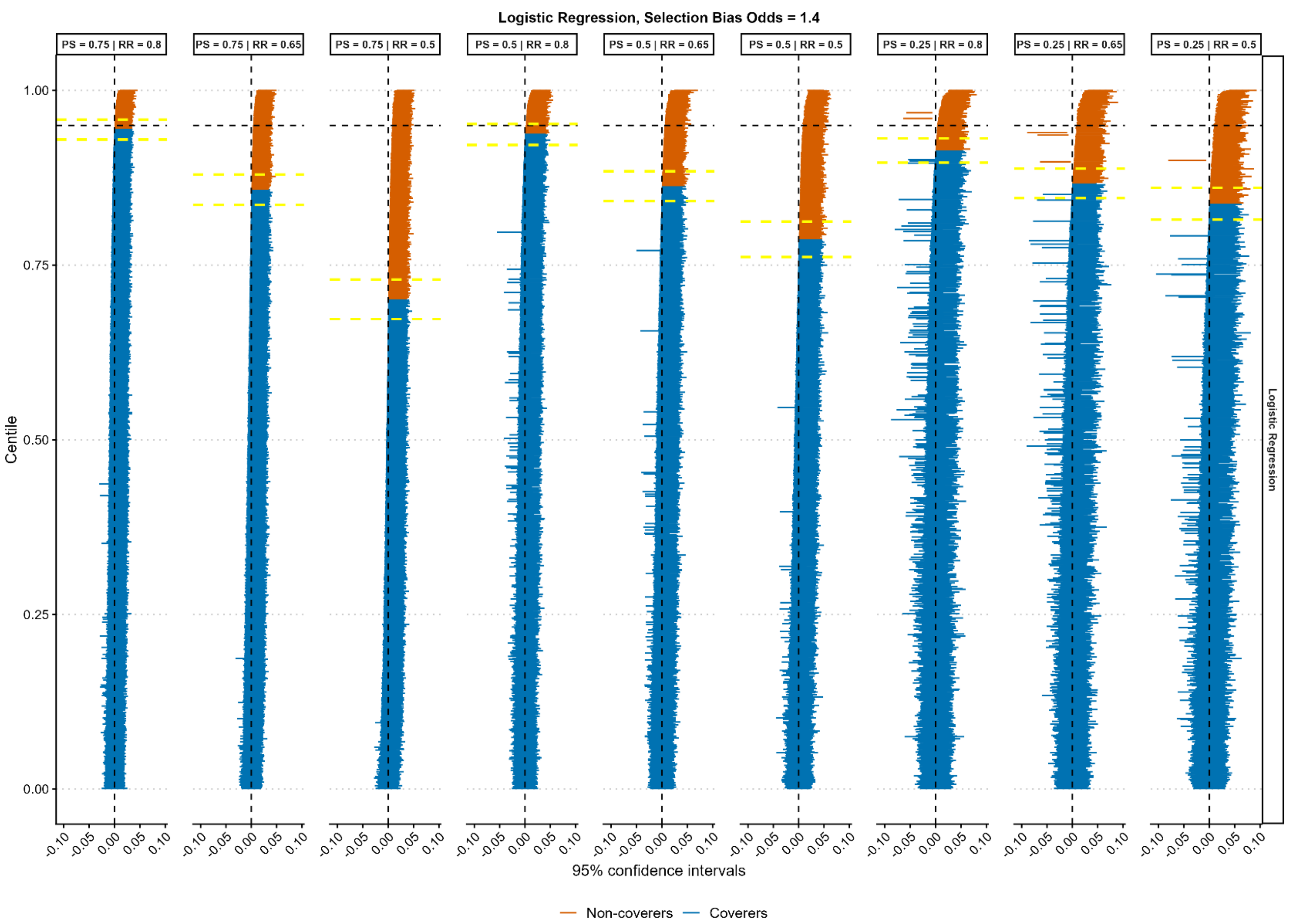

* Intervals ordered thinnest at the bottom to widest at the top. Orange intervals do not cover the bias of 0. PS = Percentage of villages sampled, RR = Response rate.

***Zipper plot of empirical coverage (95% CI) selection bias odds of 1.5, Logistic Regression***

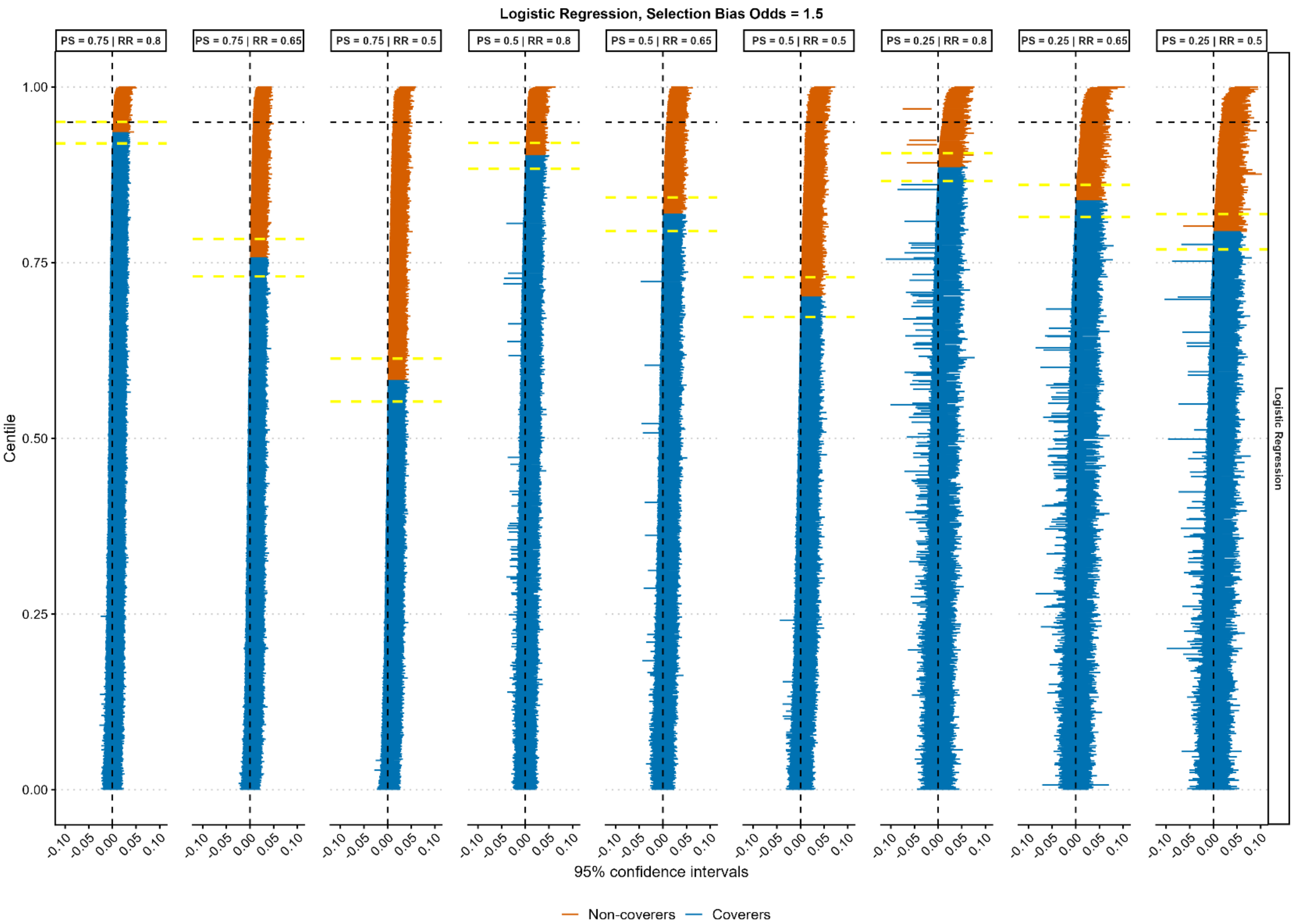

* Intervals ordered thinnest at the bottom to widest at the top. Orange intervals do not cover the bias of 0. PS = Percentage of villages sampled, RR = Response rate.

*Zipper plot of empirical coverage (95% CI) ignorable selection bias, Calibrated Weights*

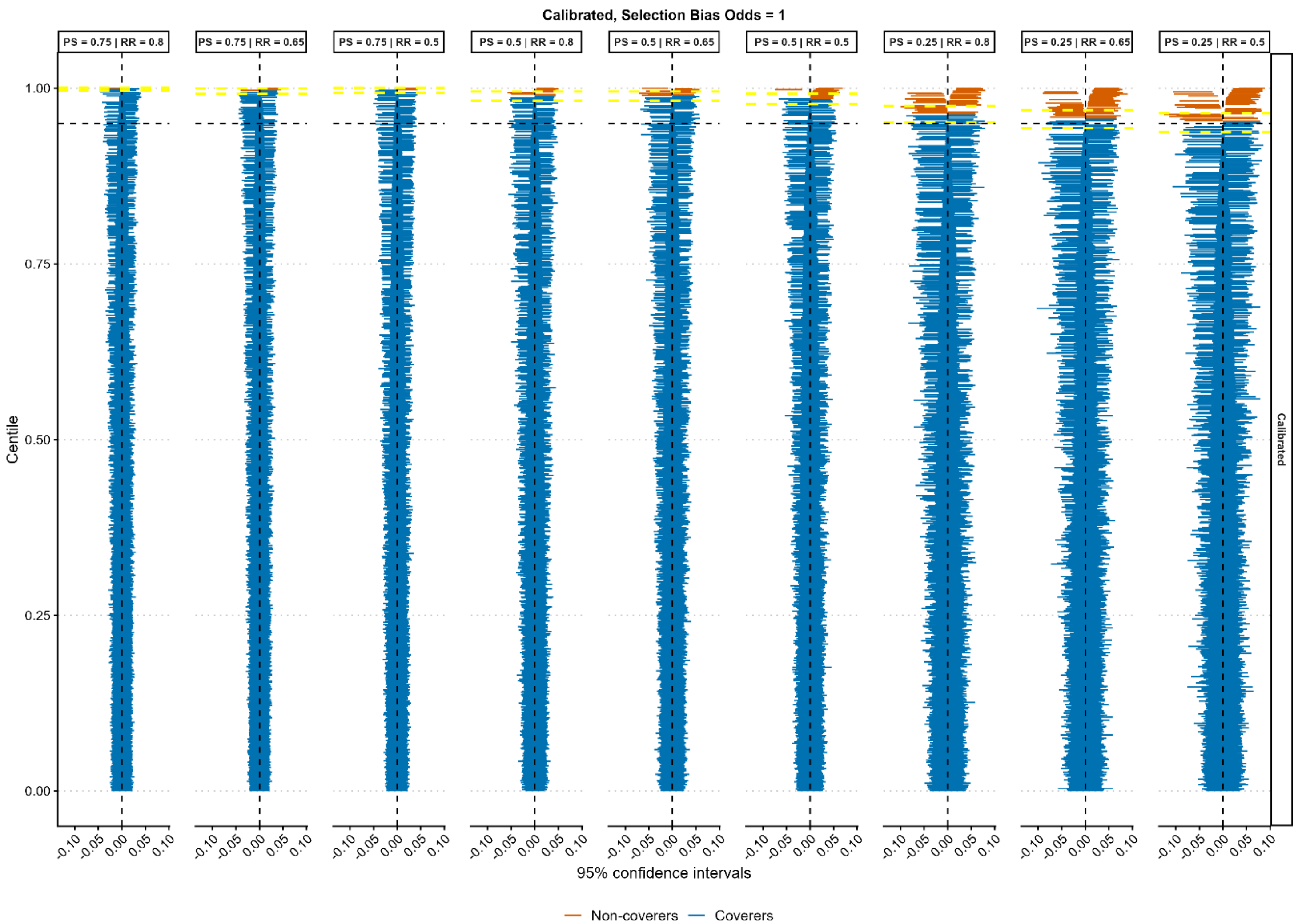

* Intervals ordered thinnest at the bottom to widest at the top. Orange intervals do not cover the bias of 0. PS = Percentage of villages sampled, RR = Response rate.

## *Zipper plot of coverage (95% CI) selection bias odds of 1.1, Calibrated weights*

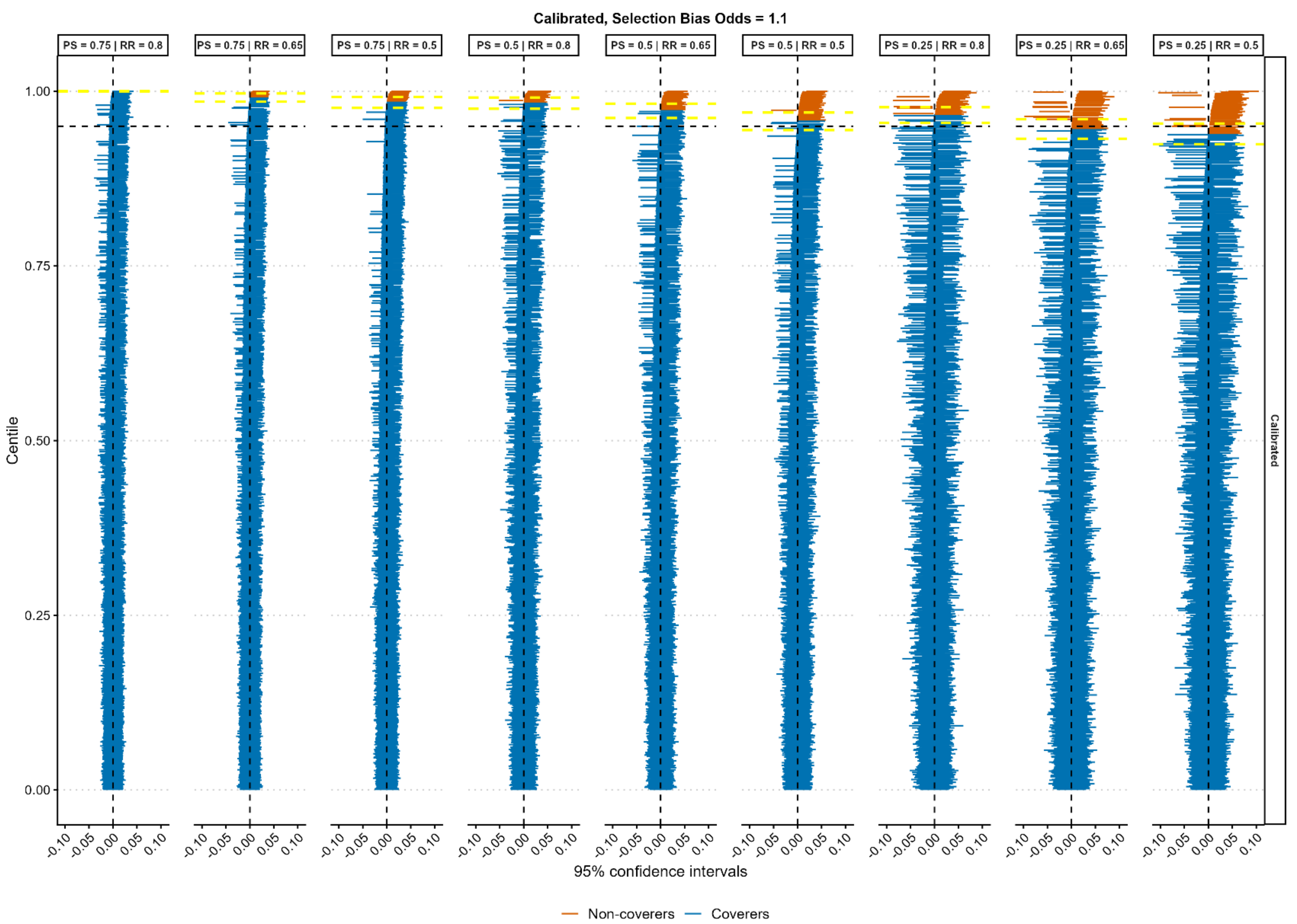

*Zipper plot of coverage (95% CI) selection bias odds of 1.2, Calibrated weights*

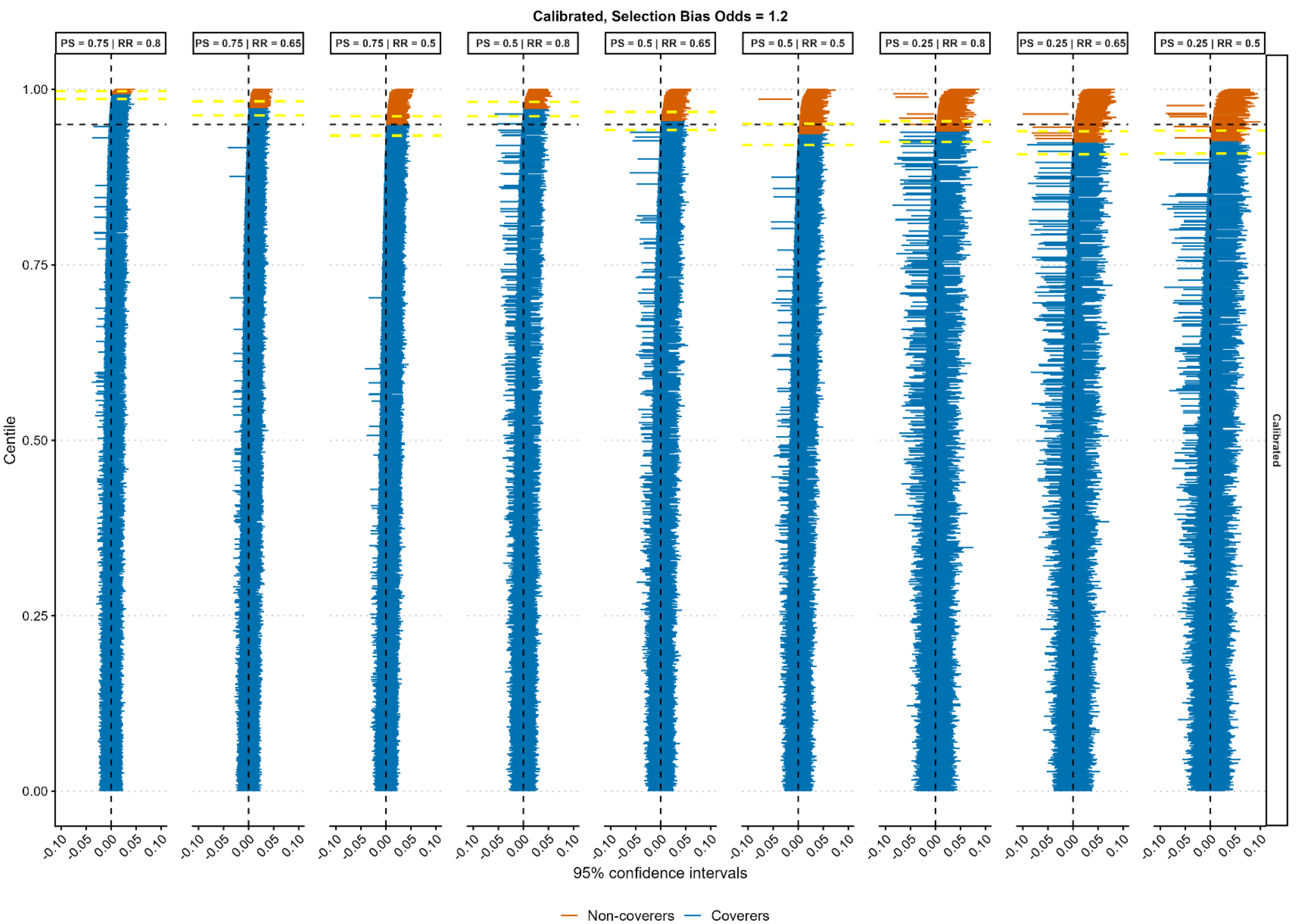

*Zipper plot of coverage (95% CI) selection bias odds of 1.3, Calibrated weights*

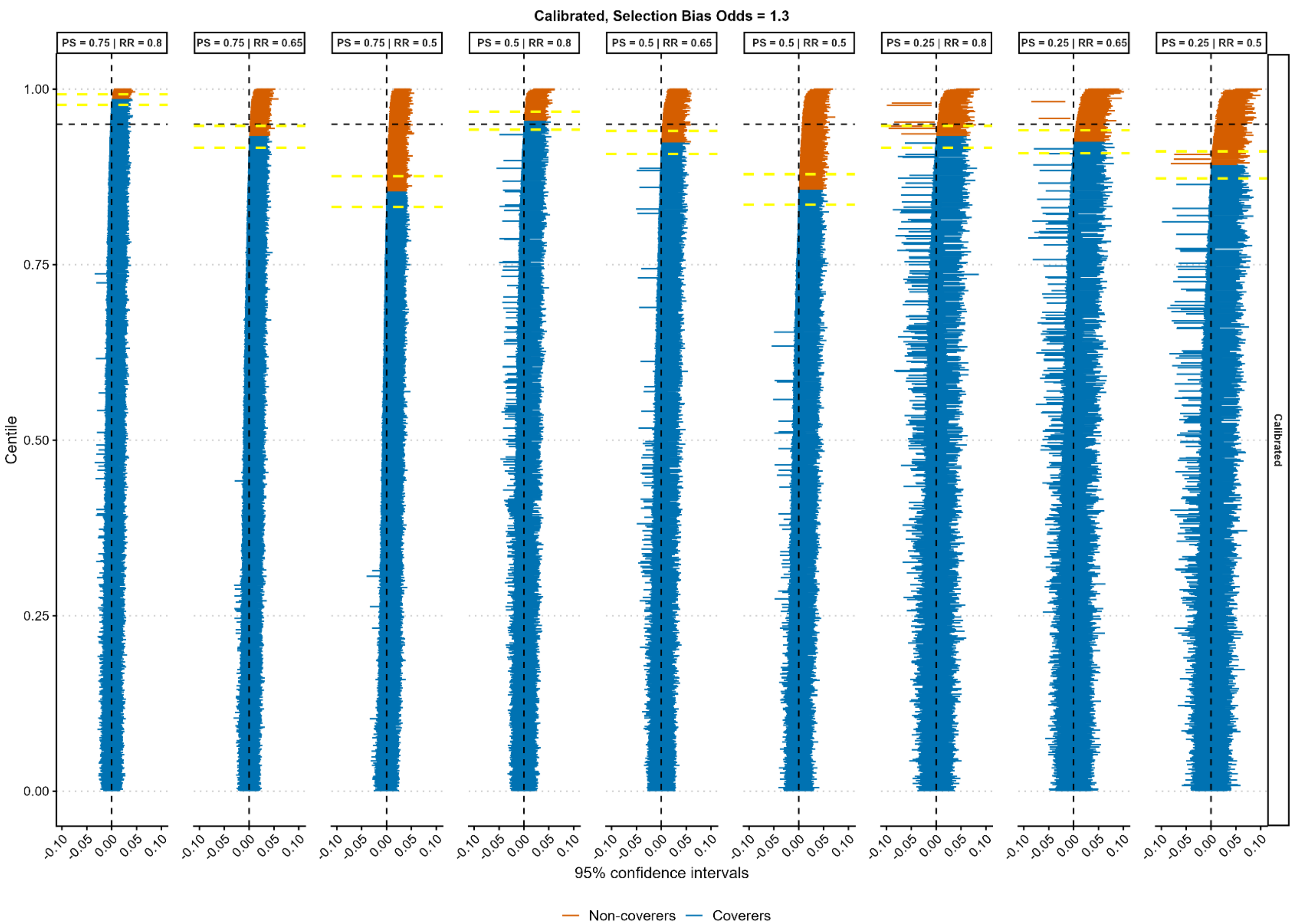

***Zipper plot of coverage (95% CI) selection bias odds of 1.4, Calibrated weights***

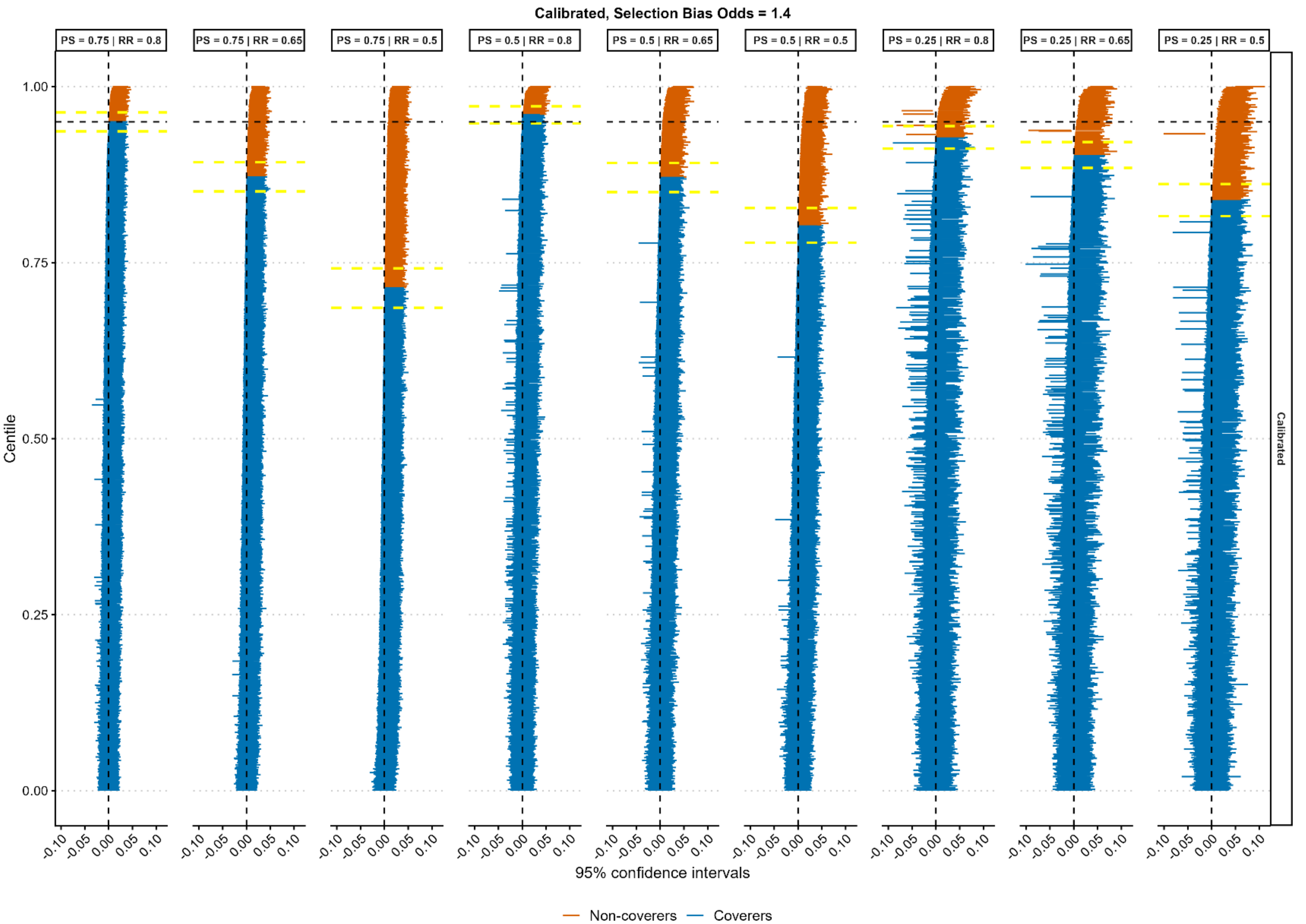

## *Zipper plot of coverage (95% CI) selection bias odds of 1.5, Calibrated weights*

**Calibrated, Selection Bias Odds = 1.5**

PS = 0.75 | RR = 0.8
PS = 0.75 | RR = 0.65
PS = 0.75 | RR = 0.5
PS = 0.5 | RR = 0.8
PS = 0.5 | RR = 0.65
PS = 0.5 | RR = 0.5
PS = 0.25 | RR = 0.8
PS = 0.25 | RR = 0.65
PS = 0.25 | RR = 0.5

Calibrated

Centile

95% confidence intervals

Non-coverers
Coverers

* Intervals ordered thinnest at the bottom to widest at the top. Orange intervals do not cover the bias of 0. PS = Percentage of villages sampled, RR = Response rate.

***TOST equivalence ±0.05 confidence intervals, ignorable selection, Logistic Regression***

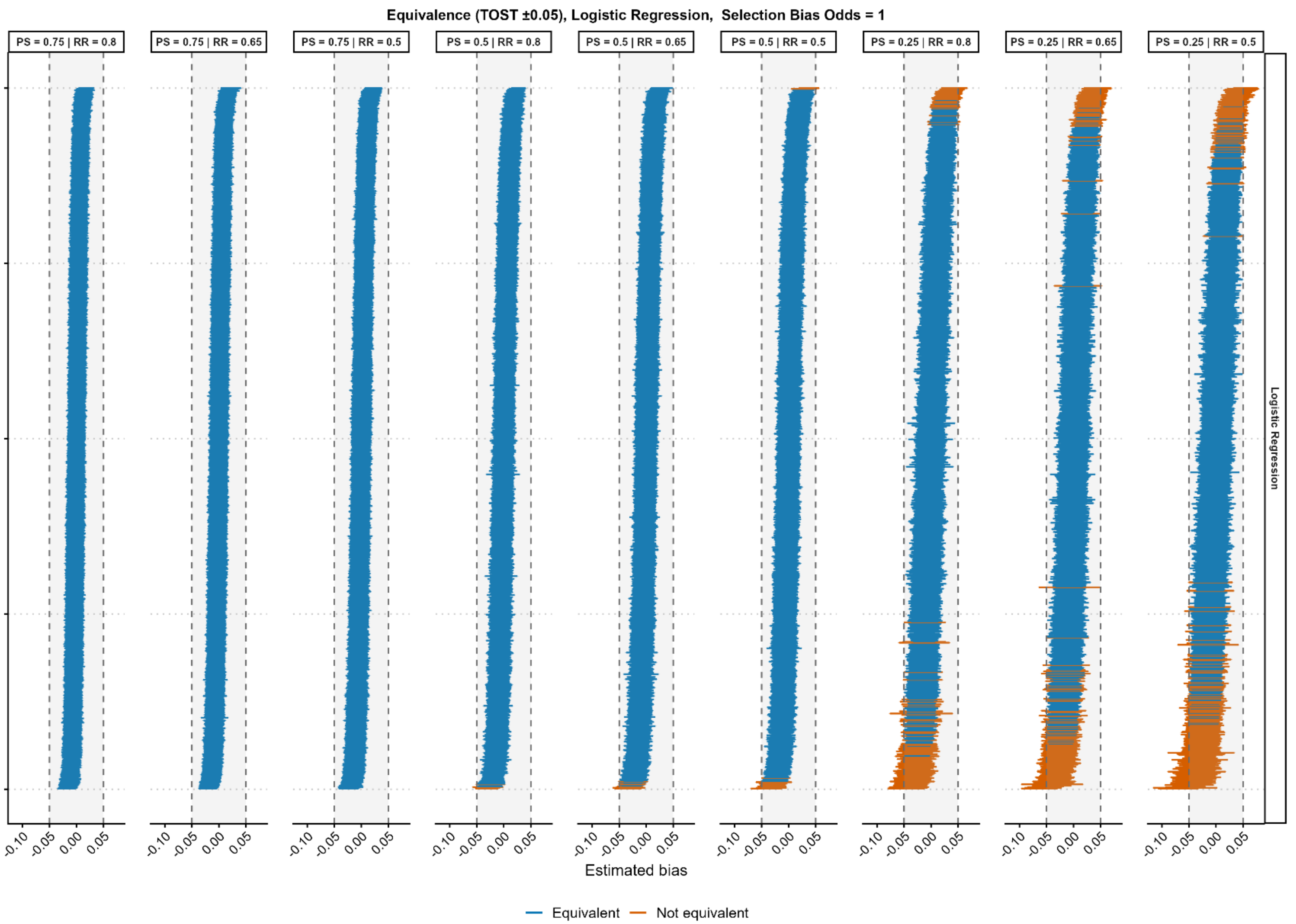

***TOST equivalence ±0.05 confidence intervals, selection bias odds of 1.1, Logistic Regression***

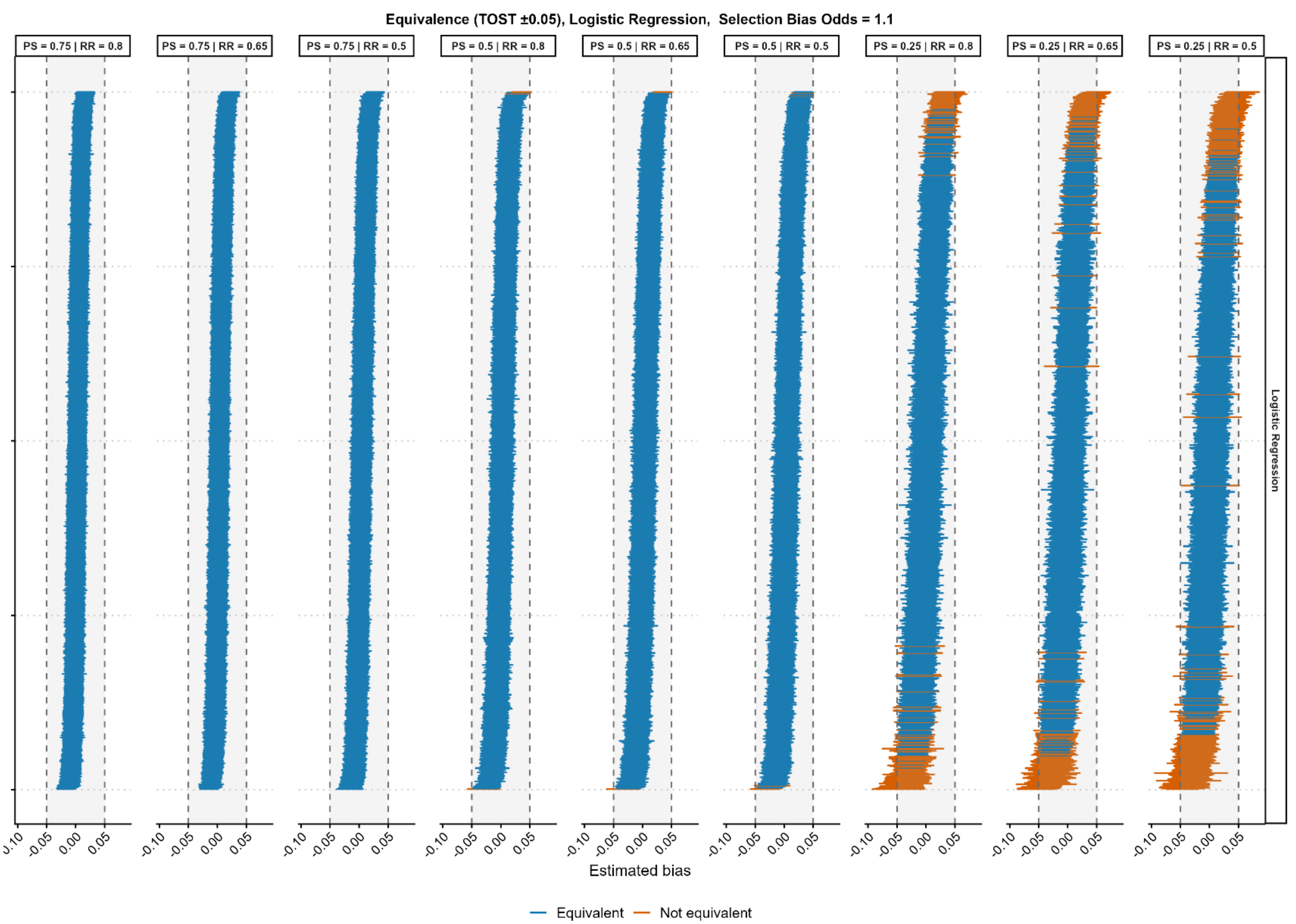

***TOST equivalence ±0.05 confidence intervals, selection bias odds of 1.2, Logistic Regression***

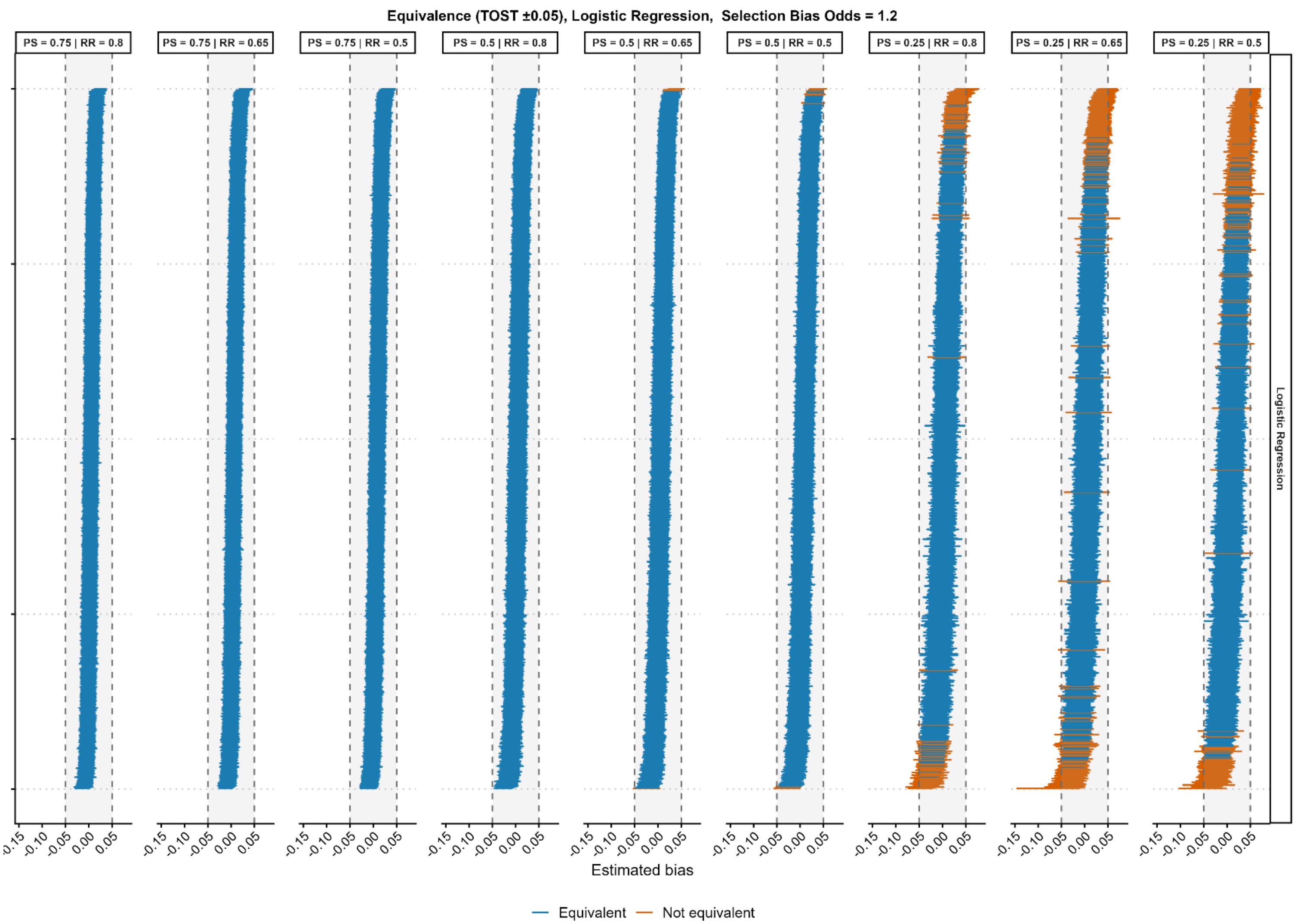

* Intervals ordered negative to positive bias. Orange intervals do not fall within the tolerance region. PS = Percentage of villages sampled, RR = Response rate.

***TOST equivalence ±0.05 confidence intervals, selection bias odds of 1.3, Logistic Regression***

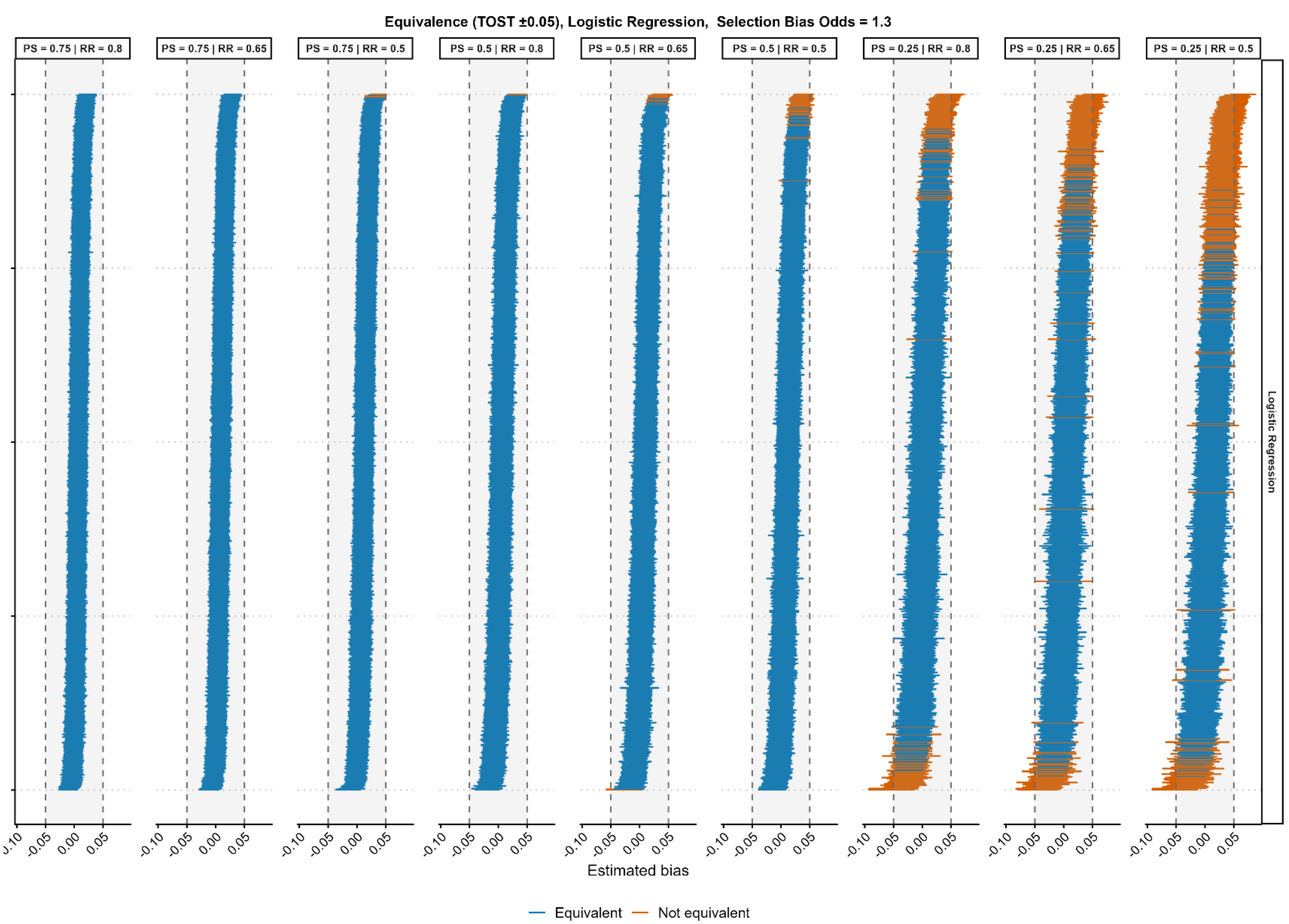

* Intervals ordered negative to positive bias. Orange intervals do not fall within the tolerance region. PS = Percentage of villages sampled, RR = Response rate.

***TOST equivalence ±0.05 confidence intervals, selection bias odds of 1.4, Logistic Regression***

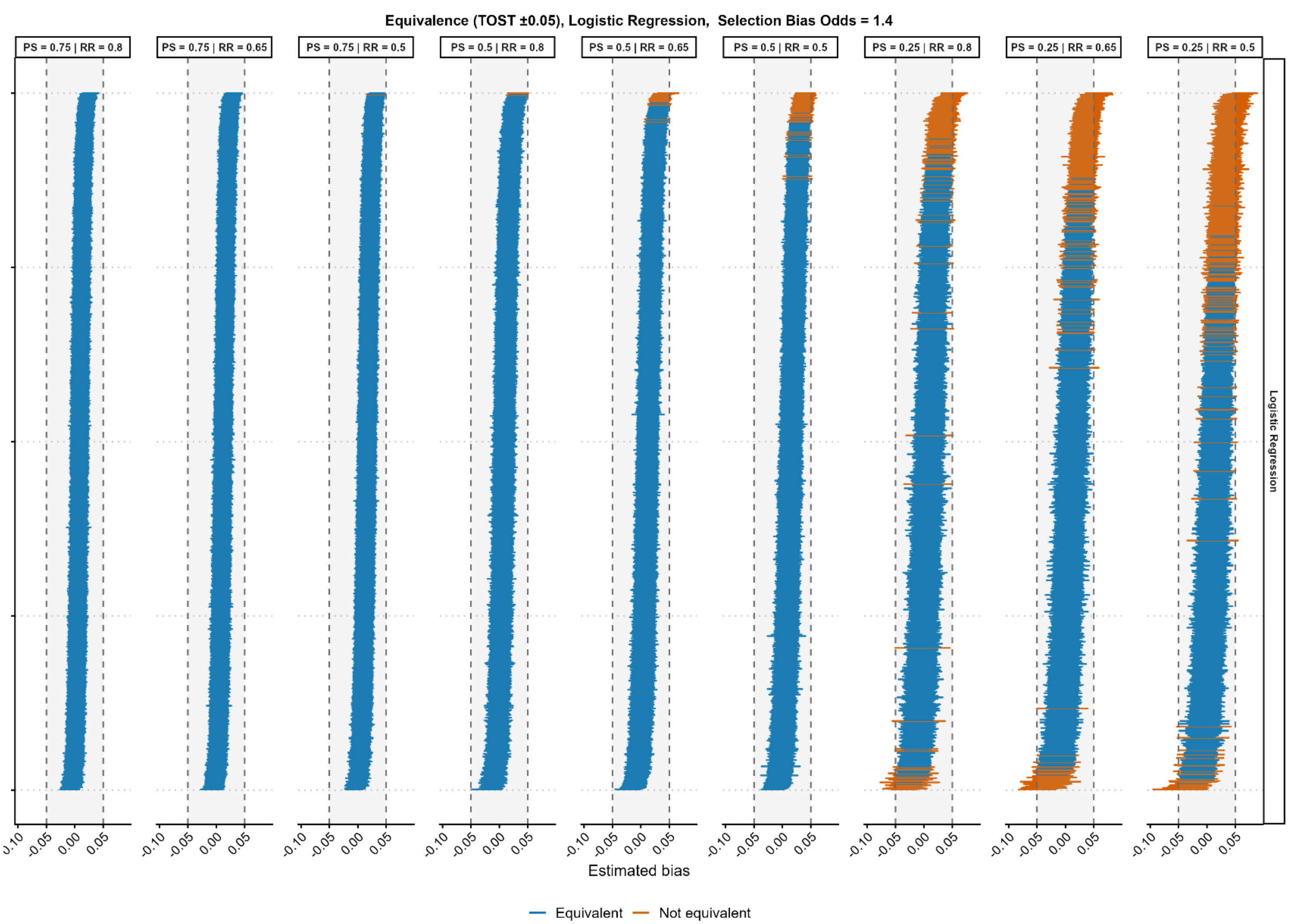

* Intervals ordered negative to positive bias. Orange intervals do not fall within the tolerance region. PS = Percentage of villages sampled, RR = Response rate.

***TOST equivalence ±0.05 confidence intervals, selection bias odds of 1.5, Logistic Regression***

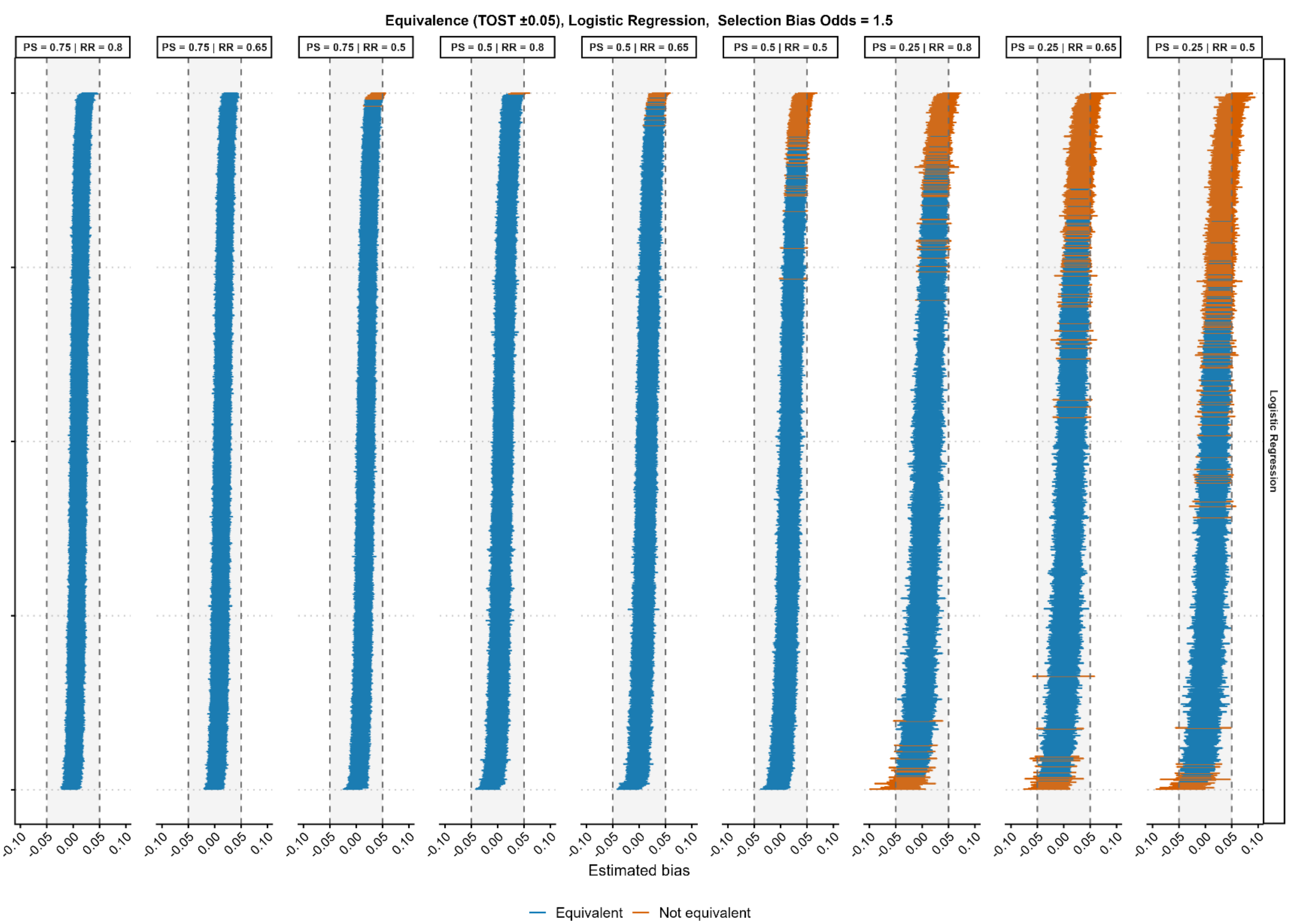

***TOST equivalence ±0.05 confidence intervals, ignorable selection, Calibrated Weights***

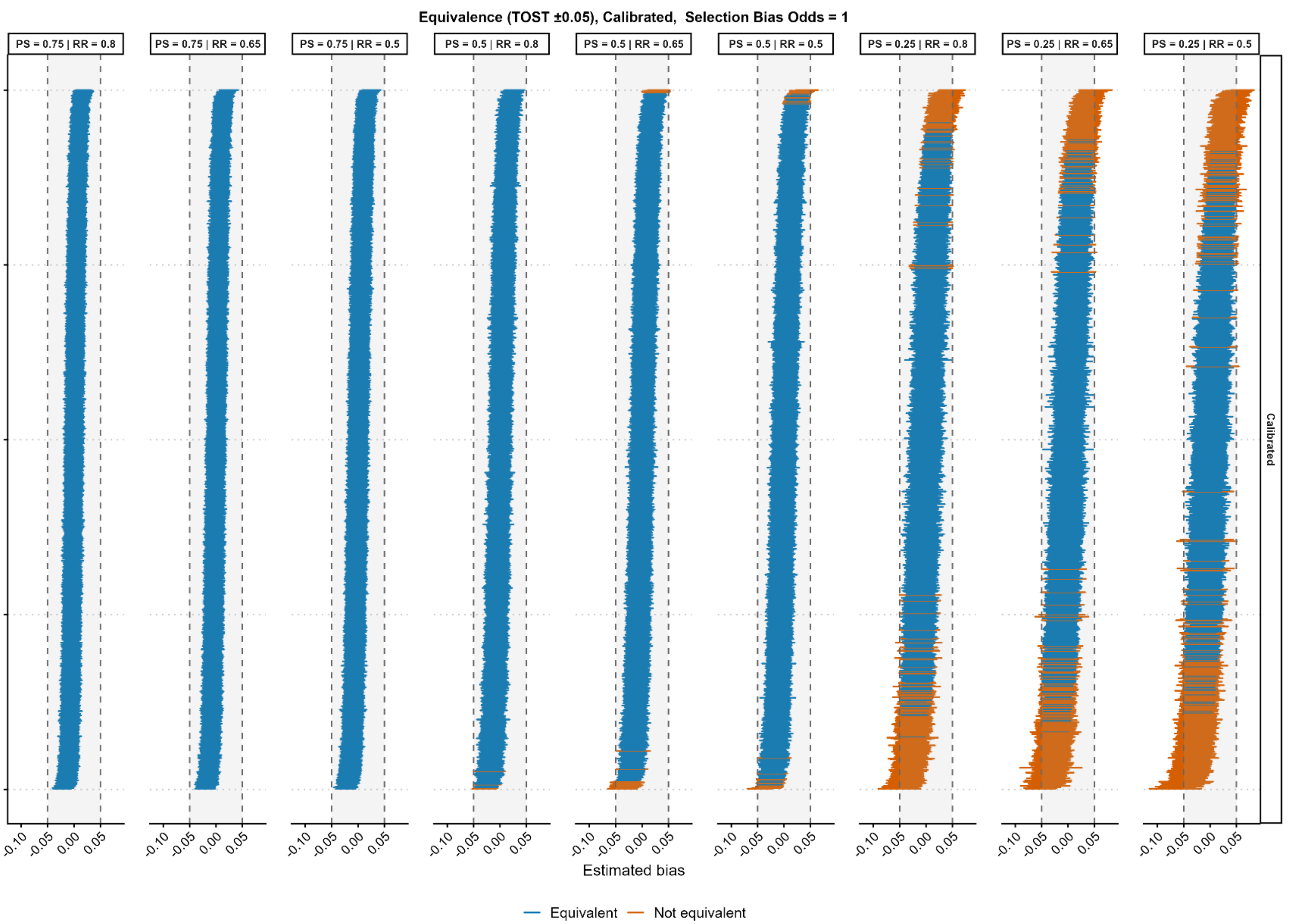

* Intervals ordered negative to positive bias. Orange intervals do not fall within the tolerance region. PS = Percentage of villages sampled, RR = Response rate.

### *TOST equivalence ±0.05 confidence intervals, selection bias odds of 1.1, Calibrated Weights*

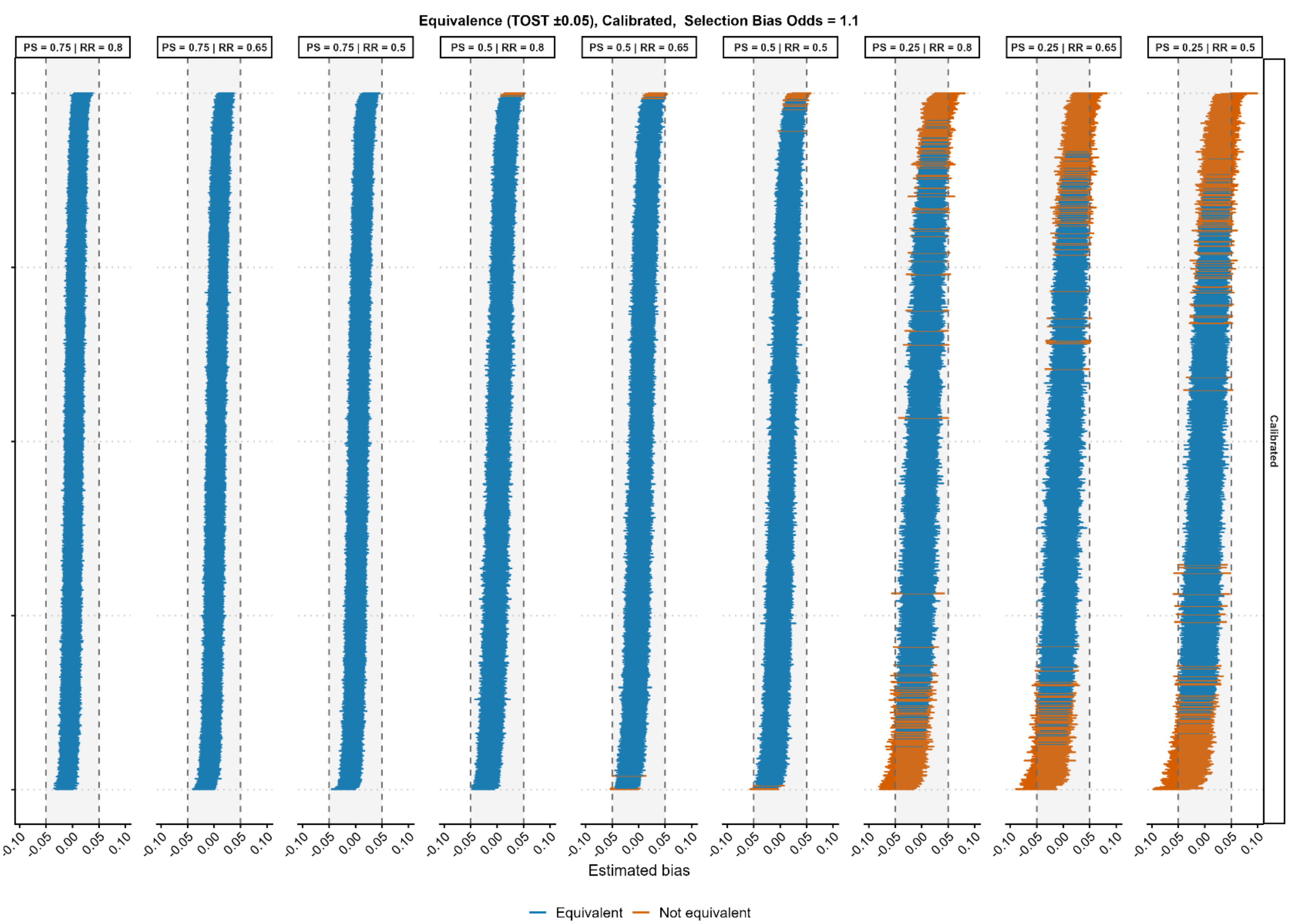

* Intervals ordered negative to positive bias. Orange intervals do not fall within the tolerance region. PS = Percentage of villages sampled, RR = Response rate.

***TOST equivalence ±0.05 confidence intervals, selection bias odds of 1.2, Calibrated Weights***

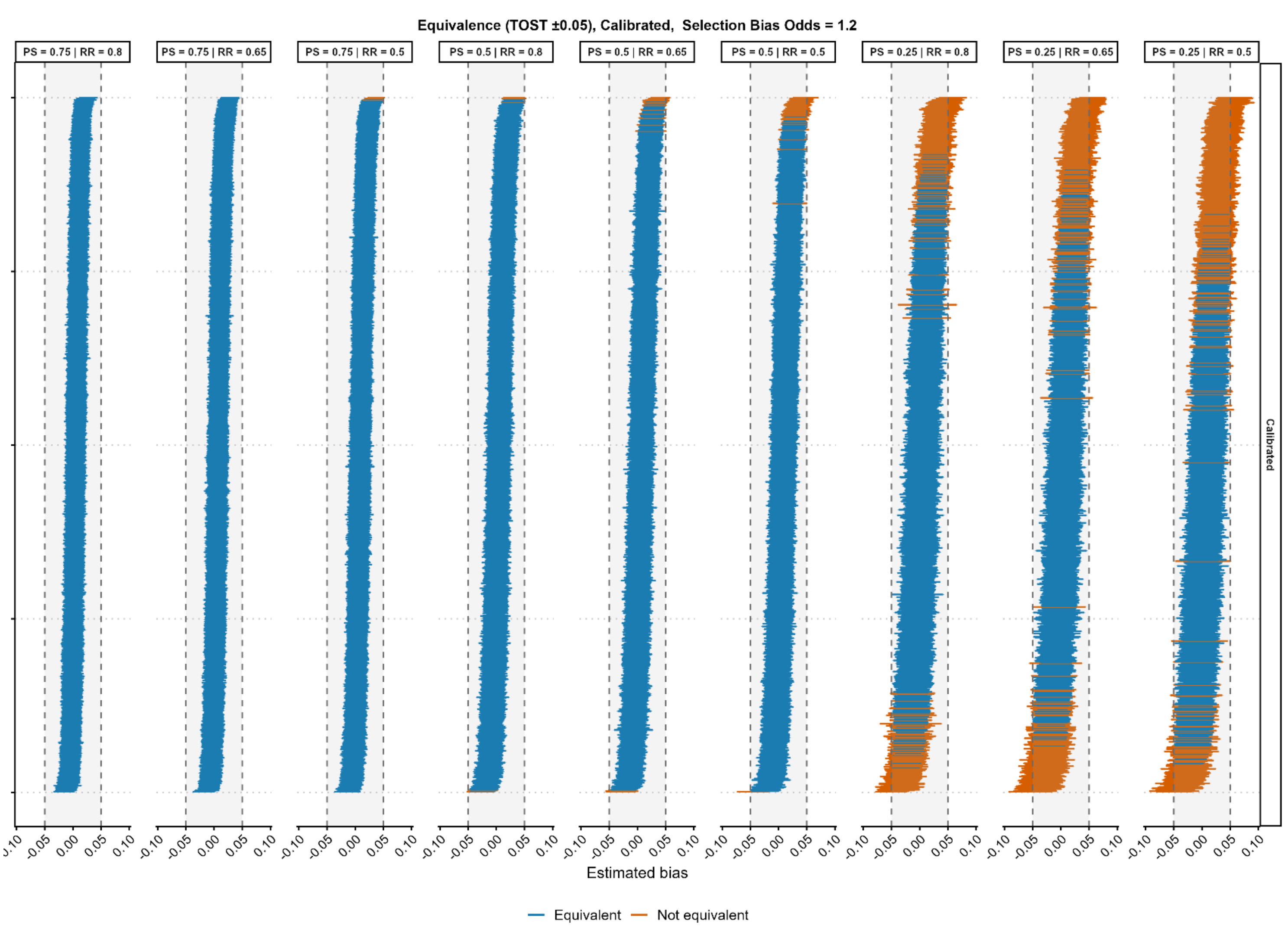

## *TOST equivalence ±0.05 confidence intervals, selection bias odds of 1.3, Calibrated Weights*

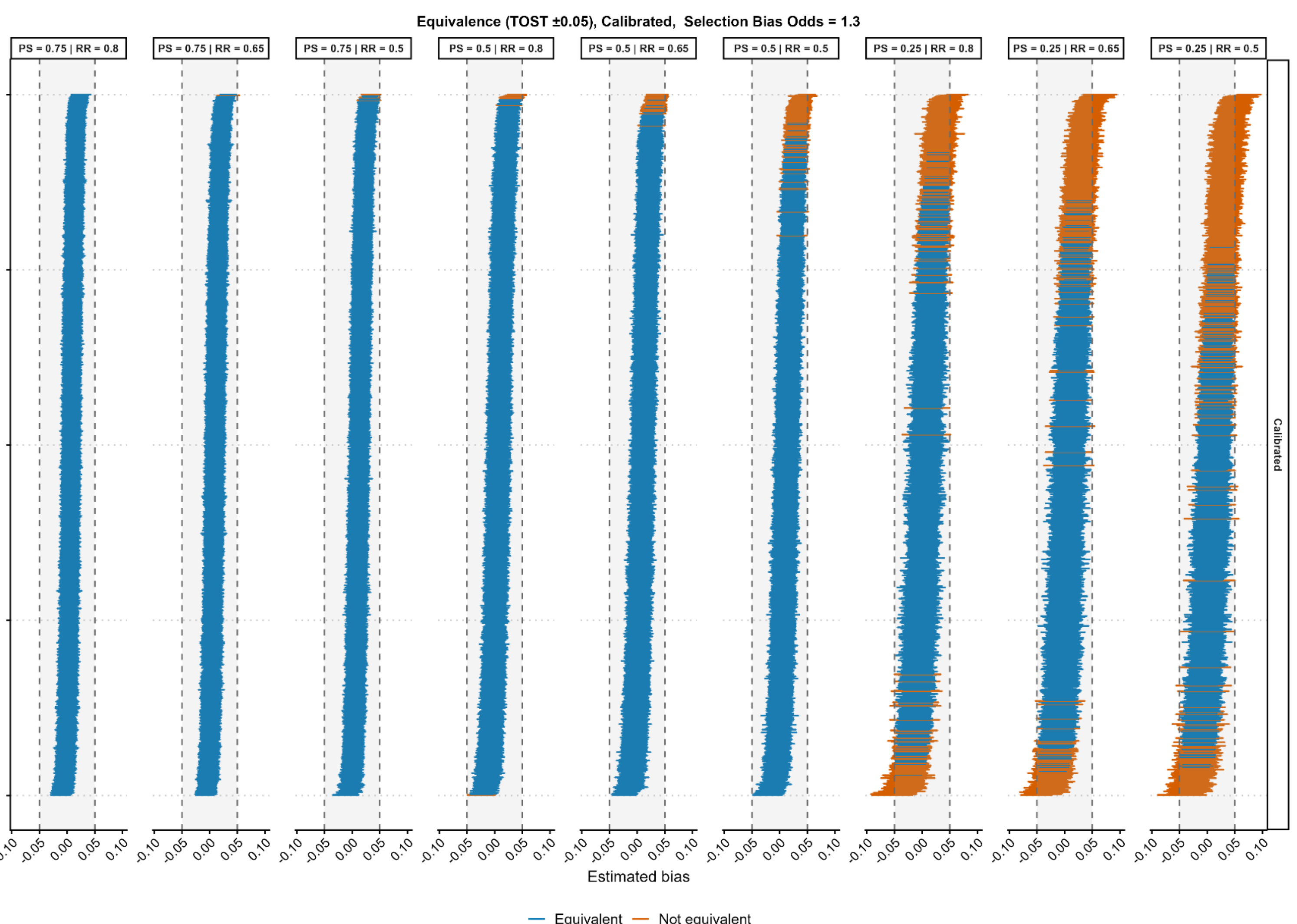

* Intervals ordered negative to positive bias. Orange intervals do not fall within the tolerance region. PS = Percentage of villages sampled, RR = Response rate.

## *TOST equivalence ±0.05 confidence intervals, selection bias odds of 1.4, Calibrated Weights*

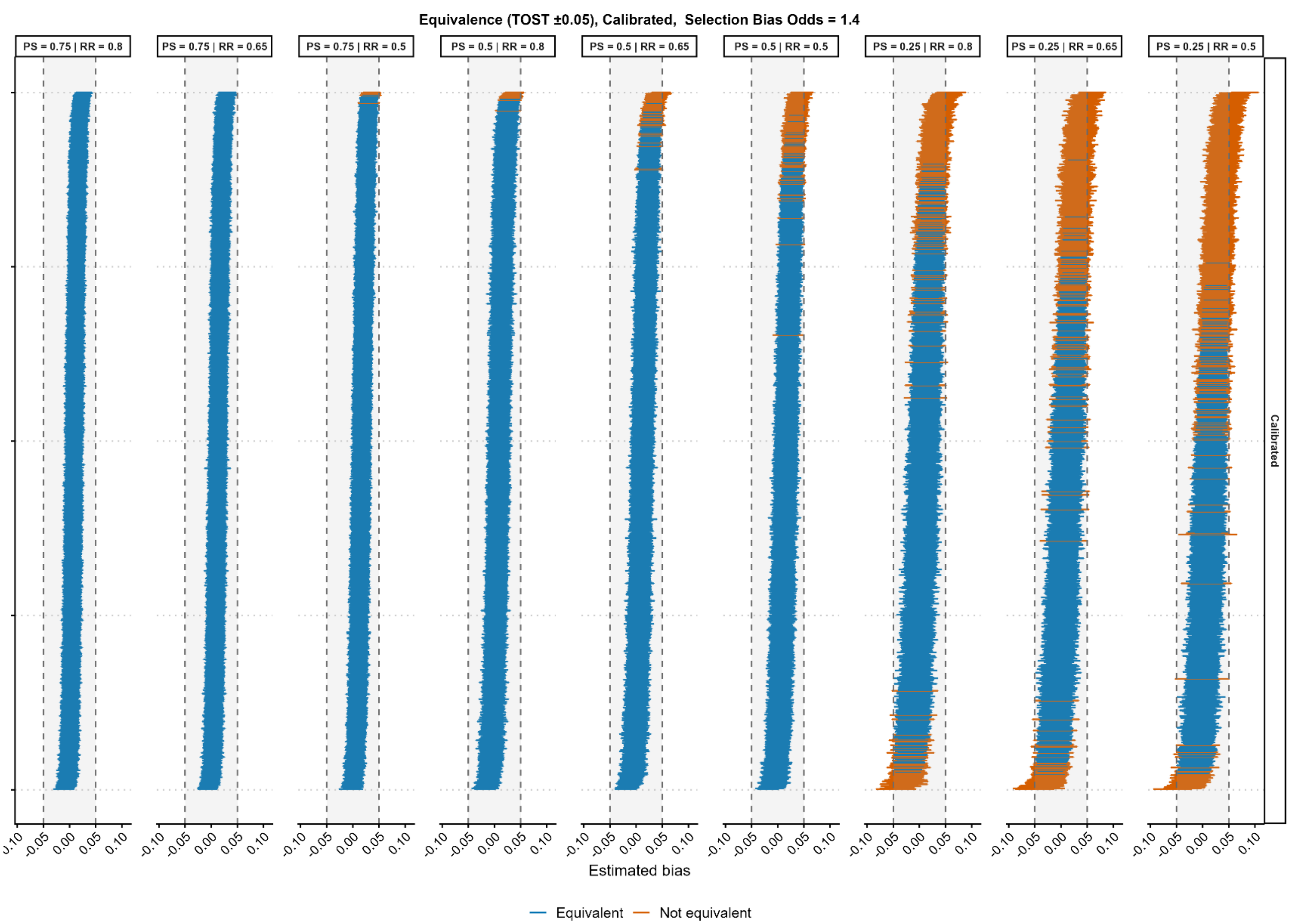

* Intervals ordered negative to positive bias. Orange intervals do not fall within the tolerance region. PS = Percentage of villages sampled, RR = Response rate.

***TOST equivalence ±0.05 confidence intervals, selection bias odds of 1.5, Calibrated Weights***

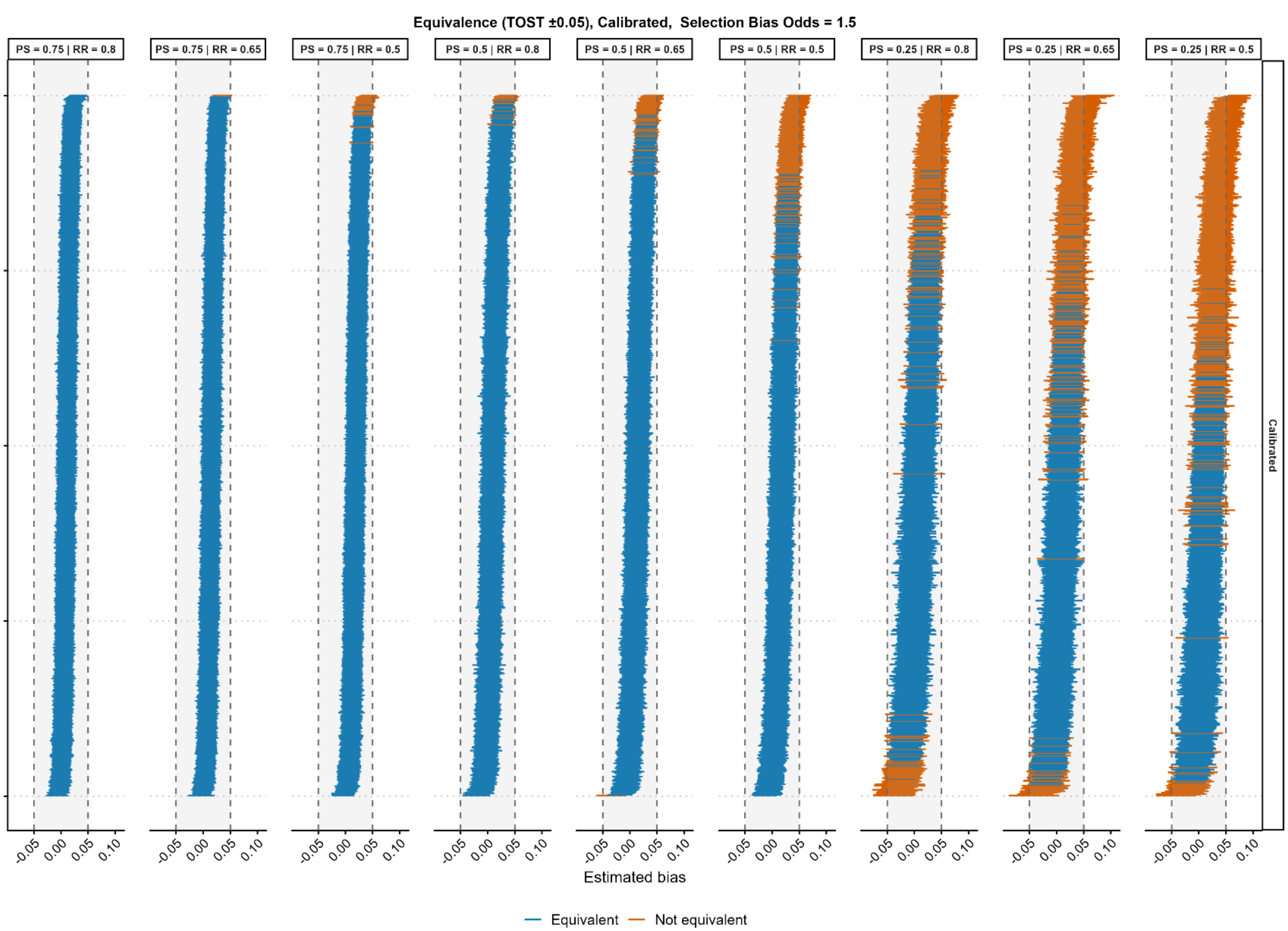

* Intervals ordered negative to positive bias. Orange intervals do not fall within the tolerance region. PS = Percentage of villages sampled, RR = Response rate.

### *TOST equivalence ±0.075 confidence intervals, ignorable selection, Logistic Regression*

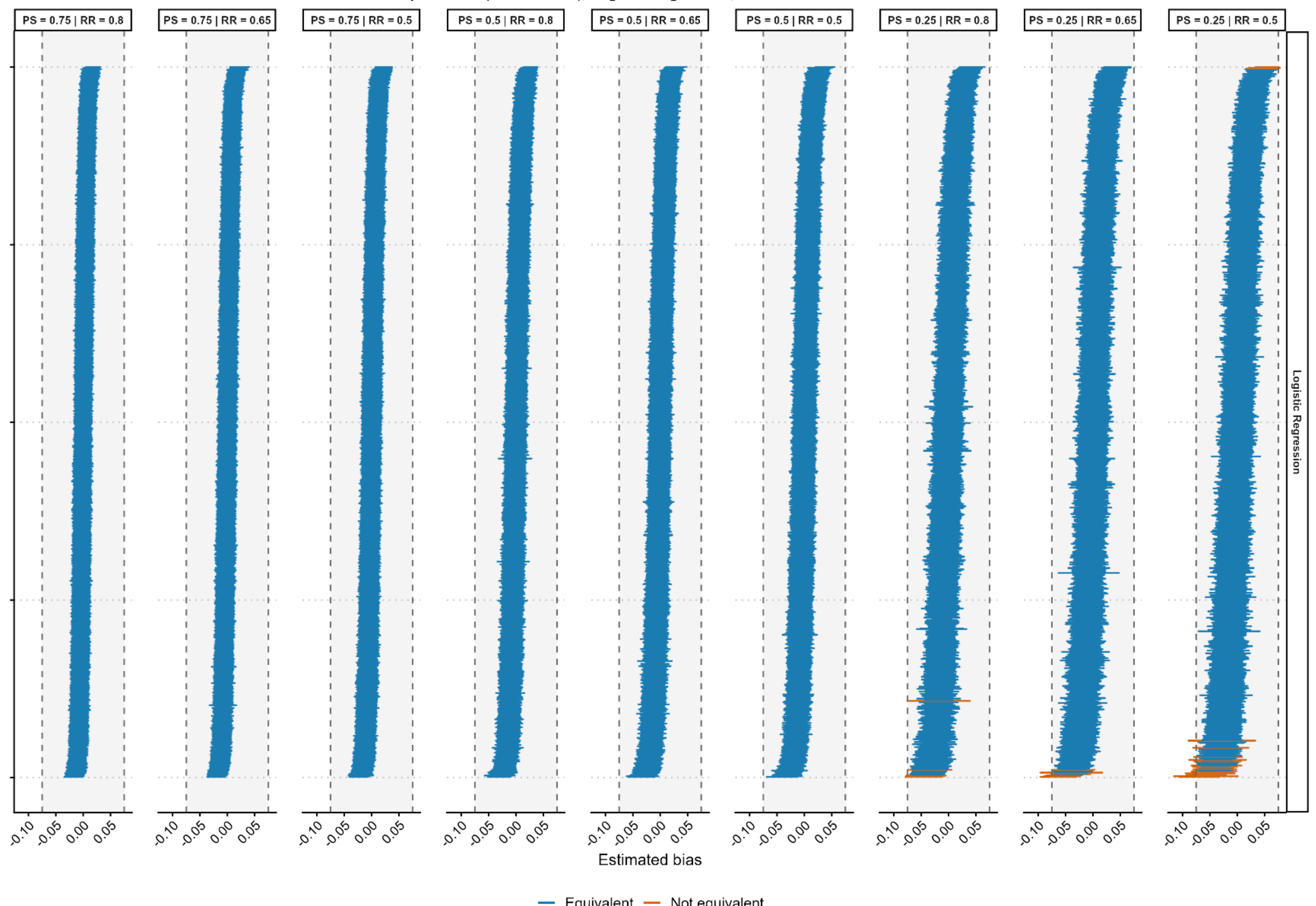

* Intervals ordered thinnest at the bottom to widest at the top. Orange intervals do not cover the bias of 0. PS = Percentage of villages sampled, RR = Response rate.

***TOST equivalence ±0.075 confidence intervals, selection bias odds of 1.1, Logistic Regression***

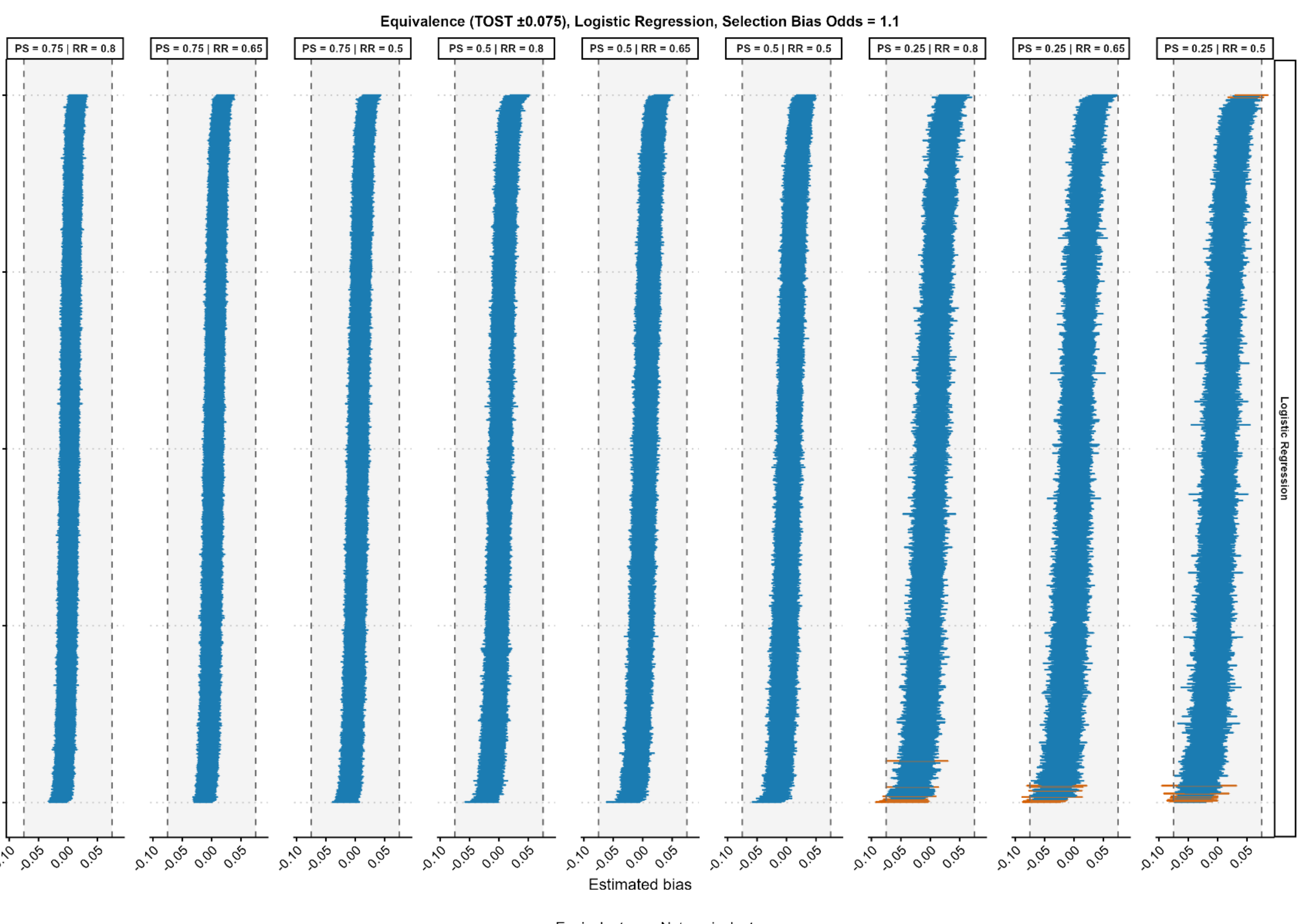

* Intervals ordered thinnest at the bottom to widest at the top. Orange intervals do not cover the bias of 0. PS = Percentage of villages sampled, RR = Response rate.

***TOST equivalence ±0.075 confidence intervals, selection bias odds of 1.2, Logistic Regression***

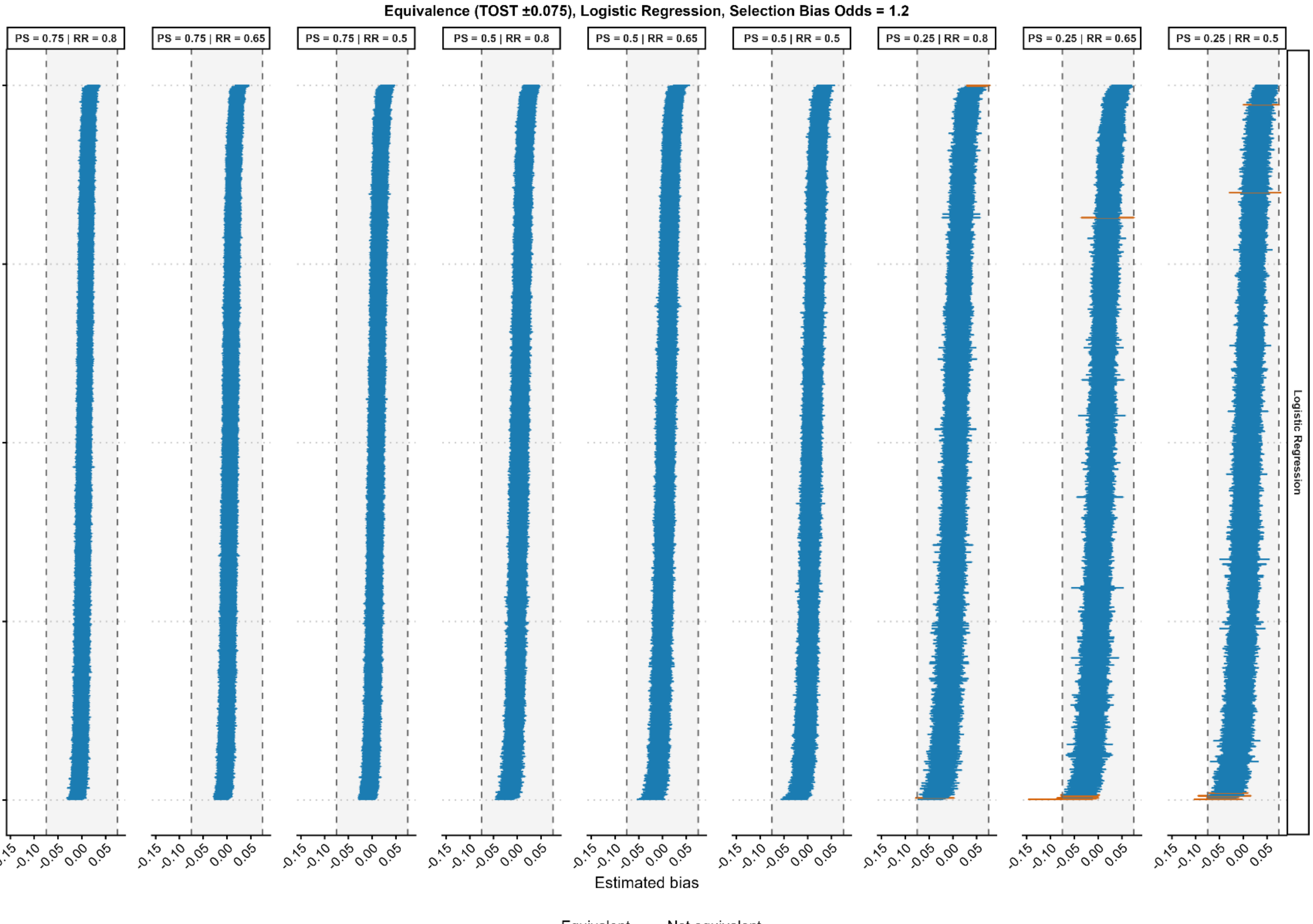

***TOST equivalence ±0.075 confidence intervals, selection bias odds of 1.3, Logistic Regression***

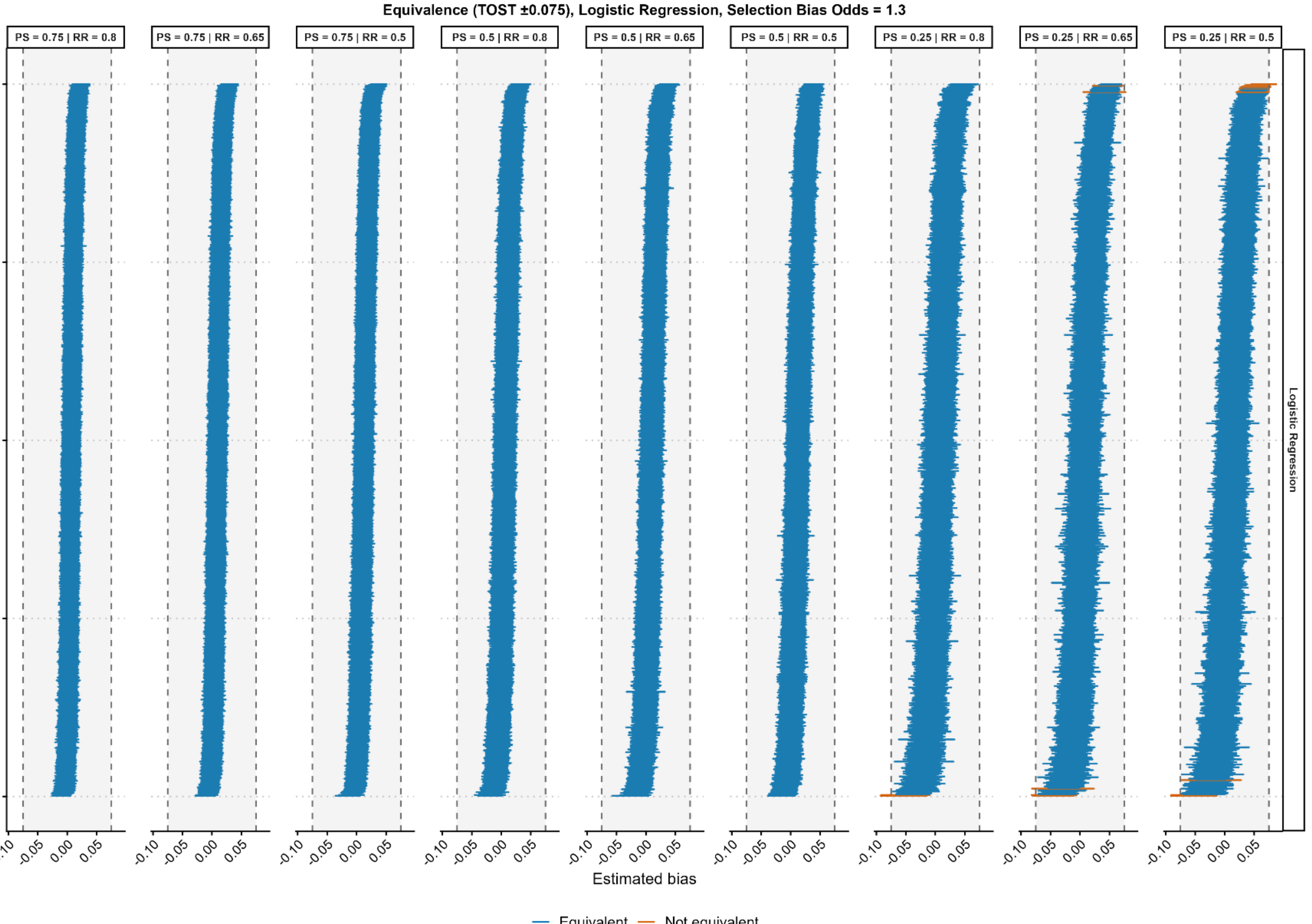

* Intervals ordered thinnest at the bottom to widest at the top. Orange intervals do not cover the bias of 0. PS = Percentage of villages sampled, RR = Response rate.

## *TOST equivalence ±0.075 confidence intervals, selection bias odds of 1.4, Logistic Regression*

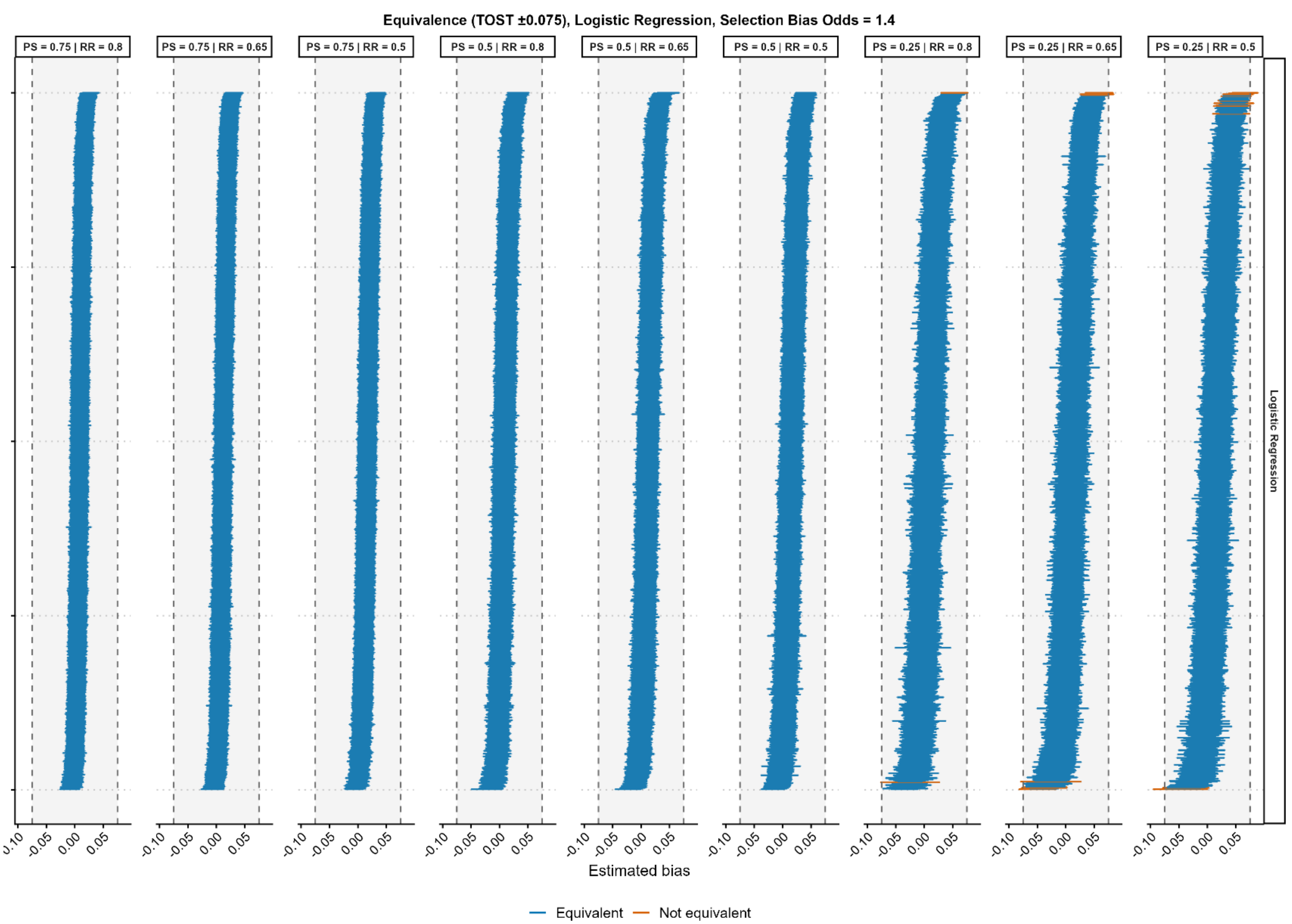

* Intervals ordered thinnest at the bottom to widest at the top. Orange intervals do not cover the bias of 0. PS = Percentage of villages sampled, RR = Response rate.

***TOST equivalence ±0.075 confidence intervals, selection bias odds of 1.5, Logistic Regression***

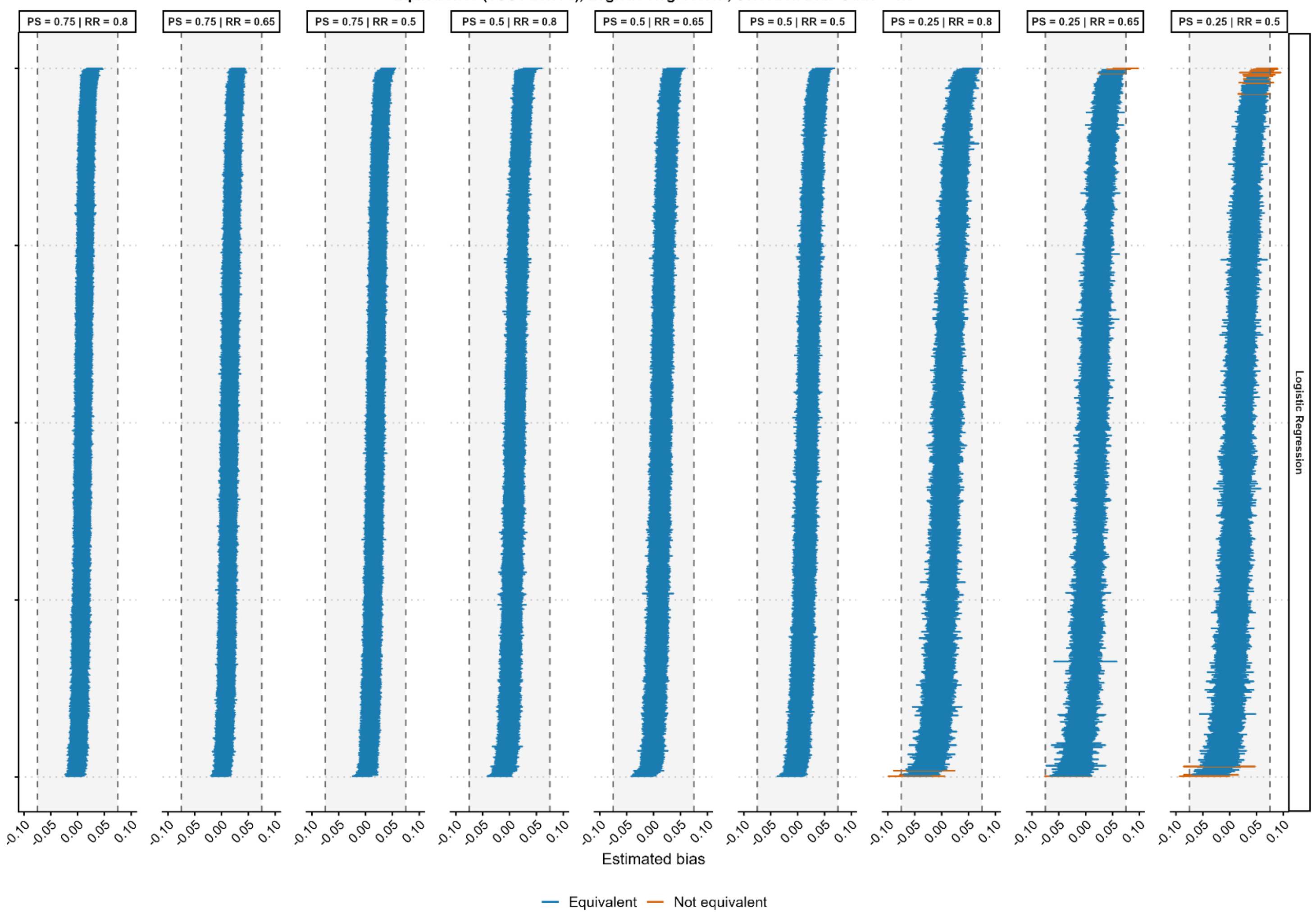

*TOST equivalence ±0.075 confidence intervals, ignorable selection, Calibrated Weights*

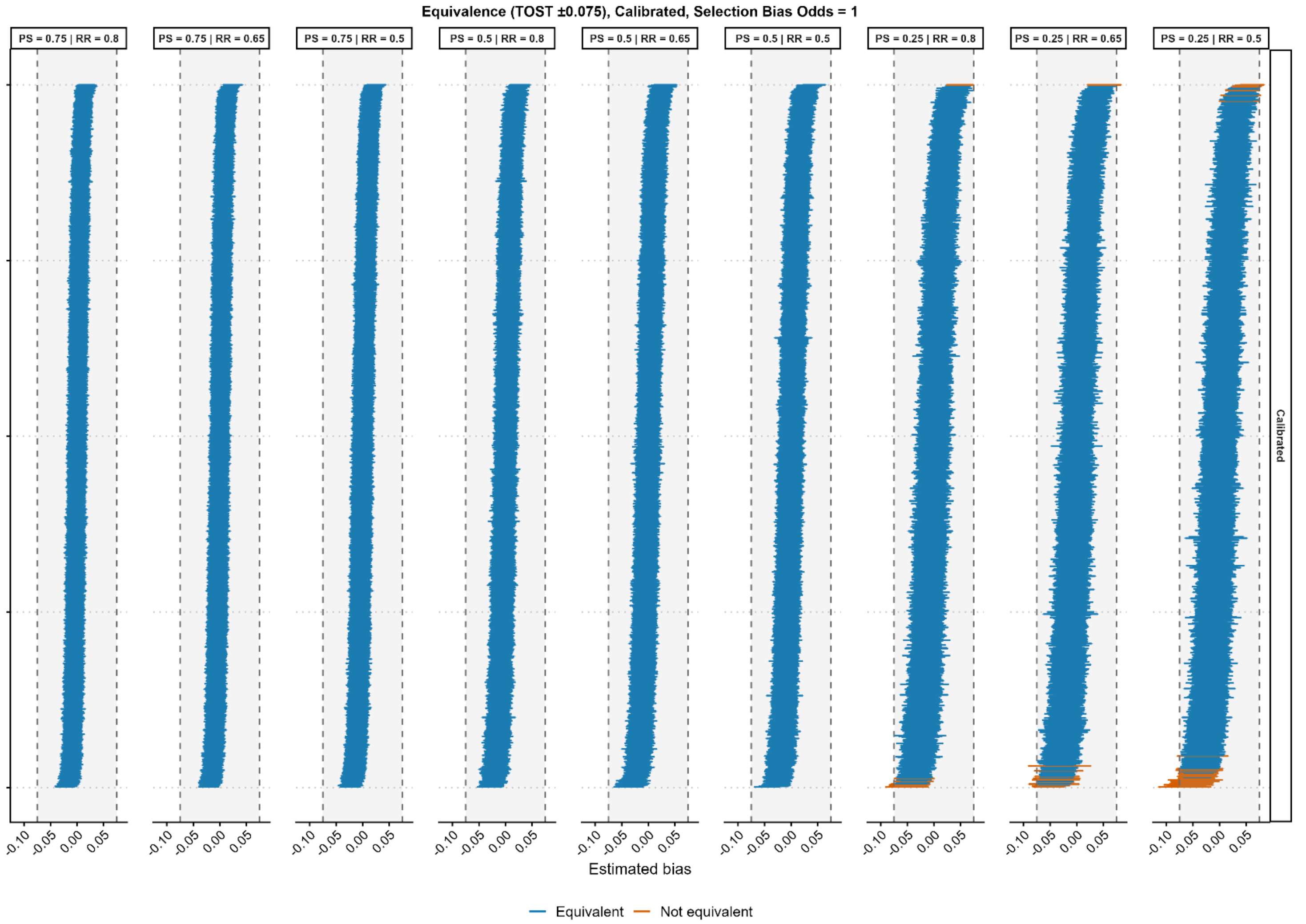

***TOST equivalence ±0.075 confidence intervals, selection odds of 1.1, Calibrated Weights***

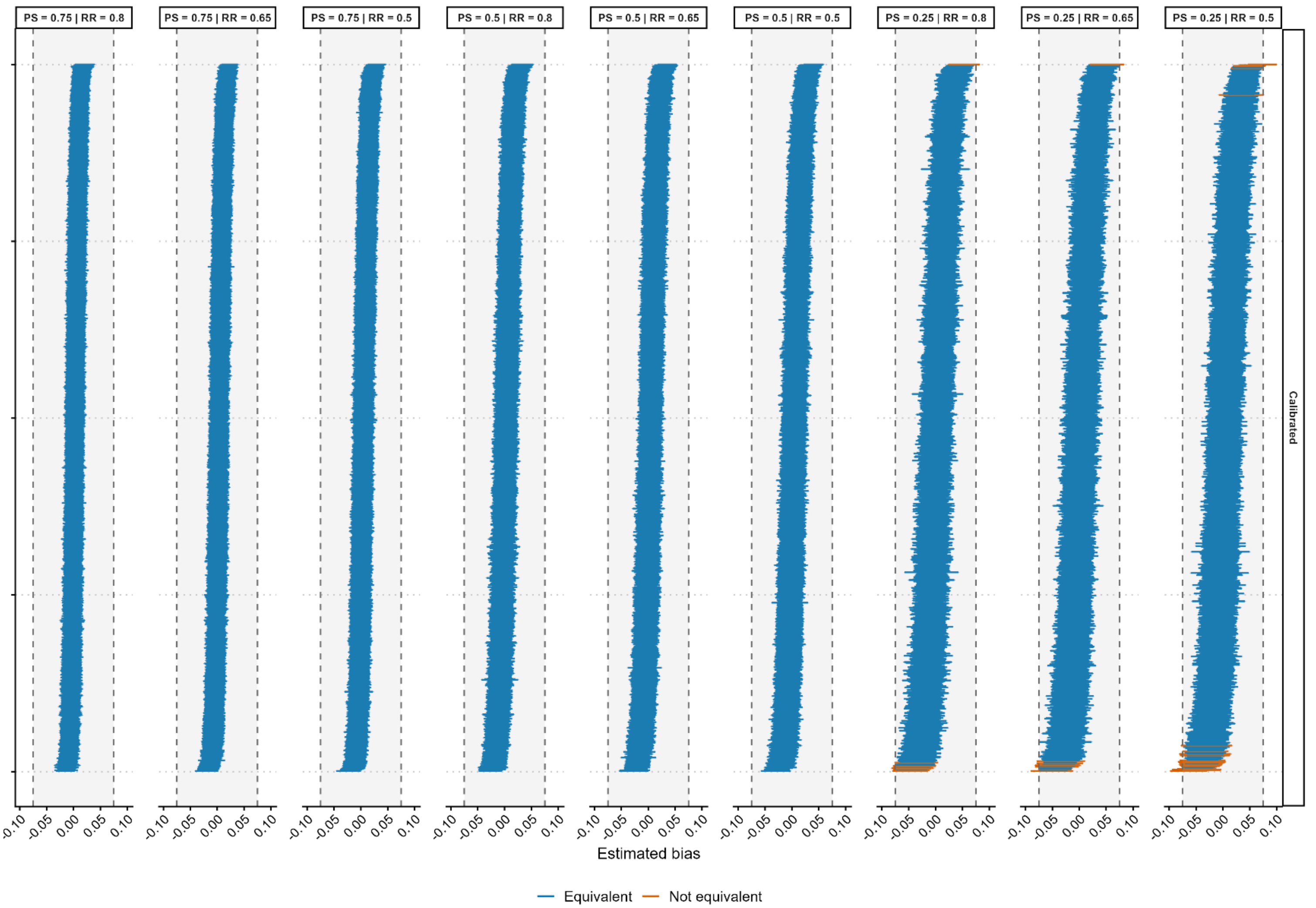

* Intervals ordered thinnest at the bottom to widest at the top. Orange intervals do not cover the bias of 0. PS = Percentage of villages sampled, RR = Response rate.

***TOST equivalence ±0.075 confidence intervals, selection odds of 1.2, Calibrated Weights***

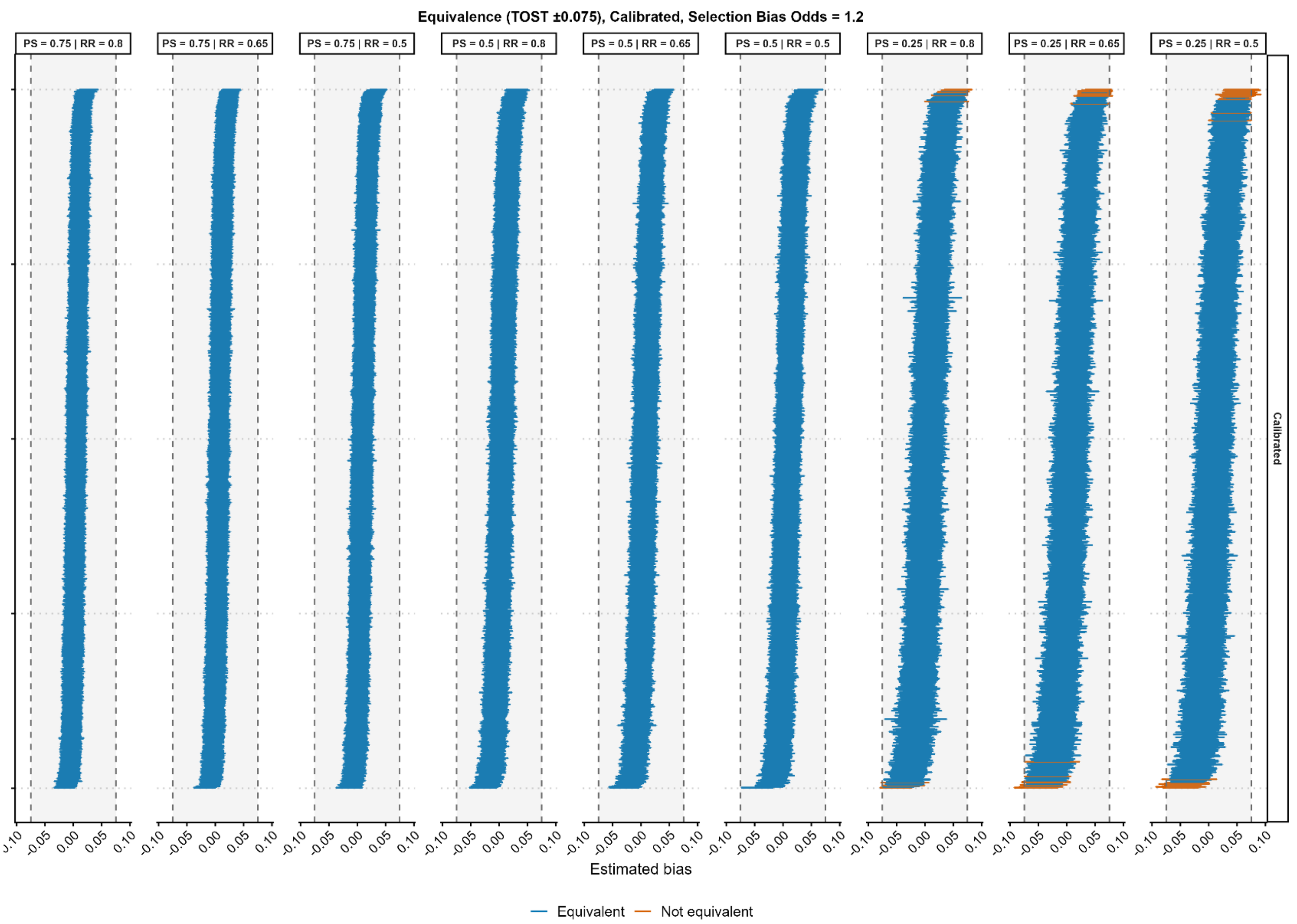

## *TOST equivalence ±0.075 confidence intervals, selection odds of 1.3, Calibrated Weights*

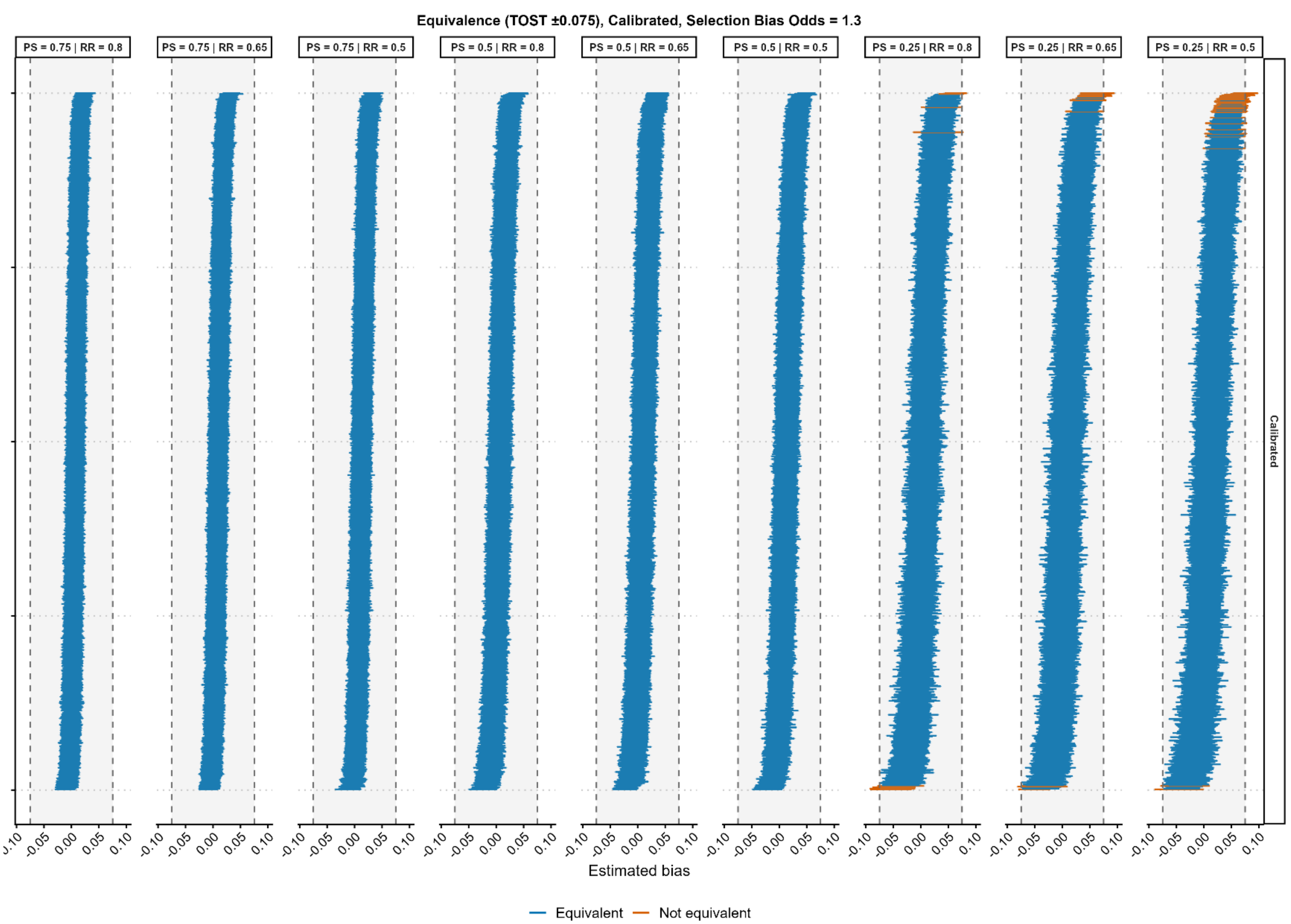

***TOST equivalence ±0.075 confidence intervals, selection odds of 1.4, Calibrated Weights***

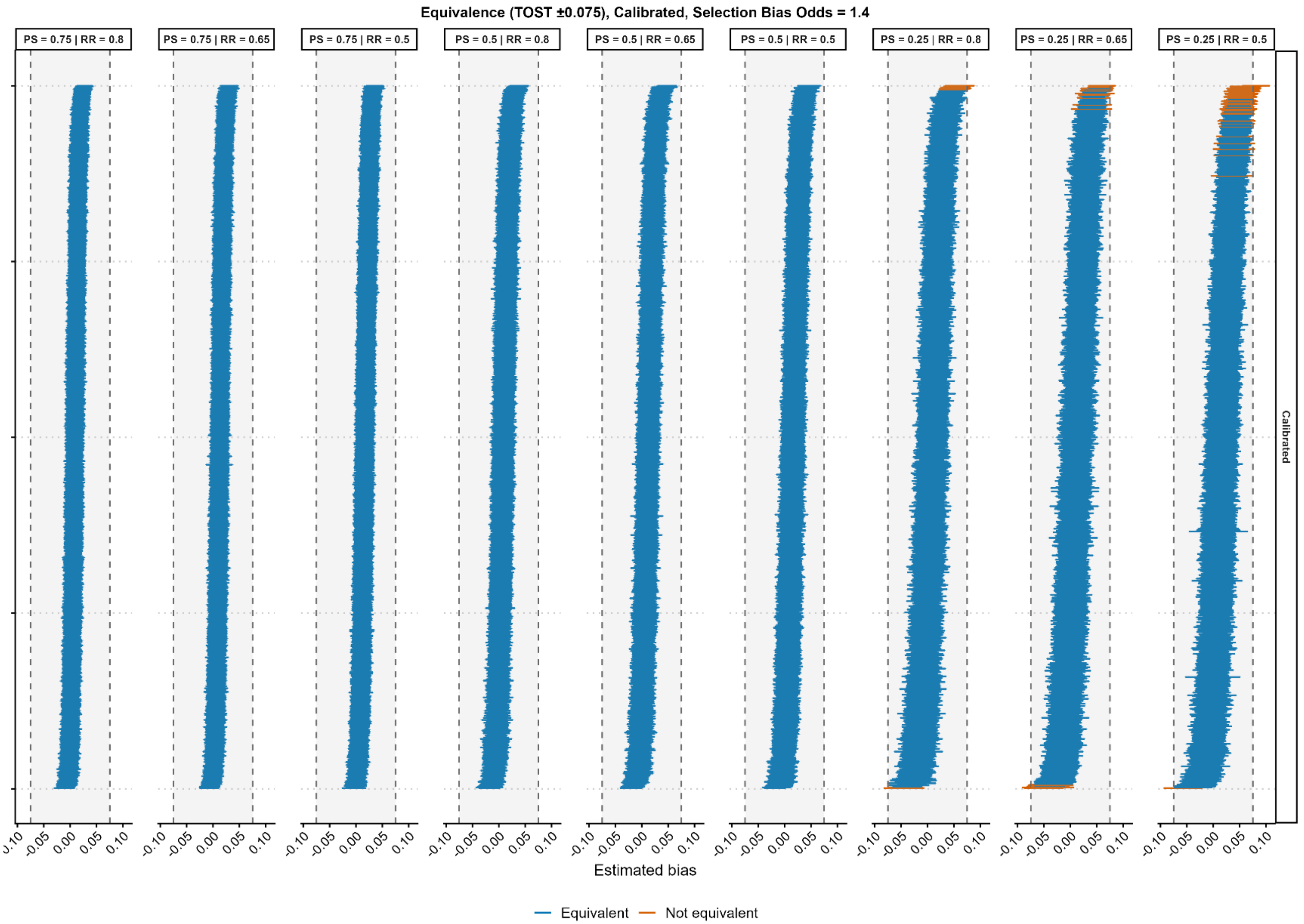

### *TOST equivalence ±0.075 confidence intervals, selection odds of 1.5, Calibrated Weights*

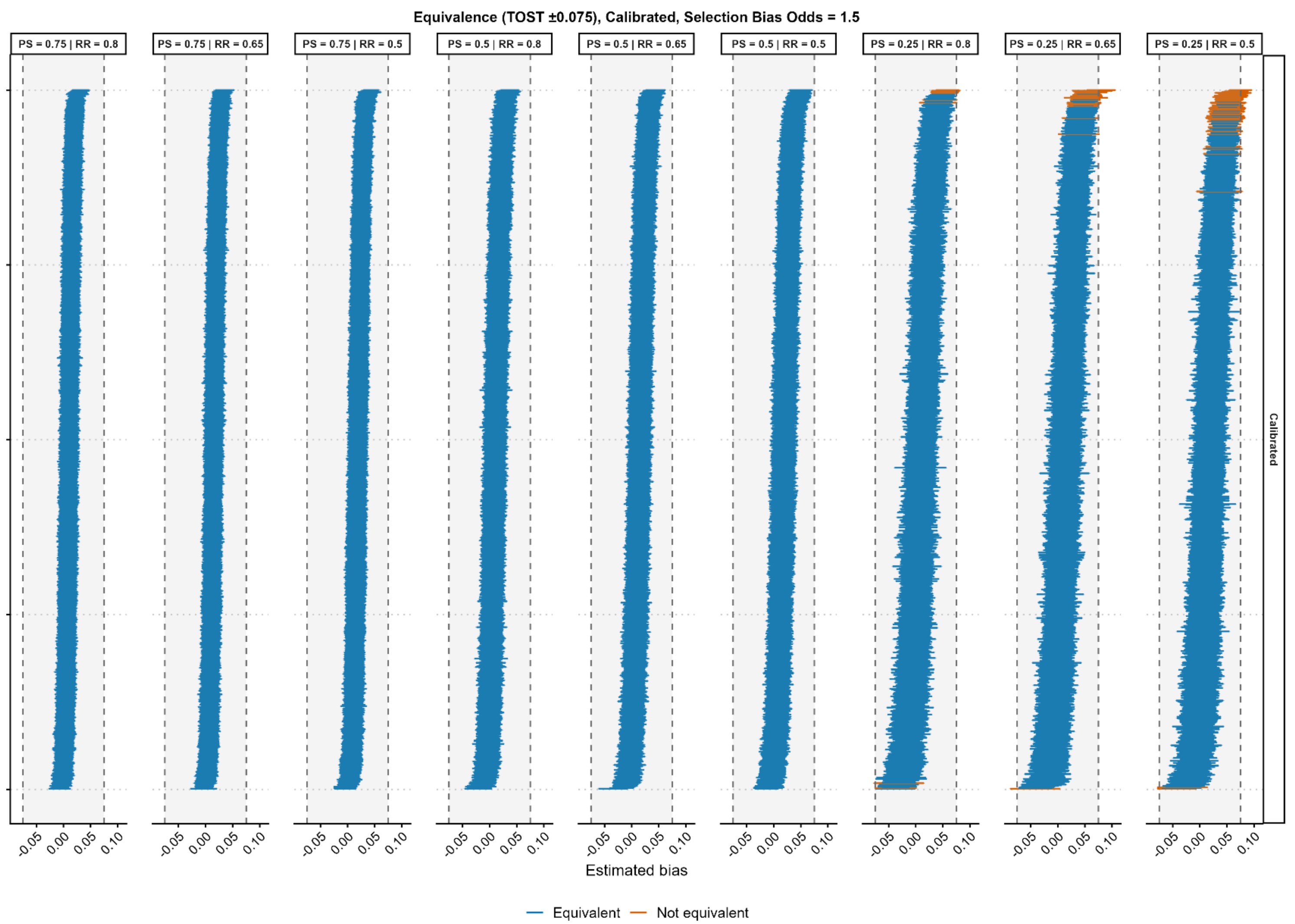

* Intervals ordered thinnest at the bottom to widest at the top. Orange intervals do not cover the bias of 0. PS = Percentage of villages sampled, RR = Response rate.

**Simulation Protocol for Anchoring Convenience Survey Samples to a Baseline Census for Vaccine Coverage Monitoring in Global Health**

Prepared by Core Clinical Sciences

| Simulation Protocol Version | Version 2.0 |
| --- | --- |
| Simulation Protocol Date | Jan 16 2025 |

**Abbreviations**

| Abbreviation | Definition |
|---|---|
| CCS | Core Clinical Sciences |
| ALIMA | Alliance of International Medical Action |
| MCSE | Monte Carlo Standard Error |
| DGM | Data generating Mechanism |
| MCV1 | Measles Containing Vaccine (at least one dose) |

# 1. Introduction

## 1.1 Purpose of this document

This document provides a comprehensive description of the simulation design and implementation to evaluate the feasibility of utilizing convenience samples to assess vaccine coverage in accordance with the ADEMP protocol (Aims, Data-generating Mechanisms, Estimands, Methods, and Performance measures) [1].

## 1.2 Project description

The goal of this project is to assess the feasibility of using a convenience sample to monitor the MCV1 (at least one dose of the Measles Containing Vaccine) vaccination rates in two rural regions: Mirriah, Niger and Ngouri, Chad. The convenience sample will be anchored by a baseline census that includes all relevant demographic variables and an initial baseline vaccination rate. If we can effectively correct the estimates of the convenience sample from a past baseline census survey, then a cost-efficient vaccine coverage method can be established. We assess two main sample survey correction methods, assuming ignorable selection bias. We also assess these methods against plausible levels of non-ignorable selection bias.

# 2. Aims

The aim of the simulation study is to assess the feasibility and performance of using a follow-up convenience survey to measure the vaccination rate when anchored by a baseline census under various levels of non-ignorable selection bias.

# 3. Data-Generating Mechanism

## 3.1 Data Generation

We will use the results of the baseline survey as the anchor for our estimates of the follow-up convenience survey(s).

Let $Y_{ij}^{t}$ be a binary variable indicating if child $i$ is vaccinated in village $j$ at time $t$. At baseline $(t = 0)$, the total number of vaccinated children in each village is known with certainty.

## 3.2 Village level covariates

Let $pop_j$ be the total population of village $j$, and $d_j$ be the distance of village $j$ to a medical centre.

## 3.3 Individual level covariates

- $cg_{ij}$ is the gender of child $i$ in village $j$ (1 = male)

- $ca_{ij}$ is the age of child $i$ in village $j$ (in months)
- $aa_{ij}$ is the age of the care giving adult of child $i$ in village $j$ (in years)
- $ag_{ij}$ is the gender of the care giving adult of child $i$ in village $j$ (1 = male)

Our data generating mechanism is as follows:

$Y_{ij}^{t} \sim Bernoulli(p_{ij}^{t})$ ,

then our followup survey is generated by

$$\text{logit}(p_{ij}^{1}) = \beta_0 + \beta_1 pop_j + \beta_2 d_j + \beta_3 ca_{ij} + \beta_4 aa_{ij} + \beta_5 cg_{ij} + \beta_6 ag_{ij} + logit(p_0) + \alpha_j,$$

with $\alpha_j \sim N(0, \sigma^2)$ being a cluster specific random effect. The Intra-cluster correlation (ICC) for the village level for binary outcome provided by [2] with the following formula: $\text{ICC}_v = \frac{\sigma^2}{\sigma^2 + \frac{\pi^2}{3}}$, where. $\pi^2/3$ is the constant variance of the logistic function.

### *3.4 Selection Bias*

We will use the results of the baseline survey as the anchor for our estimates of the selection bias of the follow-up convenience survey(s). Define $R_{ij} = 1$ as the indicator variable for whether child $i$ in village $j$ is included in the sample. Then $R_{ij} \sim Bernoulli(p_{\text{inc}\ ij})$, and the selection is generated from the following logistic mixed effects model.

$$\text{logit}(p_{\text{inc}\ ij}) = \gamma_0 + \gamma_1 pop_j + \gamma_2 d_j + \gamma_3 ca_{ij} + \gamma_4 aa_{ij} + \gamma_5 cg_{ij} + \gamma_6 ag_{ij} + \mu_j + Y_{ij}^{1} \log(\xi).$$

with $\mu_j \sim N(0, \epsilon^2)$ is a cluster random effect. $\xi$ represents the degree of non-ignorable selection bias induced in the sample. In the ignorable case, $\xi = 1$.

### *3.5 Factors of the DGM*

We define the variable `didAttendPrevious` as whether a guardian in the census baseline survey answers yes to either question a04 or a06b, which represent whether they attended a previous MUAC training as informed by the ALIMA team.

The key factors used in the data-generating mechanism are summarized below:

| Factor | Value | Justification |
|---|---|---|
| Effect of distance of village to medical centre | $\beta_1$ = to be determined | Obtained by performing logistic regression on the proportion of children aged 12-24 months with a vaccine |

| Factor | Value | Justification |
|---|---|---|
| Effect of total population of village | $\beta_2$ to be determined | Obtained by performing mixed-effects logistic regression on the proportion of children aged 12-24 months with a vaccine |
| Effect of the child's gender | $\beta_3$ to be determined | Obtained by performing mixed-effects logistic regression on the proportion of children aged 12-24 months with a vaccine |
| Effect of the child's age | $\beta_4$ to be determined | Obtained by performing mixed-effects logistic regression on the proportion of children aged 12-24 months with a vaccine |
| Effect of the guardian's age | $\beta_5$ to be determined | Obtained by performing mixed-effects logistic regression on the proportion of children aged 12-24 months with a vaccine |
| Effect of the guardian's gender | $\beta_6$ to be determined | Obtained by performing mixed-effects logistic regression on the proportion of children aged 12-24 months with a vaccine |
| Intra-cluster correlation (Vaccination) | $ICC_v$ | Informed by the baseline census data |
| Effect of distance of village to medical centre | $\gamma_1$ = to be determined | Obtained by performing logistic regression on the proportion of children aged 12-24 months who attended the previous training |

| Factor | Value | Justification |
|---|---|---|
| Selection effect of total population of village | $\gamma_2$ to be determined | Obtained by performing logistic regression on the proportion of children aged 12-24 months who attended the previous training |
| Selection effect of the child's gender | $\gamma_3$ to be determined | Obtained by performing logistic regression on the proportion of children aged 12-24 months who attended the previous training |
| Selection effect of the child's age | $\gamma_4$ to be determined | Obtained by performing logistic regression on the proportion of children aged 12-24 months who attended the previous training |
| Selection effect of the guardian's age | $\gamma_5$ to be determined | Obtained by performing logistic regression on the proportion of children aged 12-24 months who attended the previous training |
| Selection effect of the guardian's gender | $\gamma_6$ to be determined | Obtained by performing logistic regression on the proportion of children aged 12-24 months who attended the previous training |
| Selection Intra-cluster correlation | $ICC_s = \frac{1}{3}$ | WHO recommendation for planning vaccination surveys [3] |
| Non-ignorable selection bias | $\xi \in \{1,1.1,1.2,1.3,1.4,1.5\}$ | Informed by the ALIMA team |
| Proportion of of villages sampled | 0.25,0.50,0.75 | informed via expert opinion |
| Response Rate | 0.5,0.65,0.80 | informed via expert opinion |

** $\gamma_0$ will be tuned to achieve relevant response rates.

Based on all the possible number of parameters, there will be a total of 54 possible selections, and for each selection there will have 1,000 repetitions.

## 4. Estimands and Targets

Our estimand of interest is $p^1$ or, moving forward, simply $p$ which is the finite proportion of children aged 12-24 months that are vaccinated with MCV1 at the time of the convenience survey in the finite-population of Mirrah, Niger and Ngouri, Chad.

## 5. Methods

We consider several candidate models for adjusting selection bias.

### *5.1 Calibration Weights*

Our first and primary candidate method is to use a propensity score approach (design-based) by adjusting the sampling weights $d_{ij}$ according to the population totals. This approach is referred to as calibrating estimation [4]. We first assume that the first stage (villages) is sampled via a simple random sample, such that $d_j = \frac{m}{M}$, where $M$ is the total number of villages, and $m$ is the number of villages sampled, thus $w_j = \frac{1}{d_j}$. We also assume (incorrectly) that the second stage is sampled via a simple random sample of size $v_j$ out of $V_j$ in each village $j$, thus $w_{i|j} = \frac{V_j}{v_j}$, and $d_{ij} = d_{i|j} d_j$.

If we know the population totals $t_{\mathbf{x}}$ for all the relevant covariates $\mathbf{x}_{ij}$, we can adjust the design weights $d_{ij}$ such that the new weights $w_{ij}$ match the population totals: $\sum_j \sum_i w_{ij}\, \mathbf{x}_{ij} = t_{\mathbf{x}}$. This is done by a distance minimizing procedure between $w_{ij}$ and $d_{ij}$. The details can be found in [4], and the survey package [5] handles this internally.

### *5.2 Model based correction*

Let $R_i$ be an indicator variable with $R_i = 1$ representing whether individual $i$ is included in the sample. Here we take a model based approach imputation approach described in Wu (2022) and Kim et al., (2021) [6] [7]. This involves fitting a model to predict or impute the vaccination statuses of the census. We use a logistic regression to model the relationship between the demographic covariates and the vaccine statuses.

Our estimate of the vaccine status is then

$$E[Y] = E\big[E[Y_i|X_i]\big] \approx E\big[E\widehat{[Y_i|X_i]}\big] \approx \frac{1}{N}\sum_{i=1}^{N} E\widehat{[Y_i|X_i]},$$

where $E\widehat{[Y_i|X_i]}$ is the estimate from the logistic regression with $R_i = 1$. The cluster-robust variance-covariance matrix from the sandwich package is used to account for clustering [8]. The variance will be estimated using the delta method. This involves taking the super-population approach where the finite population proportion $p$ is a random variance and includes additional variation; see Chapter 5 of Wu & Thompson (2020) [9].

## 6. Performance Measures

### *6.1 Performance Measures of Interest*

We are interested in 3 performance measures for each model: Bias, Coverage, and Equivalence. Let $n_{rep}$ be the number of repetitions done for the Monte-Carlo simulation. For each of the performance metrics we will assess and calculate the monte carlo uncertainty.

### *6.2 Bias*

The bias is estimated as the expected difference between the estimated proportion of vaccination of children aged 12-24 months ($\hat{p}_k$) and the true proportion ($p_k$). Let $k$ index the simulation, then our estimated bias is

$$\widehat{Bias} = \frac{\sum_{i=1\ rep}^{n} \hat{p}_k}{n_{rep}} - p_k.$$

The Monte Carlo Standard Error (MCSE) for the bias follows the formula [1]. We compute the MCSE of the estimated bias given by

$$MCSE_{\widehat{Bias}} = \sqrt{\frac{1}{n_{rep}\left(n_{rep} - 1\right)} \sum_{s=1}^{n} (\hat{p}_k - \bar{p})^2},$$

where $\bar{p} = \frac{\sum_{k=1}^{n_{rep}} \hat{p}}{n_{rep}}$. A small adjustment can be made to if $p_k$ changes for each $k$. Set $d_k = \hat{p}_k - p_k$, then $MCSE_{\widehat{Bias}} = \sqrt{\frac{1}{n_{rep}\left(n_{rep}-1\right)} \sum_{k=1}^{n_{rep}} \left(\hat{d}_k - \bar{d}\right)^2}$.

### *6.3 Coverage*

Our second performance measure is the coverage of the vaccination proportion estimator. Let the 95% confidence interval for the vaccinated proportion be $\hat{p}_{0.025}$, $\hat{p}_{0.975}$, then our coverage estimate is

$$\widehat{Coverage} = \frac{\sum_{k=1}^{n_{\text{rep}}} 1\left(\hat{p}_{0.025k} \le p \le \hat{p}_{0.975k}\right)}{n_{\text{rep}}}.$$

The Monte Carlo standard error for the coverage is given by [1] ’

$$MCSE_{\widehat{Coverage}} = \sqrt{\frac{\widehat{Coverage}\left(1 - \widehat{Coverage}\right)}{n_{rep}}}.$$

### *6.4 Equivalence Tests*

Our last Performance metric is an equivalence test This is equivalent to two one-sided tests at $\alpha = 0.05$, with the specified equivalence margin of $\delta = 0.05{,}0.075$.

$$\widehat{\text{Equivalence}} = \frac{\sum_{k=1}^{n_{\text{rep}}} 1\left(\{\hat{p}_{0.05k}, \hat{p}_{0.95k}\} \in \{p_k - \delta, p_k + \delta\}\right)}{n_{\text{rep}}}$$

The Monte Carlo standard error of the equivalence test is the same as above for the coverage

$$MCSE_{\widehat{\text{Equivalence}}} = \sqrt{\frac{\widehat{\text{Equivalence}}\left(1 - \widehat{\text{Equivalence}}\right)}{n_{rep}}}$$

### *6.5 Other*

### *6.6 Statistical software and packages*

We used the sandwich package for the cluster robust variance co-variance matrix [8], the survey package for the calibrated weights [10], and lme4 for fitting logistic mixed effects model [11].